\documentclass{aastex63}
\usepackage[encapsulated]{CJK}
\usepackage{threeparttable}
\usepackage{mathrsfs}
\usepackage{epstopdf}
\usepackage{amsmath}
\usepackage{multirow}
\usepackage{multirow}
\usepackage{hyperref}
\usepackage{subfigure}
\usepackage{rotating}

\newcommand{\co}{$^{12}$CO }        
\newcommand{\xco}{$^{13}$CO }       
\newcommand{\xxco}{C$^{18}$O }      
\newcommand{\hco}{HCO$^{+}$ }       
\newcommand{\jx}{($J = 1 - 0$)}    
\newcommand{\jxx}{($J = 2 - 1$)}   
\newcommand{\jxs}{($J = 1 - 0$) }    
\newcommand{\kms}{km s$^{-1}$ }     
\newcommand{\kmss}{km s$^{-1}$} 
\newcommand{\msun}{$M_{\odot}$ }

\newcommand{\coo}{$^{12}$CO}        
\newcommand{\xcoo}{$^{13}$CO}       
\newcommand{\xxcoo}{C$^{18}$O}      
\newcommand{\hcoo}{HCO$^{+}$}       

\shorttitle{Outflows in nine sources}
\shortauthors{Liu et al.}

\begin{document}
\begin{CJK*}{UTF8}{gbsn}

\title{High-sensitivity millimeter imaging of molecular outflows in nine nearby high-mass star-forming regions }

\correspondingauthor{Ye Xu}
\email{xuye@pmo.ac.cn}

\author{De-Jian Liu}
\affiliation{ College of Science, China Three Gorges University, Yichang 443002, Peopleʼs Republic of China;}
\affiliation{ Purple Mountain Observatory, Chinese Academy of Sciences, Nanjing 210023, Peopleʼs Republic of China;}

\author{Ye Xu}
\affiliation{ Purple Mountain Observatory, Chinese Academy of Sciences, Nanjing 210023, Peopleʼs Republic of China;}

\author{Ying-Jie Li}
\affiliation{ Purple Mountain Observatory, Chinese Academy of Sciences, Nanjing 210023, Peopleʼs Republic of China;}

\author{Sheng Zheng}
\affiliation{ College of Science, China Three Gorges University, Yichang 443002, Peopleʼs Republic of China;}

\author{Deng-Rong Lu}
\affiliation{ Purple Mountain Observatory, Chinese Academy of Sciences, Nanjing 210023, Peopleʼs Republic of China;}

\author{Chao-Jie Hao}
\affiliation{ Purple Mountain Observatory, Chinese Academy of Sciences, Nanjing 210023, Peopleʼs Republic of China;}
\affiliation{ School of Astronomy and Space Science, University of Science and Technology of China, Hefei 230026, Peopleʼs Republic of China;}

\author{Ze-Hao Lin}
\affiliation{ Purple Mountain Observatory, Chinese Academy of Sciences, Nanjing 210023, Peopleʼs Republic of China;}
\affiliation{ School of Astronomy and Space Science, University of Science and Technology of China, Hefei 230026, Peopleʼs Republic of China;}

\author{Shuai-Bo Bian}
\affiliation{ Purple Mountain Observatory, Chinese Academy of Sciences, Nanjing 210023, Peopleʼs Republic of China;}
\affiliation{ School of Astronomy and Space Science, University of Science and Technology of China, Hefei 230026, Peopleʼs Republic of China;}

\author{Li-Ming Liu}
\affiliation{ College of Science, China Three Gorges University, Yichang 443002, Peopleʼs Republic of China;}

\begin{abstract}

We present a study of molecular outflows using six molecular lines (including \coo/\xcoo/\xxcoo/\hco \jxs and CS/SiO \jxx) toward nine nearby high-mass star-forming regions with accurate known distances. This work is based on the high-sensitivity observations obtained with the 14-m millimeter telescope of Purple Mountain Observatory Delingha (PMODLH) observatory. The detection rate of outflows (including \coo, \xcoo, \hcoo, and CS) is 100\%. However, the emission of SiO was not detected for all sources. The full line widths ($\Delta V$) at 3$\sigma$ above the baseline of these molecular lines have the relationship $\Delta V_{\rm ^{12}CO} > \Delta V_{\rm HCO^{+}} > \Delta V_{\rm CS} \approx \Delta V_{\rm ^{13}CO} > \Delta V_{\rm ^{18}CO}$. \co and \hco can be used to trace relatively high-velocity outflows, while \xco and CS can be employed to trace relatively low-velocity outflows. The dynamical timescales of the \xco and CS outflows are longer than those of the \co and \hco outflows. The mechanical luminosities, masses, mass-loss rates and forces of all outflows (including \coo, \xcoo, \hcoo, and CS) are correlated with the bolometric luminosities of their central IRAS sources.

\end{abstract}


\section{Introduction} \label{sec:intro}
The physical processes associated with low-mass star formation are reasonably well understood \cite[e.g., ][]{Shu+etal+1987,McKee+Ostriker+2007,Kennicutt+Evans+2012}. However, details of the forming mechanism(s) of high-mass stars remain poorly understood, so creating a clear map of high-mass star formation is extremely challenging work. To achieve this goal, it is vital to study molecular outflows, which arise during an important phase in high-mass star formation and are a ubiquitous phenomenon during the earliest stage of formation for stars of all masses \citep{Beuther+etal+2002,Arce+etal+2007}. 

Since molecular outflows were first detected in Orion KL by \citet{Kwan+Scoville+1976}, outflows in high-mass star-forming regions (HMSFRs) have been investigated by many researchers \citep{Shepherd+Churchwell+1996,Zhang+etal+2001,Zhang+etal+2005,Xu+etal+2006b,Li+etal+2018,Li+etal+2019}. In each of these studies, typically only one outflow tracer was used. However, there is no perfect outflow tracer, and a complete characterization of outflow phenomena requires observations made with many tracers \citep{Bally+2016}. Hence, to provide a more comprehensive understanding of the physical properties of outflows, it is necessary to conduct systematic studies of different molecular outflows using multiple molecular tracers toward HMSFRs.  

Accurate distances over a relatively narrow distance range are important to statistically analyze the physical properties of outflows. Typical selection criteria in terms of distance have been commonly used: (1) the distances are measured by the trigonometric parallaxes of masers, which are typically more accurate than kinematic distances \citep{Xu+etal+2006a,Reid+etal+2019}, so more accurate physical quantities can be ascertained; (2) the sources are nearby and within a relatively narrow distance range (i.e., 0.5--1~kpc). In the context of PMODLH, the linear resolution can be better than 0.27~pc with beam size of $\sim 55''$. This resolution is usually sufficient to visually depict the morphology of outflows, where the typical length of an outflow is about 1~pc. 

This work is the first high-sensitivity systematic research of outflows, which was conducted using six molecular lines (i.e., SiO \jx, CS \jxx, \hco \jx, \co \jx, \xco \jx, and \xxco \jx) towards nine nearby HMSFRs (see Table~\ref{tab:list}). SiO was used as a shock tracer and the dense cores were mapped with \xxcoo. Meanwhile, the molecular outflows were imagined using \coo, \xcoo, \hcoo, and CS, which were utilized to trace areas of different densities in HMSFRs. These lines were observed by the same telescope with high-sensitivity observations and binned to the same pixel scales, which allowed us to make direct comparisons between the different line tracers. Additionally, the high-sensitivity observations were helpful for unvieling high-velocity outflow gas. 

The remainder of the paper is organized as follows. We present the observations and data-reduction techniques in Section~\ref{sect:Obs}. In Section~\ref{sect:Res}, we summarize the detected outflows and calculate the relevant physical quantities and the relationships between them. Finally, a summary of our work is given in Section~\ref{sect:Sum}.

\begin{deluxetable}{cccccccc}
	\centering
	\setlength\tabcolsep{15pt}
	\tablecolumns{4}
	\tabletypesize{\normalsize}
	\tablewidth{15cm}
	\tablecaption{List of objects. \label{tab:list}}
	\tablehead{
		\colhead{ID} & \colhead{IRAS Name} & \colhead{Other Name} & \colhead{R.A.(J2000)} & \colhead{Decl.(J2000)}  & \colhead{Dist (kpc)}& \colhead{Ref(s)}\\		
		\colhead{(1)}   & \colhead{(2)} & \colhead{(3)} & \colhead{(4)} & \colhead{(5)} & \colhead{(6)}   & \colhead{(7)}
	}
	\startdata
	{1} & {00338+6312} & {L1287} 		 	& {00:36:47.4} & {63:29:02}   & {0.9} & {1}\\
	{2} & {05345+3157} & {G176.51+00.20} 	& {05:37:52.1} & {32:00:03}   & {1.0} & {2}\\
	{3} & {06053-0622} & {Mon R2} 		& {06:07:47.9} & {-06:22:56}  & {0.8} & {2, 3}\\
	{4} & {06384+0932} & {NGC2264}    	& {06:41:09.9} & {09:29:14}   & {0.7} & {4, 5}\\
	{5} & {21007+4951} & {G090.21+02.32} 	& {21:02:22.7} & {50:03:08}   & {0.7} & {6}\\ 
	{6} & {21418+6552} & {G105.41+09.87} 	& {21:43:06.5} & {66:06:55}   & {0.9} & {3, 6}\\
	{7} & {22198+6336} & {...}& {22:21:26.7} & {63:51:37}   & {0.8} & {7}\\ 
	{8} & {22272+6358 A} & {L1206} 			& {22:28:51.4} & {64:13:41}   & {0.8} & {1}\\
	{9} & {22543+6413} & {Cep A} 			& {22:56:18.1} & {62:01:49}   & {0.7} & {3, 8}\\ 
	\enddata
	\tablewidth{15cm}
	\tablecomments{ (1) ID. (2) and (3) Source name(s). (4) and (5) Position of the source (R.A.: hh:mm:ss.s, Decl.: dd:mm:ss). (6): Distance to the source. (7): Parallax reference. }		
	\tablerefs{(1) \citet{Rygl+etal+2010}; (2) \citet{Xu+etal+2016}; (3) \citet{Plume+etal+1992}; (4) \citet{Kamezaki+etal+2014}; (5) \citet{Schreyer+etal+1997}; (6) \citet{Xu+etal+2013}; (7) \citet{Hirota+etal+2008}; (8) \citet{Moscadelli+etal+2009} }
\end{deluxetable}

\section{Observations}
\label{sect:Obs}
The observations of nine HMSFRs were carried out from July 2019 to May 2020 using the PMODLH 14-m millimeters-wavelength telescope. The detailed observations of the six molecular lines are: (1)\co \jx (115.271~GHz), \xco \jx (110.201~GHz) and \xxco \jx (109.782~GHz) were simultaneously observed from July to November 2019; (2) SiO \jxx (86.847~GHz) was observed from August 2019 to January 2020; (3) \hco \jx (89.189~GHz) was observed from November 2019 to February 2020; and (4) CS \jxx (97.981~GHz) was observed in May 2020. All lines were observed with the nine-beam Superconducting Spectroscopic Array Receiver system in the sideband separation mode \citep{Shan+etal+2012}, where the lines of SiO, \xcoo, and \xxco were in the lower sideband, and \coo, \hcoo, and CS were in the upper sideband. 

The lines of SiO were obtained via the single pointing observation mode with integration times over 40~min. Other lines were observed with the on-the-fly (OTF) mode, using typical sample steps of ${10}''$--${15}''$. The OTF raw data were gridded in a FITS cube with a pixel size of ${30}''$ using the GILDAS software package. \footnote{\url{http://www.iram.fr/IRAMFR/GILDAS} \label{GILDAS}} The typical integration time for each position was about 15~min. Each fast Fourier transform (FFT) spectrometer with a bandwidth of 1~GHz provided 16384 channels, producing a spectral resolution of 61~kHz. The parameters of the antenna and the velocity resolutions of the spectrometers are listed in Table~\ref{tab:obs}. \footnote{\url{http://www.radioast.nsdc.cn/english/zhuangtaibaogao.php} \label{radioast}}

All results presented in this work are expressed as brightness temperatures, $T^{*}_{R}$ = ${T^{*}_{A}}/{\eta_{mb}}$, where $T^{*}_{A}$ is the antenna temperature and $\eta_{mb}$ is the main beam efficiency, which are listed in Table~\ref{tab:obs}. The main beam root-mean-square (RMS) noises of the molecular lines for all observed sources are listed in Table~\ref{tab:noise}. 

 \begin{deluxetable}{ccccccc}
 	\centering
 	\setlength\tabcolsep{15pt}
 	\tablecolumns{4}
 	\tabletypesize{\normalsize}
 	\tablewidth{15cm}
 	\tablecaption{Basic observation parameters at the observed frequencies. \label{tab:obs}}
 	\tablehead{
 		\colhead{No} & \colhead{Lines} & \colhead{$T_{\rm sys}$ (K)} & \colhead{$\eta_{mb}$}  & \colhead{HPBW ($''$)}& \colhead{Velocity resolution (\kmss)}\\		
 		\colhead{(1)}   & \colhead{(2)} & \colhead{(3)} & \colhead{(4)} & \colhead{(5)} & \colhead{(6)} 
 	}
 	\startdata
 	1     & \co    & 250-300 & 0.49  & 49   & 0.159  \\
 	2     & \xco   & 150-200 & 0.54  & 51   & 0.166  \\
 	3     & \xxco  & 150-200 & 0.54  & 52   & 0.166  \\
 	4     & SiO    & 100-150 & 0.58  & 62   & 0.205  \\
 	5     & \hco   & 100-150 & 0.60  & 61   & 0.187  \\
 	6     & CS     & 120-170 & 0.67  & 55   & 0.212  \\
 	\enddata
 	\tablewidth{15cm}
 	\tablecomments{ (1) No. (2) Molecular line. (3) Typical system temperature ($T_{\rm sys}$). (4) Main beam efficiency ($\eta_{mb}$). (5) Half power beam width (HPBW).  (6) Velocity resolution. }		
 	
 \end{deluxetable}

\section{Results}
\label{sect:Res}

\subsection{Emission Peak spectra}
We successfully detected \coo, \xcoo, \xxcoo, \hcoo, and CS emission from all sources; unfortunately, we did not detect SiO emission for all sources.

\begin{figure}
	\centering
	\vspace{-0.3cm}
	\subfigbottomskip=2pt
	\subfigure[L1287]{\includegraphics[width=4.5cm]{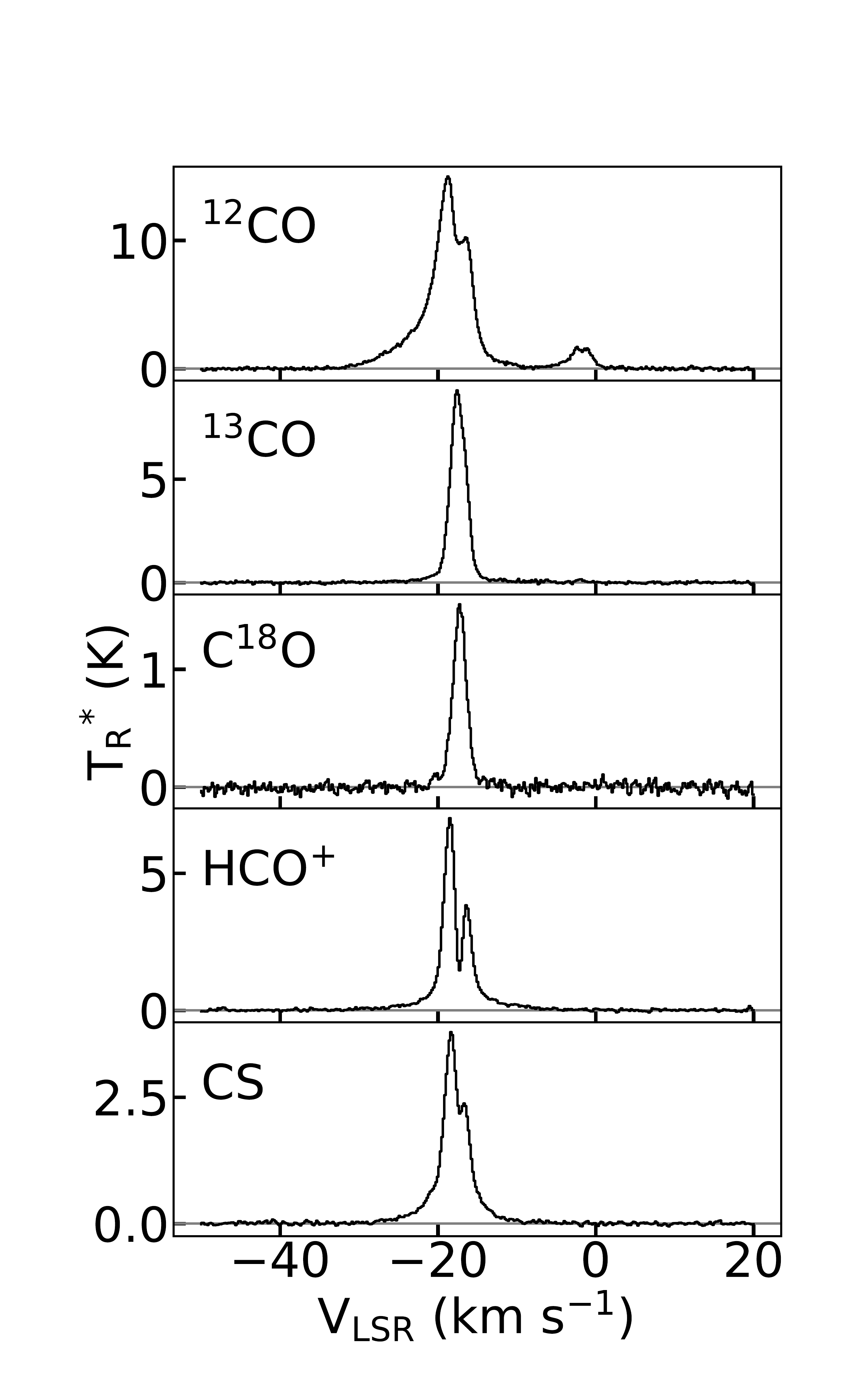}}
	\subfigure[G176.51+00.20]{\includegraphics[width=4.5cm]{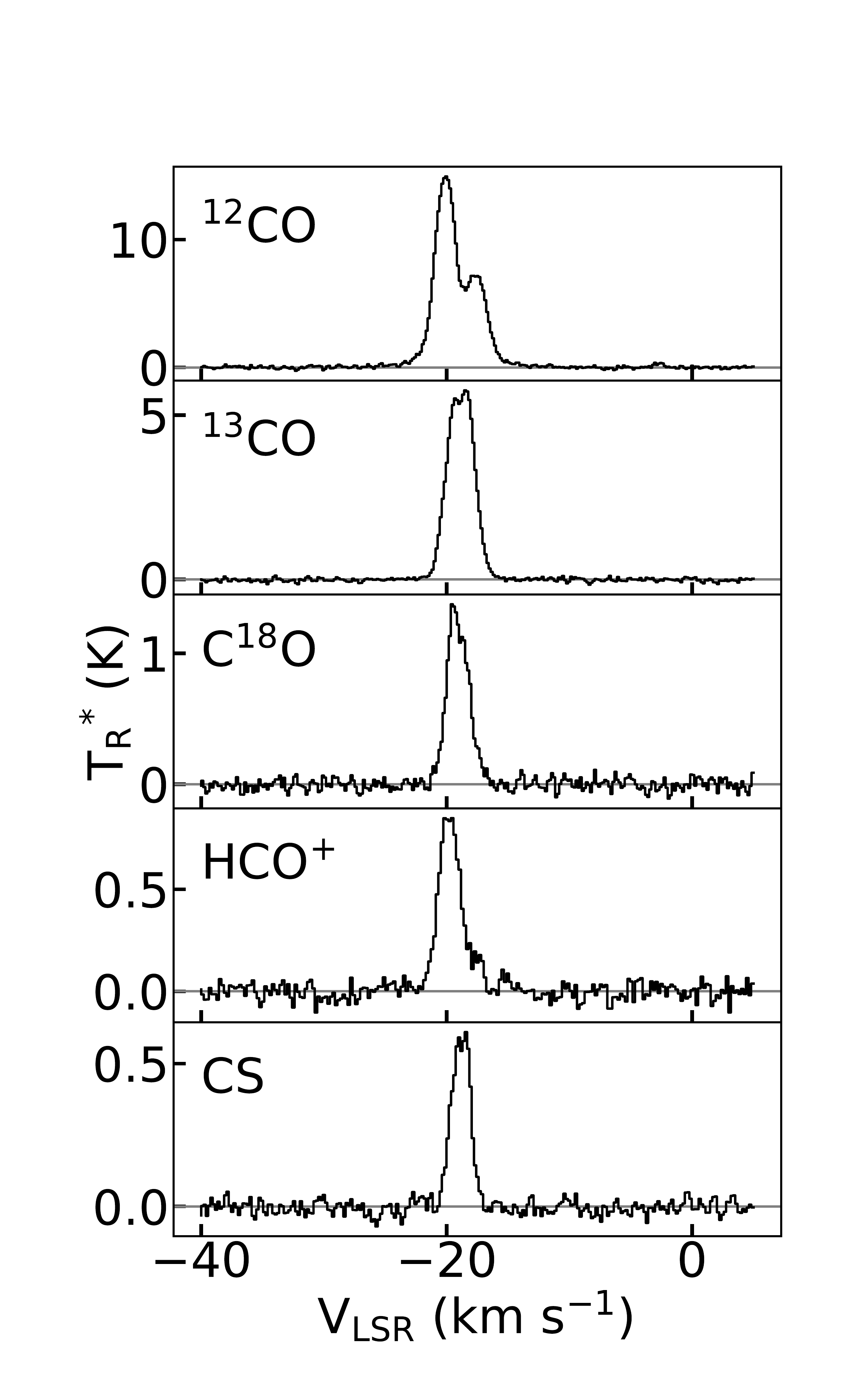}}
	\subfigure[Mon R2]{\includegraphics[width=4.5cm]{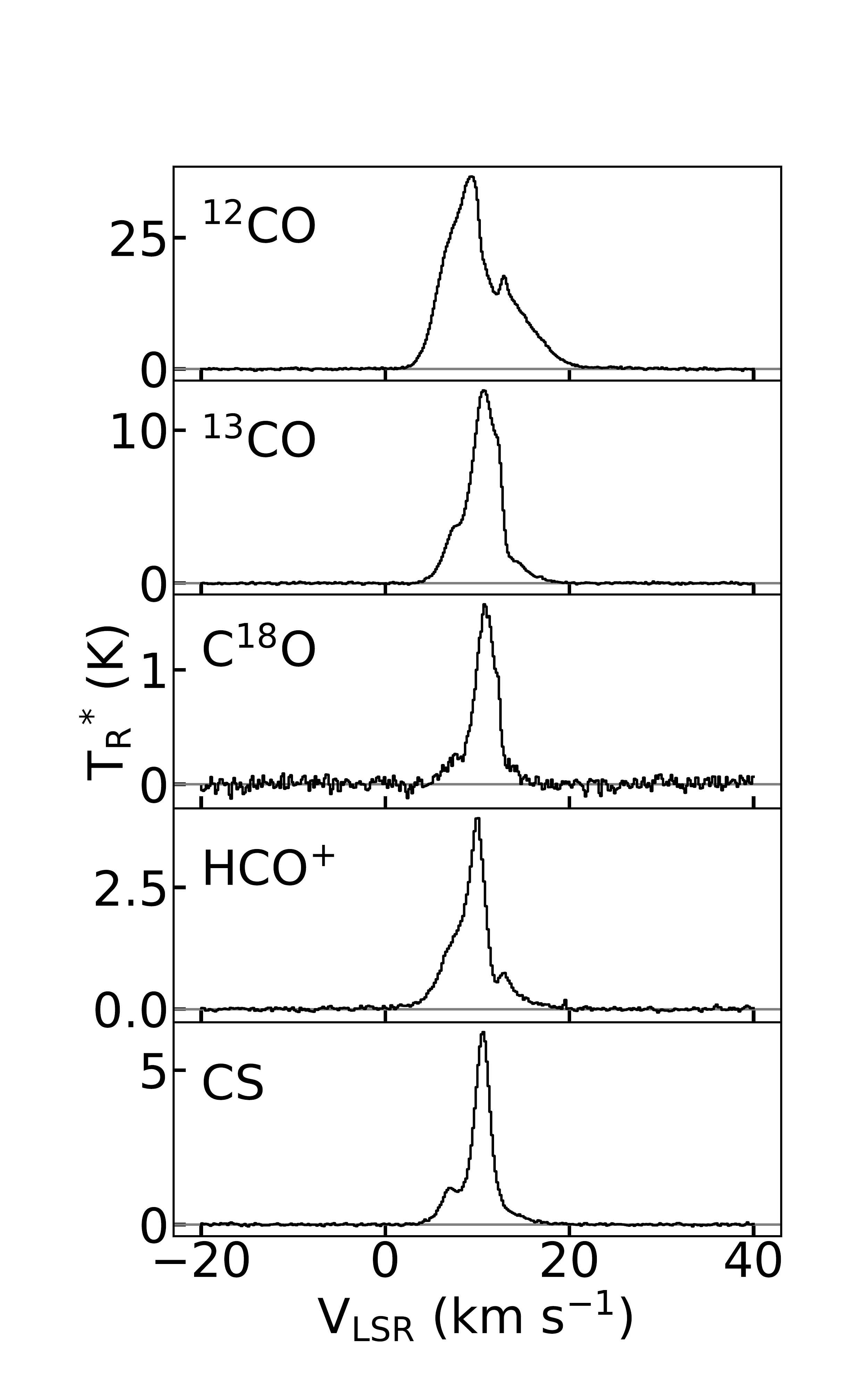}} \vskip -0.3cm
	
	\subfigure[NGC2264]{\includegraphics[width=4.5cm]{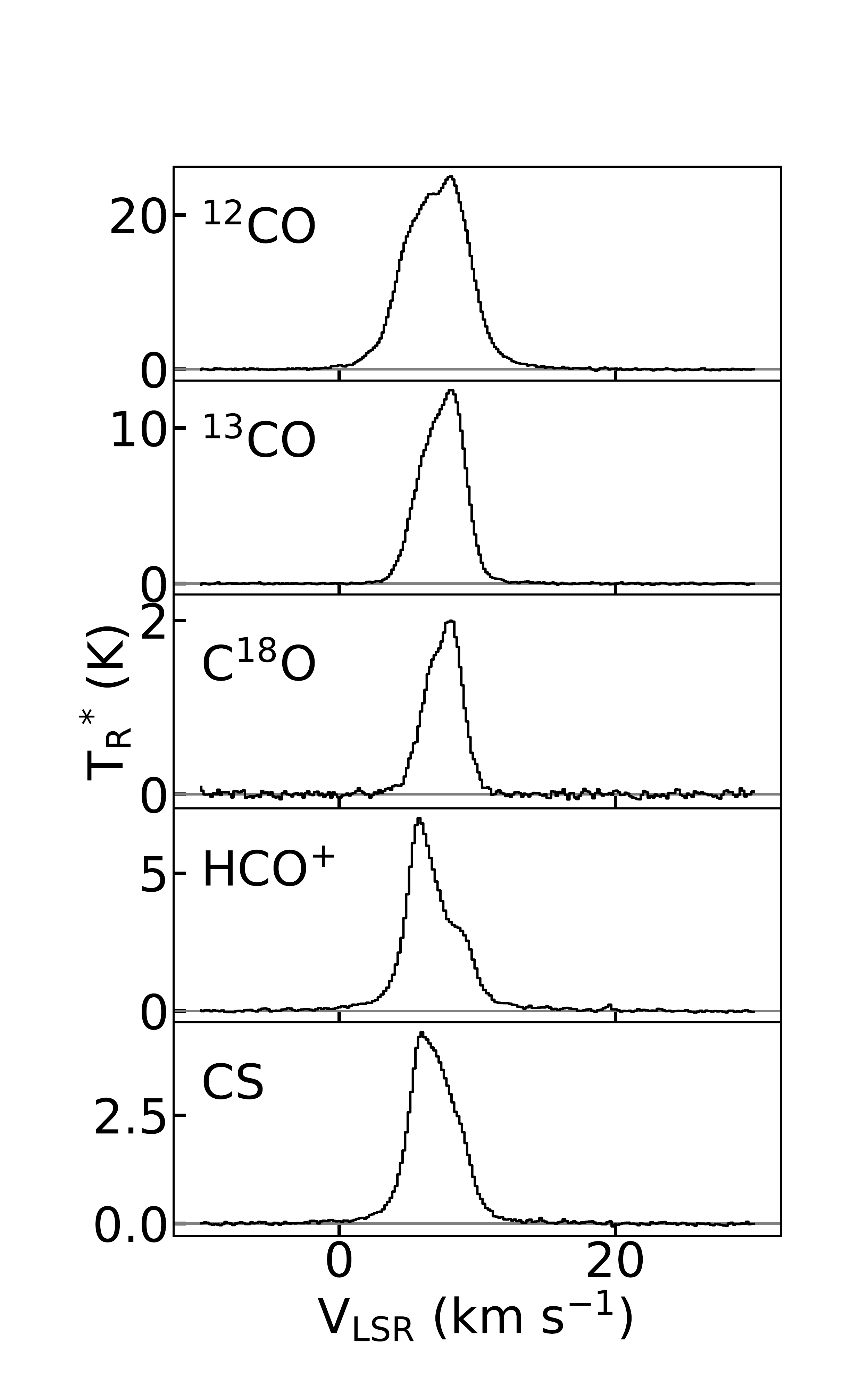}} 
	\subfigure[G090.21+02.32]{\includegraphics[width=4.5cm]{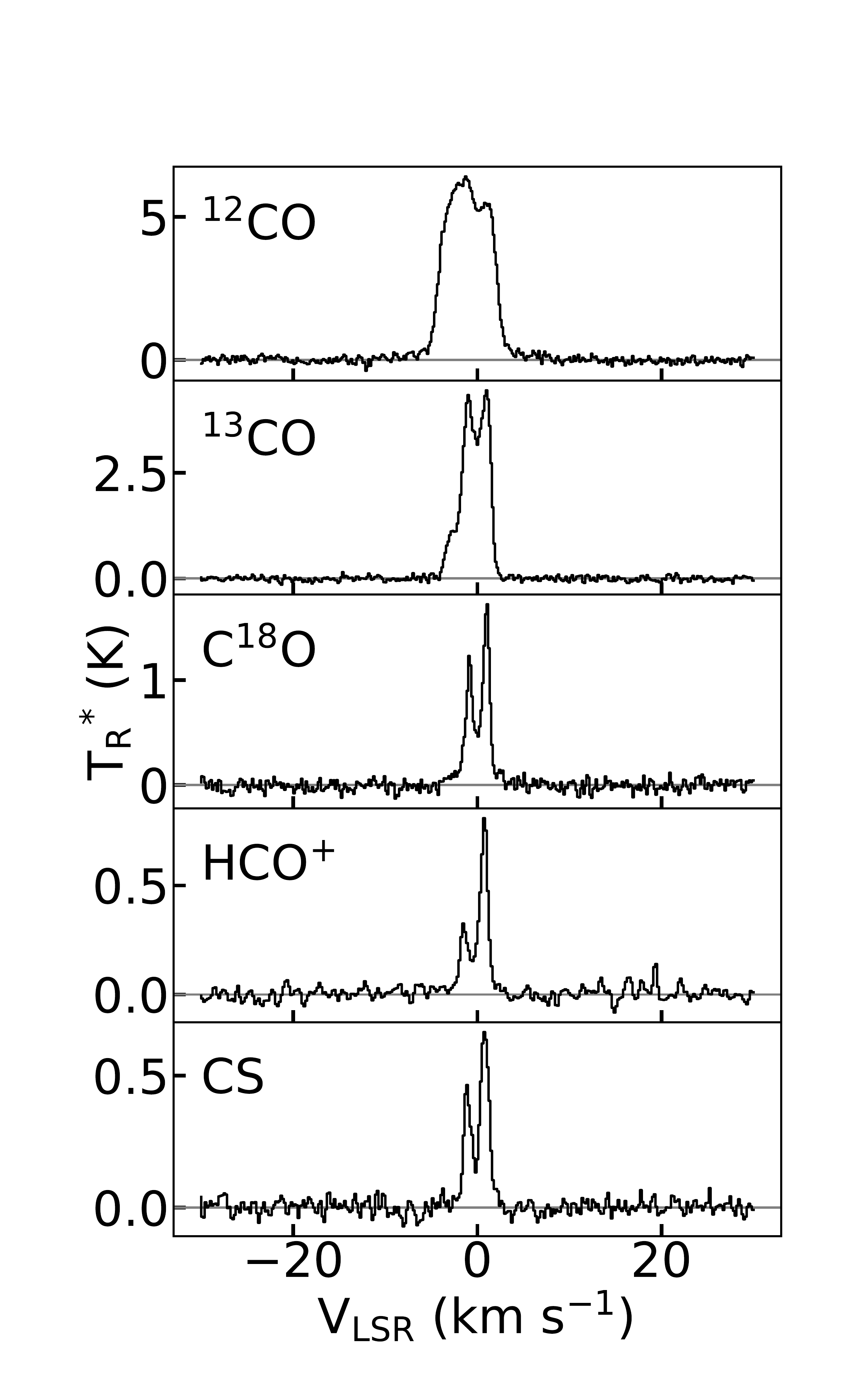}}
	\subfigure[G105.41+09.87]{\includegraphics[width=4.5cm]{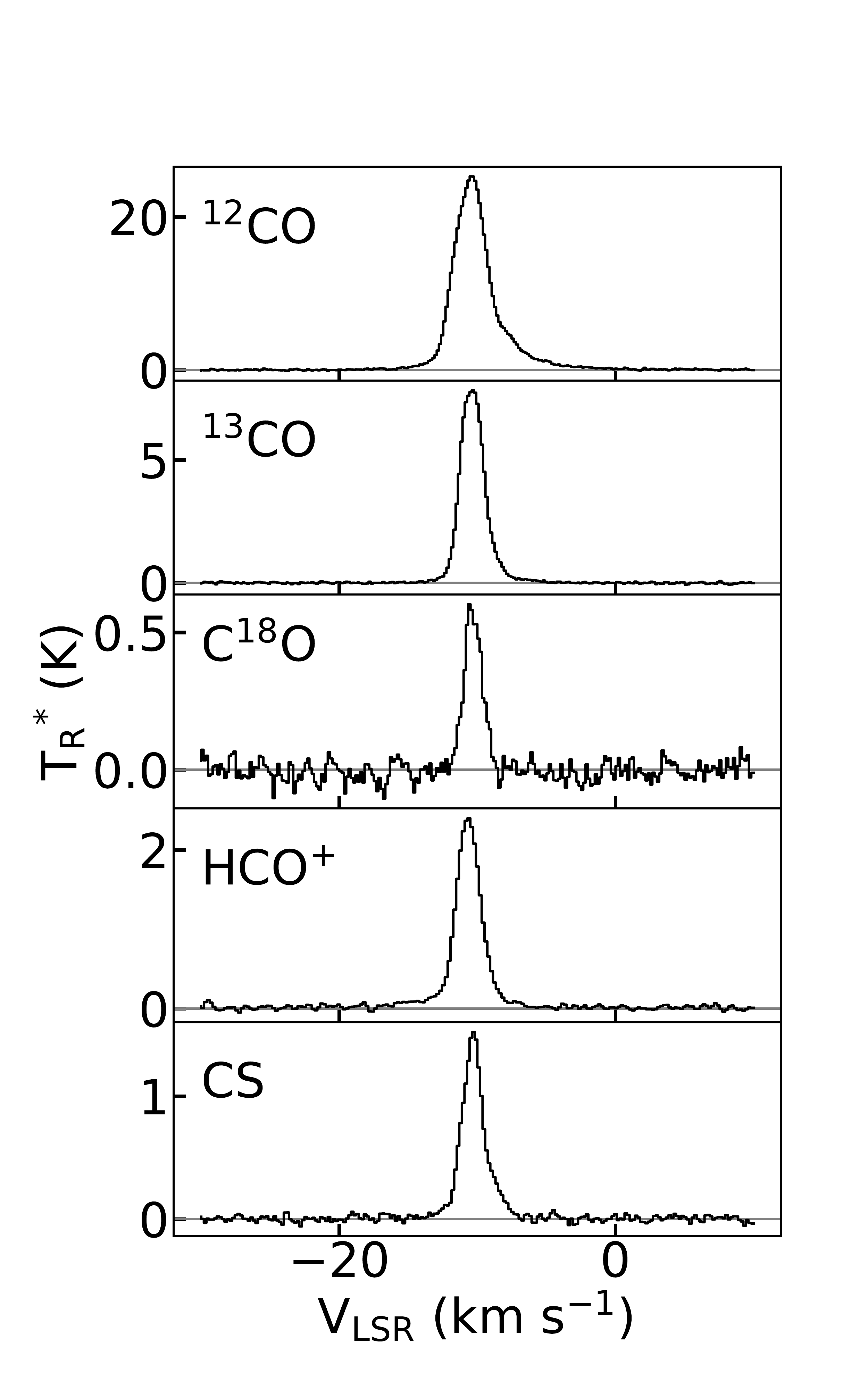}} \vskip -0.3cm
	
	\subfigure[IRAS 22198+6336]{\includegraphics[width=4.5cm]{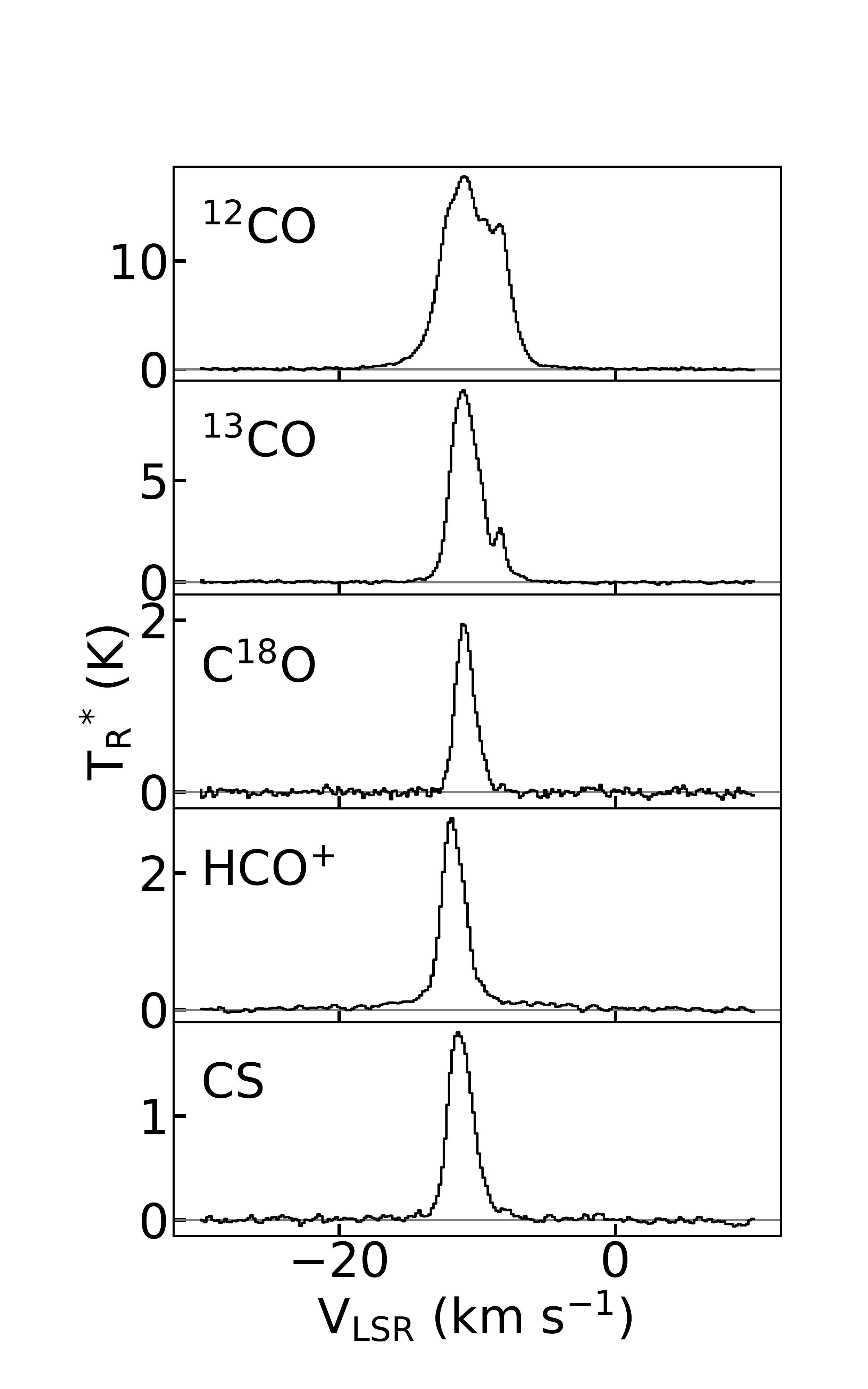}}
	\subfigure[L1206]{\includegraphics[width=4.5cm]{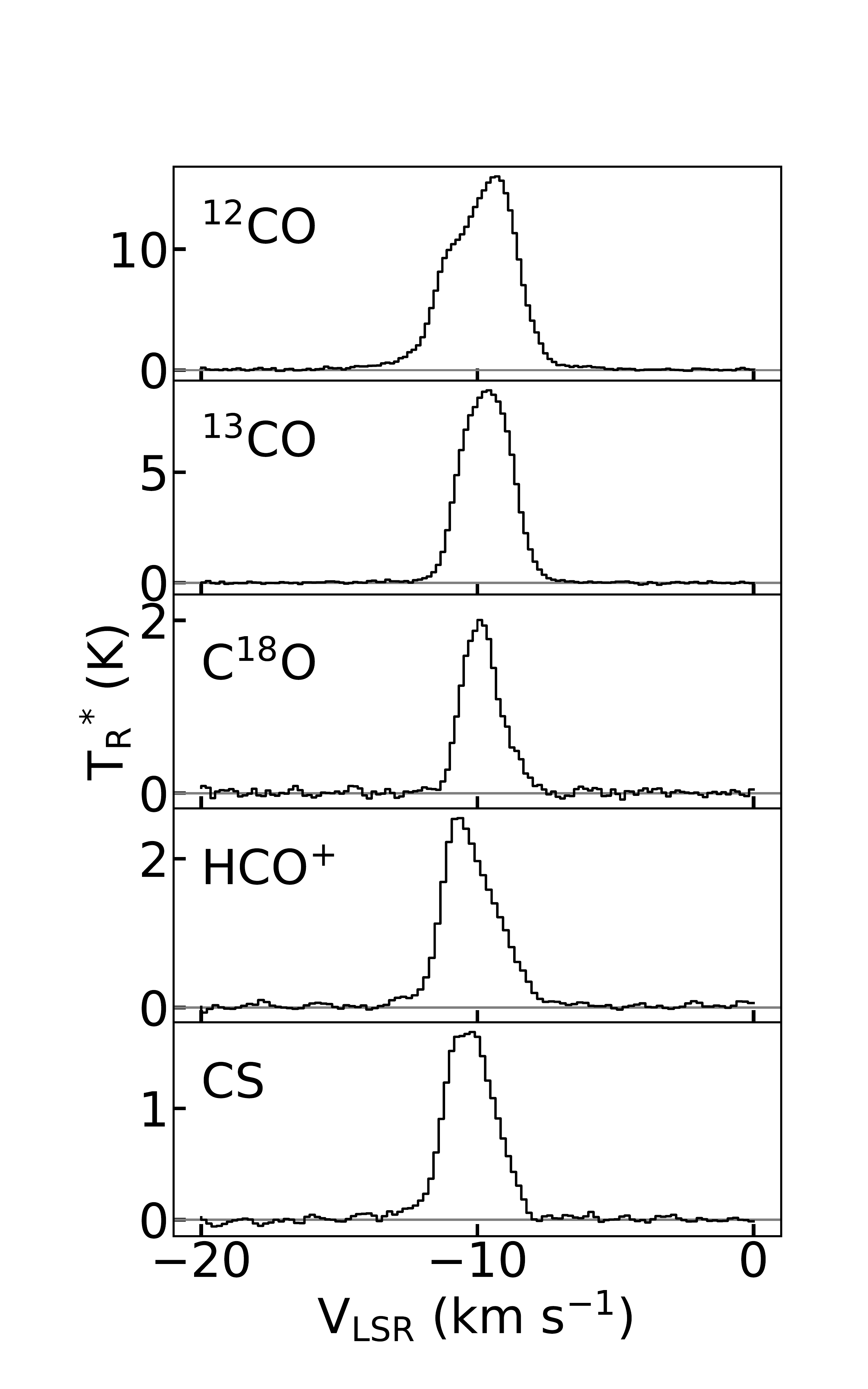}}
	\subfigure[Cep A]{\includegraphics[width=4.5cm]{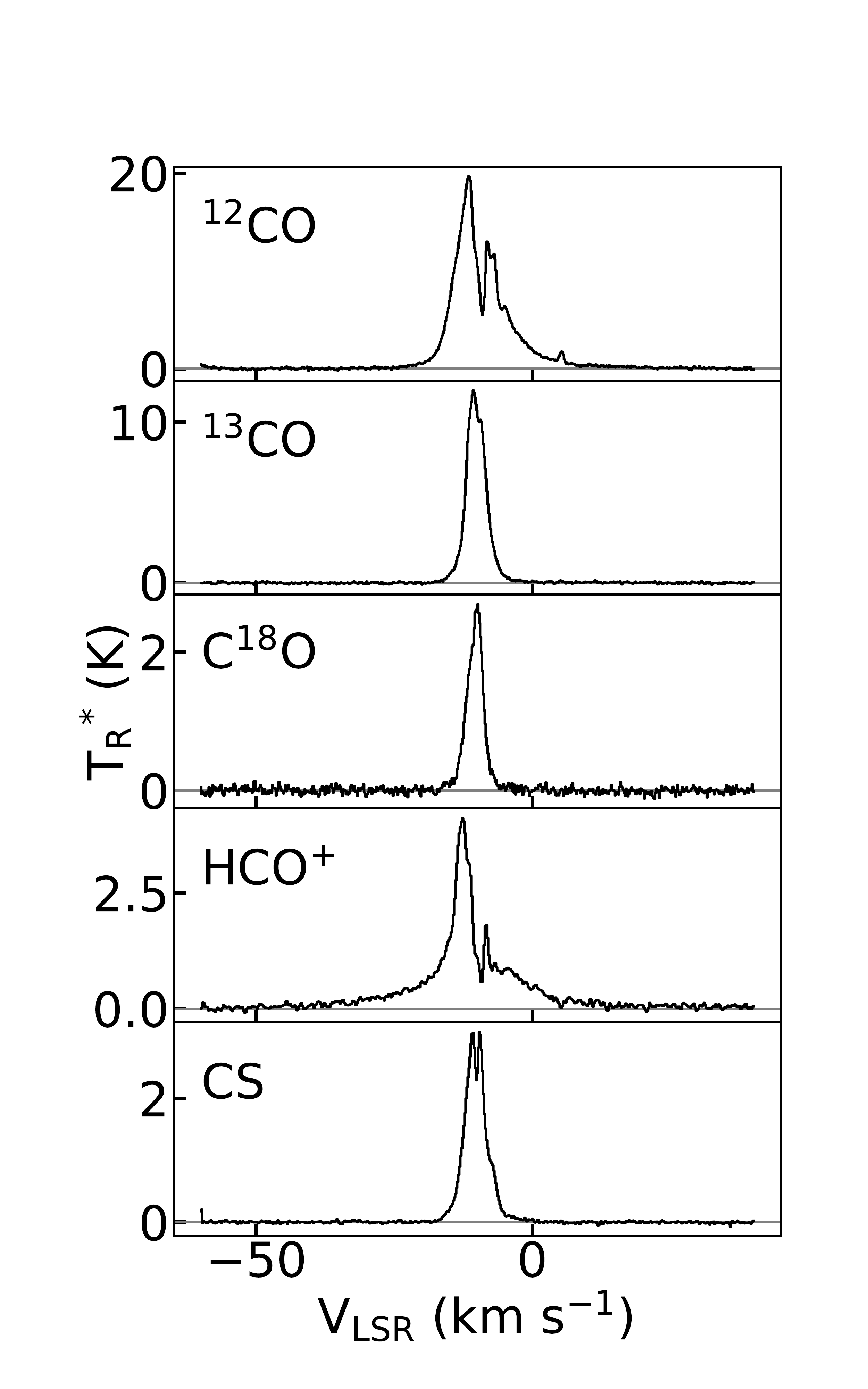}}
	\caption{ Line profiles of the molecular transitions for the nine HMSFR sources. The spectra were obtained at the position of the \xxco emission peak. The name of each source is given at the bottom of each panel. From top to bottom the lines are \coo, \xcoo, \xxcoo, \hcoo, and CS. }
	\label{fig:spectra}
\end{figure}

The basic parameters of the molecular lines at the respective emission peaks are listed in Table~\ref{tab:specPmts}. Meanwhile, the molecular line profiles of the emission peak of \xxco for each source are shown in Figure~\ref{fig:spectra}. To select the most suitable sensitivity to achieve maximum telescope efficiency, the following tests were performed for the CS observation of Cep A. When the integration time of each position was about 15~min, the RMS noise was about 15.9 mK, and the full width ($\Delta V$) at 3$\sigma$ above the baseline was 16.8~\kmss. Then we doubled the integration time (i.e., to 30~min), finding an RMS noise of about 10.2~mK, and a full width of 17.7~\kmss. There was no significant extension of its full width. Hence, we chose 15~min as the typical integration time of each position. Our results indicate the following relationship: $\Delta V_{\rm ^{12}CO} > \Delta V_{\rm HCO^{+}} > \Delta V_{\rm CS} \approx \Delta V_{\rm ^{13}CO} > \Delta V_{\rm ^{18}CO}$. Meanwhile, multiple velocity features were visible in some line profiles, particularly those of \co and \hcoo, while \xcoo, \xxcoo, and CS were quite smooth.

\begin{figure}
	\centering
	\includegraphics[width=12.8cm]{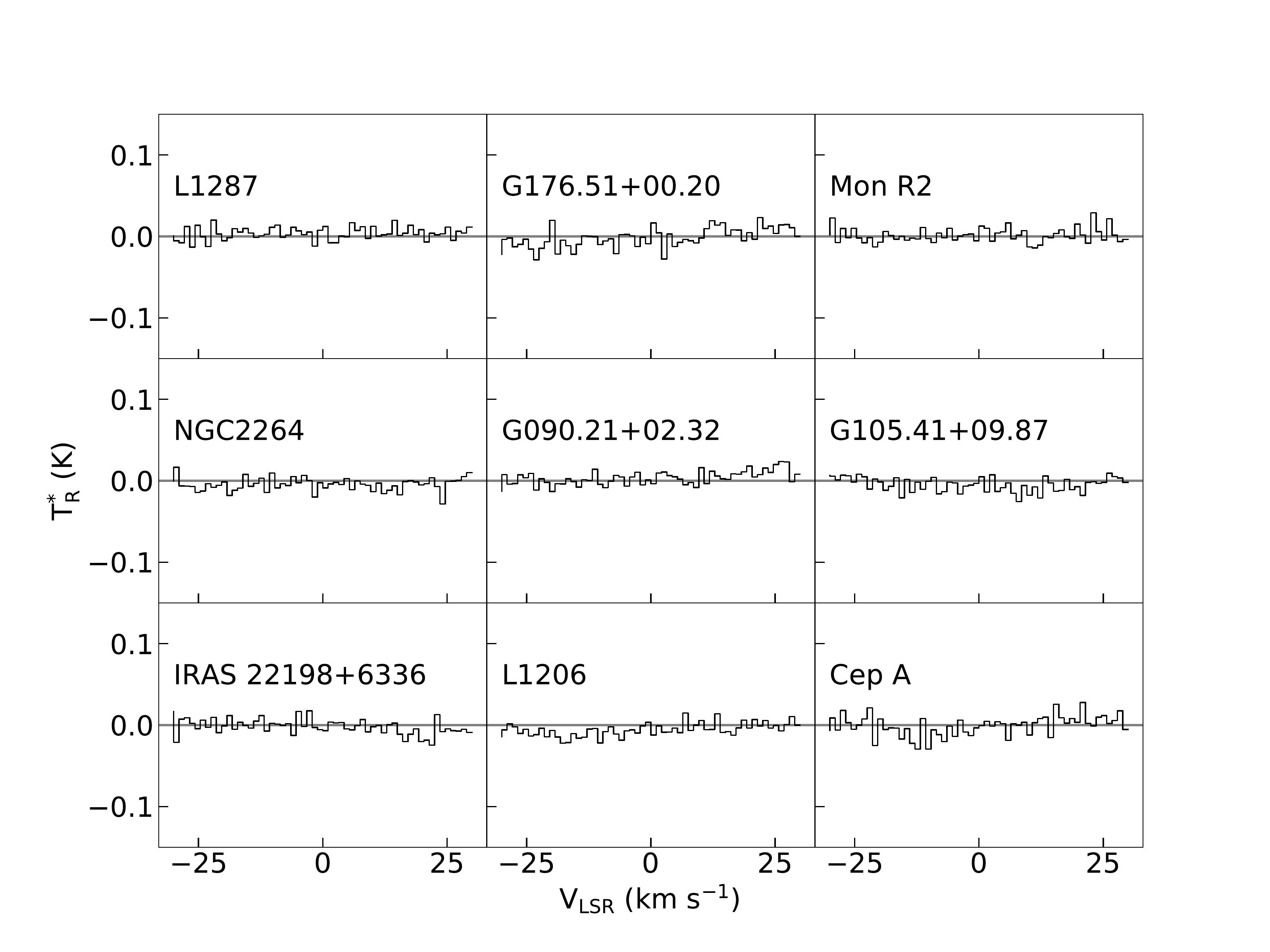}
	\caption{SiO spectra of the nine sources, including null detections. The name of each source is given in the top left-hand corner of each panel.}
	\label{fig:SiOs}
\end{figure}

SiO emission, which was used as a shock tracer, was not detected for all sources (see Figure~\ref{fig:SiOs}). These sources had also been observed by \citet{Harju+etal+1998}, who attempted to correlate the SiO emission with maser characteristics and with ultra-compact (UC) H$_{\rm II}$ regions. In their work, \citet{Harju+etal+1998} used the 15-m telescope Swedish-ESO Submillimetre Telescope (SEST) and the 20-m radio telescope of the Onsala Space Observatory in Sweden. Of the nine sources considered here, \citet{Harju+etal+1998} only detected SiO emission for L1287 with Onsala, whose velocity resolution was 0.17~\kms and RMS was 20~mK, which were similar to what we achieved with our observations. Thus, the velocity resolution and sensitivity of PMODLH might not be the reason why we did not detect any SiO emission from L1287. Whilst the half power beam width (HPBW) of Onsala (i.e., 43$''$) was better than that of PMODLH (i.e., 62$''$). \citet{Harju+etal+1998} pointed out that SiO emission can be better traced by telescopes with higher resolutions (e.g. a 30-m telescope, which has a larger aperture). Therefore, resolution might be the reason why we did not detect SiO emission for all sources.

\begin{deluxetable}{ccccccc}
	\centering
	\setlength\tabcolsep{15pt}
	\tablecolumns{4}
	\tabletypesize{\normalsize}
	\tablewidth{15cm}
	\tablecaption{Main beam RMS noise. \label{tab:noise}}
	\tablehead{
		\colhead{Source} & \multicolumn{6}{c}{RMS Noise (mK)} \\ \cline{2-7}
		\colhead{} & \colhead{SiO} & \colhead{CS} & \colhead{\hcoo} & \colhead{\coo} & \colhead{\xcoo} & \colhead{\xxcoo}\\
		\colhead{(1)}   & \colhead{(2)} & \colhead{(3)} & \colhead{(4)} & \colhead{(5)} & \colhead{(6)} & \colhead{(7)}
	}
	\startdata
	{L1287}                  & {14.3} &{15.1}  & {16.9}   & {51.2}   & {29.8} & {28.1}\\
	{G176.51+00.20}          & {18.1} &{15.3}  & {17.2}   & {60.9}   & {32.8} & {31.5}\\
	{Mon R2}                 & {15.3} &{15.6}  & {18.3}   & {51.1}   & {25.7} & {24.2}\\
	{NGC2264}                & {14.5} &{16.1}  & {16.3}   & {47.4}   & {27.5} & {25.9}\\
	{G090.21+02.32}          & {16.8} &{13.4}  & {17.7}   & {66.9}   & {36.4} & {32.8}\\
	{G105.41+09.87}          & {17.8} &{13.2}  & {17.2}   & {50.0}   & {27.5} & {27.0}\\
	{IRAS 22198+6336}        & {16.4} &{14.7}  & {13.5}   & {44.0}   & {23.7} & {26.1}\\
	{L1206}                  & {14.8} &{16.7}  & {17.9}   & {43.6}   & {27.4} & {25.8}\\
	{Cep A}                  & {15.3} &{10.2}  & {21.4}   & {50.8}   & {32.0} & {29.4}\\
	\enddata
	\tablewidth{15cm}
	\tablecomments{ (1) Source name. (2), (3), (4), (5), (6) and (7) RMS noise of SiO, CS, \hcoo, \coo, \xcoo, and \xxcoo, respectively.}		
\end{deluxetable}

\begin{longrotatetable}
	\begin{deluxetable*}{cccccccccccccccccccc}
		\tablecaption{Basic parameters of the molecular lines. \label{tab:specPmts}}
		\tabletypesize{\scriptsize}
		\tablehead{
			\colhead{} & \multicolumn{3}{c}{\co} & \colhead{} & \multicolumn{3}{c}{\xco} &{}  & \multicolumn{3}{c}{\xxco} &{}	& \multicolumn{3}{c}{\hco}  &{} & \multicolumn{3}{c}{CS}     \\ \cline{2-4} \cline{6-8} \cline{10-12} \cline{14-16} \cline{18-20}
			\colhead{Source} & \colhead{$T_{\rm R}^{*}$} & \colhead{$V_{\rm peak}$} & \colhead{$\Delta V$} & \colhead{} & \colhead{$T_{\rm R}^{*}$} & \colhead{$V_{\rm peak}$} & \colhead{$\Delta V$} & \colhead{} & \colhead{$T_{\rm R}^{*}$} & \colhead{$V_{\rm peak}$} & \colhead{$\Delta V$} & \colhead{} & \colhead{$T_{\rm R}^{*}$} & \colhead{$V_{\rm peak}$} & \colhead{$\Delta V$} & \colhead{} & \colhead{$T_{\rm R}^{*}$} & \colhead{$V_{\rm peak}$} & \colhead{$\Delta V$}\\
			\colhead{} & \colhead{(K)} & \colhead{(\kmss)} & \colhead{(\kmss)} & \colhead{} & \colhead{(K)} & \colhead{(\kmss)} & \colhead{(\kmss)} & \colhead{} & \colhead{(K)} & \colhead{(\kmss)} & \colhead{(\kmss)} & \colhead{} & \colhead{(K)} & \colhead{(\kmss)} & \colhead{(\kmss)} & \colhead{} & \colhead{(K)} & \colhead{(\kmss)} & \colhead{(\kmss)}\\
			\colhead{(1)} & \colhead{(2)} & \colhead{(3)} & \colhead{(4)} & \colhead{} & \colhead{(5)} & \colhead{(6)} & \colhead{(7)} & \colhead{} & \colhead{(8)} & \colhead{(9)} & \colhead{(10)} & \colhead{} & \colhead{(11)} & \colhead{(12)} & \colhead{(13)} & \colhead{} & \colhead{(14)} & \colhead{(15)} & \colhead{(16)}
		} 
		\startdata
		{L1287}                  & {19.0} &{-18.6}  &{31.8} &{} &{9.3} &{-17.6} &{11.6} & {}   & {1.6}   & {-17.2} & {3.3} & {} & {7.0} & {-18.5} & {17.4} & {} & {3.8} & {-18.2} & {16.8} \\
		{G176.51+00.20}          & {14.9} &{-17.5}  &{22.2} &{} &{8.3} &{-19.2} &{5.8} & {}   & {1.4}   & {-19.5} & {3.3} & {} & {3.2} & {-18.1} & {18.5} & {} & {3.2} & {-18.2} & {7.5} \\
		{Mon R2}                 & {36.0} &{9.5}  &{20.6} &{} &{13.0} &{10.8} &{14.1} & {}   & {1.6}   & {10.8} & {8.3} & {} & {5.9} & {9.3} & {18.5} & {} & {6.2} & {10.8} & {12.1} \\
		{NGC2264}                & {24.7} &{7.4}  &{23.8} &{} &{12.4} &{8.3} &{9.1} & {}   & {3.5}   & {8.2} & {5.8} & {} & {7.6} & {8.9} & {18.5} & {} & {5.3} & {7.7} & {14.0} \\
		{G090.21+02.32}          & {8.0} &{0.6}  &{15.9} &{} &{5.1} &{0.9} &{6.7} & {}   & {1.7}   & {1.2} & {5.0} & {} & {1.4} & {0.7} & {4.1} & {} & {0.8} & {1.0} & {2.8} \\
		{G105.41+09.87}          & {14.5} &{-11.3}  &{17.5} &{} &{9.9} &{-10.0} &{9.1} & {}   & {2.8}   & {-9.2} & {2.5} & {} & {3.7} & {-10.6} & {16.4} & {} & {1.8} & {-10.1} & {6.5} \\
		{IRAS 22198+6336}        & {17.4} &{-11.0}  &{15.9} &{} &{9.3} &{-10.5}  & {7.5} & {} &{2.0} & {-11.0}   & {3.3}   & {} & {2.8} & {-11.7} & {11.3} & {} & {1.9} & {-11.6} & {6.5}  \\
		{L1206}                  & {17.0} &{-9.2}  &{23.3} &{} &{8.7} &{-9.5} &{5.0} & {}   & {2.0}   & {-9.8} & {2.5} & {} & {2.5} & {-10.7} & {5.1} & {}  & {1.7} & {-10.1} & {4.7} \\
		{Cep A}                  & {22.4} &{-11.7}  &{43.7} &{} &{14.0} &{-11.6} &{16.6} & {}   & {2.7}   & {-9.8} & {9.2} & {} & {5.1} & {-12.4} & {47.2} & {} & {3.5} & {-11.0} & {17.7} \\
		\enddata
		\tablecomments{ (1) Source name. (2), (5), (8), (11) and (14) Brightness temperature(s) of the emission peaks of \coo, \xcoo, \xxcoo, \hcoo, and CS, respectively. (3), (6), (9), (12) and (15) Central velocities of \coo, \xcoo, \xxcoo, \hcoo, and CS, respectively. (4), (7), (10), (13) and (16) Full line width at 3$\sigma$ above the baseline of \coo, \xcoo, \xxcoo, \hcoo, and CS, respectively.}
	\end{deluxetable*}
\end{longrotatetable}

\begin{longrotatetable}
	\begin{deluxetable*}{ccccccccccccc}
		\tablecaption{ Outflow parameters. \label{tab:parameters}}
		\tabletypesize{\scriptsize}
		\tablehead{
			\colhead{} & \multicolumn{2}{c}{\co} &\colhead{} & \multicolumn{2}{c}{\xco} &\colhead{} & \multicolumn{2}{c}{\hco} &\colhead{} & \multicolumn{2}{c}{CS} &\colhead{}  \\ \cline{2-3} \cline{5-6} \cline{8-9} \cline{11-12} 
			\colhead{Source} & \colhead{ $\Delta v_{\rm b} ({\rm km\ s^{-1}})$} & \colhead{$\Delta v_{\rm r} ({\rm km\ s^{-1}})$} &\colhead{} & \colhead{ $\Delta v_{\rm b} ({\rm km\ s^{-1}})$} & \colhead{$\Delta v_{\rm r} ({\rm km\ s^{-1}})$} &\colhead{} & \colhead{ $\Delta v_{\rm b} ({\rm km\ s^{-1}})$} & \colhead{$\Delta v_{\rm r} ({\rm km\ s^{-1}})$} &\colhead{} & \colhead{ $\Delta v_{\rm b} ({\rm km\ s^{-1}})$} & \colhead{$\Delta v_{\rm r} ({\rm km\ s^{-1}})$} \\
			\colhead{(1)} & \colhead{(2)} & \colhead{(3)} &\colhead{} & \colhead{(4)} & \colhead{(5)} &\colhead{} & \colhead{(6)} & \colhead{(7)} &\colhead{} & \colhead{(8)} & \colhead{(9)}
		} 
		\startdata
		{L1287}      & {(-34, -21)} & {(-15, -8)} &{} & {(-30, -20)} &{(-15, -6)} &{} & {(-33, -20)} & {(-15, -6)} &{} & {(-31, -20)} & {(-15, -6)} \\
		{G176.51+00.20}    & {(-47, -22)} & {(-15, -1)} &{} & {(-25, -22)} &{(-16, -11)} &{} & {(-39, -21)} & {(-16, 3)} &{}  & {(-27, -21)} & {(-15, -8)} \\
		{Mon R2}     & {(-2, 6)}    & {(14, 30)}  &{} & {(4, 7)}     &{(16, 21)}  &{}  & {(0, 6)}     & {(15, 22)}  &{} & {(-2, 4)}     & {(17, 21)}  \\
		{NGC 2264}   & {(-7, 0)}    & {(12, 22)}  &{} & {...}        &{(11, 15)}   &{} & {(-4, 4)}    & {(12, 21)}  &{} & {(-5, 3)}    & {(11, 17)} \\
		{G090.21+02.32}    & {...}        & {(4, 12)} &{}  & {...}        &{(2, 5)}   &{}   & {...}        & {(2, 7)}   &{}  & {...}        & {(2, 5)}  \\
		{G105.41+09.87}    & {(-20, -13)} & {(-6, 3)}  &{}  & {(-14, -12)} &{(-8, -5)}  &{}  & {(-20, -13)} & {(0, 9)}   &{}  & {(-15, -12)} & {(-8, -2)}  \\
		{IRAS 22198+6336} & {(-25, -15)} & {(-5, 0)}  &{}  & {(-16, -13)} &{...}     &{}    & {(-23, -14)} & {(6, 0)}   &{} & {(-16, -13)} & {(-9, -7)} \\
		{L1206}      & {(-21, -13)} & {(-7, -4)}  &{} & {(-13, -11)} &{(-8, -5)}  &{}  & {(-24, -12)} & {(-8, -4)}  &{} & {(-16, -12)} & {(-9, -7)} \\
		{Cep A}      & {(-42, -21)} & {(-1, 33)}  &{} & {(-21, -15)} &{(-7, -3)}   &{} & {(-49, -21)} & {(-1, 27)}   &{}& {(-19, -15)} & {(-6, 1)}  \\
		\enddata
		\tablecomments{ (1) Source name. (2)--(9) Outflow velocity ranges ($\Delta v$) for the blue and red wings of \coo, \xcoo, \hcoo, and CS, respectively. }
		
	\end{deluxetable*}
\end{longrotatetable}

\subsection{Outflow morphology}
\label{sect:morphology}
This work represents the first time that four molecular outflow tracers have been used in a single work (\coo, \xcoo, \hcoo, and CS), in addition to one shock tracer (SiO) and one dense core tracer (\xxcoo). With our observations, we successfully detected outflows associated with all nine sources. The high detection rate of outflows suggests that \coo, \xcoo, \hcoo, and CS may be all suitably used to trace the molecular outflows of HMSFRs.

Outflow maps are presented in Figures~\ref{fig:L1287}--\ref{fig:CepA}. Within these figures, we have plotted the \coo, \xcoo, \hcoo, and CS contours overlaid on the \xxco integrated images and WISE false-color images \citep{Wright+etal+2010}, \footnote{\url{https://irsa.ipac.caltech.edu/Missions/wise.html} \label{WISE}} where blue, green, and red correspond to the 4.6, 12, and 22 $\mu$m data, respectively. To determine the possible excitation sources of these outflows, we also marked the positions of the associated IRAS sources in the images, which were obtained from the IRAS Point Source Catalog 2.1. \footnote{\url{https://irsa.ipac.caltech.edu/Missions/iras.html} \label{IRAS}} The bolometric luminosities of these IRAS sources were sourced from \citet{Sanders+Mirabel+1996}. Meanwhile, the plotted spectra which have been smoothed to 5$\delta v$ (i.e., we smoothed the data by merging five channels into a one channel) at the emission peaks of the blue and red lobes, and marked out the velocity ranges of the outflows using different color shades (see Figure~\ref{fig:L1287} for details). The velocity ranges of the different outflows are listed in Table~\ref{tab:parameters}. Our main results are summarized as follows:

1. \co molecular outflows were detected for all sources. Eight sources (L1287, G176.51+00.20, Mon R2, NGC2264, G105.41+09.87, IRAS 22198+6336, L1206, and Cep A) showed clear bipolar or multiple outflow structures. Among them, the outflow of IRAS 22198+6336 was mapped for the first time, and NGC2264 was found to contain bipolar outflows for the first time. G090.21+02.32 appeared to show only a red lobe, although the source had been previously identified as a bipolar outflow \citep{Clark+1986}. The velocity ranges of the outflows of four sources (L1287, G176.51+00.20, Mon R2, and Cep A) have been extended with the improved high-sensitivity observations.

2. We detected \xco molecular outflows for the first time for all sources. Six sources (L1287, G176.51+00.20, Mon R2, G105.41+09.87, L1206, and Cep A) showed clear bipolar or multiple outflow structures. Three sources (NGC2264, G090.21+02.32, and IRAS 22198+6336) appeared to present a one-sided lobe, i.e., NGC2264 and G090.21+02.32 showed only a red lobe, and IRAS 22198+6336 presented only a blue lobe.

3. We detected \hco molecular outflows for all sources. Eight sources (L1287, G176.51+00.20, Mon R2, NGC2264, G105.41+09.87, IRAS 22198+6336, L1206, and Cep A) showed clear bipolar or multiple outflow structures. G090.21+02.32 showed only a red lobe. Except for L1287 and Cep A, the \hco outflows of the other sources were mapped for the first time.

4. We detected CS molecular outflows for the first time for all sources. Except for G090.21+02.32, which showed only a red lobe, the other eight sources (L1287, G176.51+00.20, Mon R2, NGC2264, G105.41+09.87, IRAS 22198+6336, L1206, and Cep A) presented clear bipolar or multiple outflow structures.

\subsubsection{L1287}
\begin{figure}
	\centering
	\vspace{-0.3cm}
	\subfigure[\co]{\includegraphics[width=8cm]{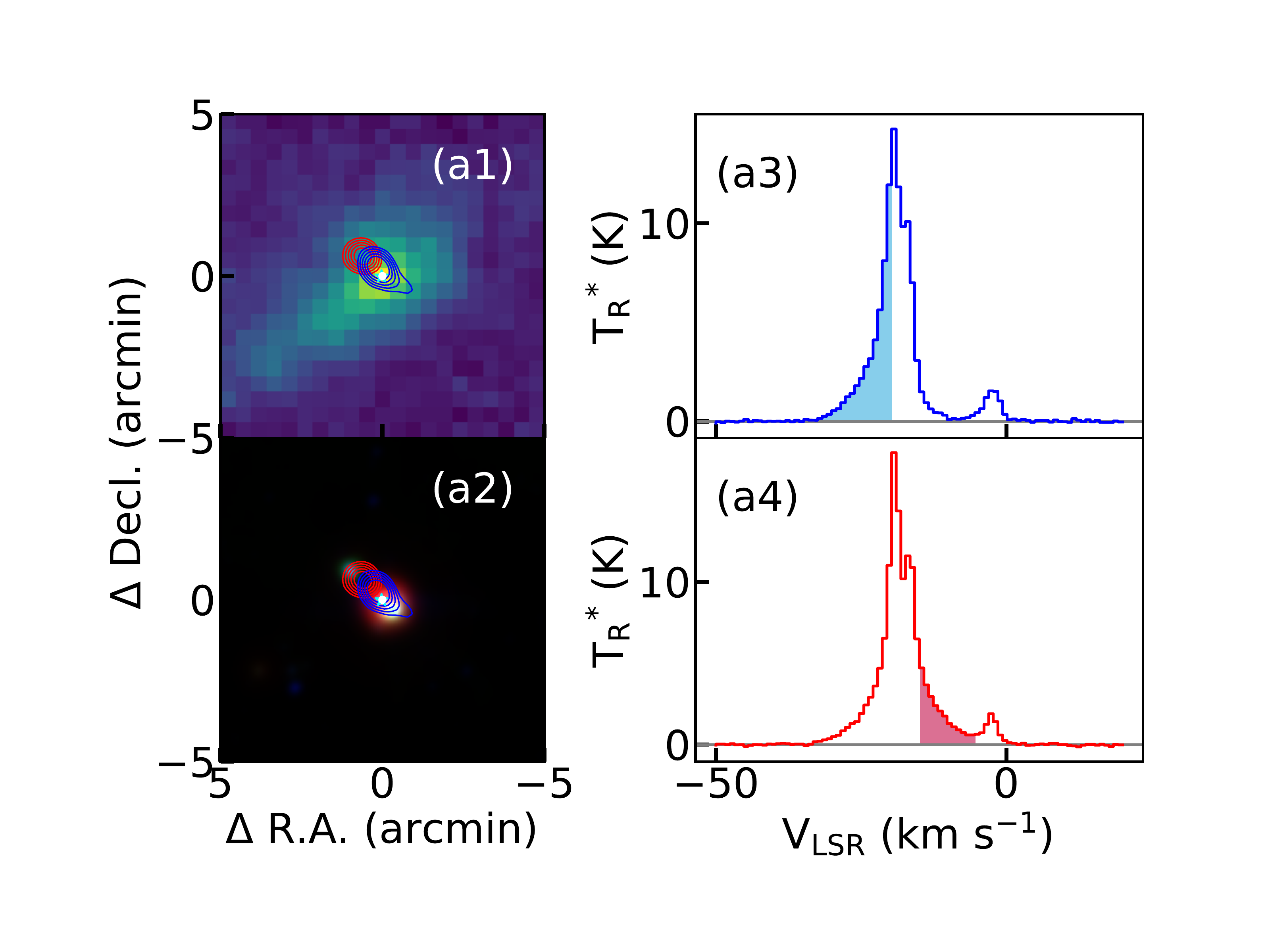}}
	\subfigure[\xco]{\includegraphics[width=8cm]{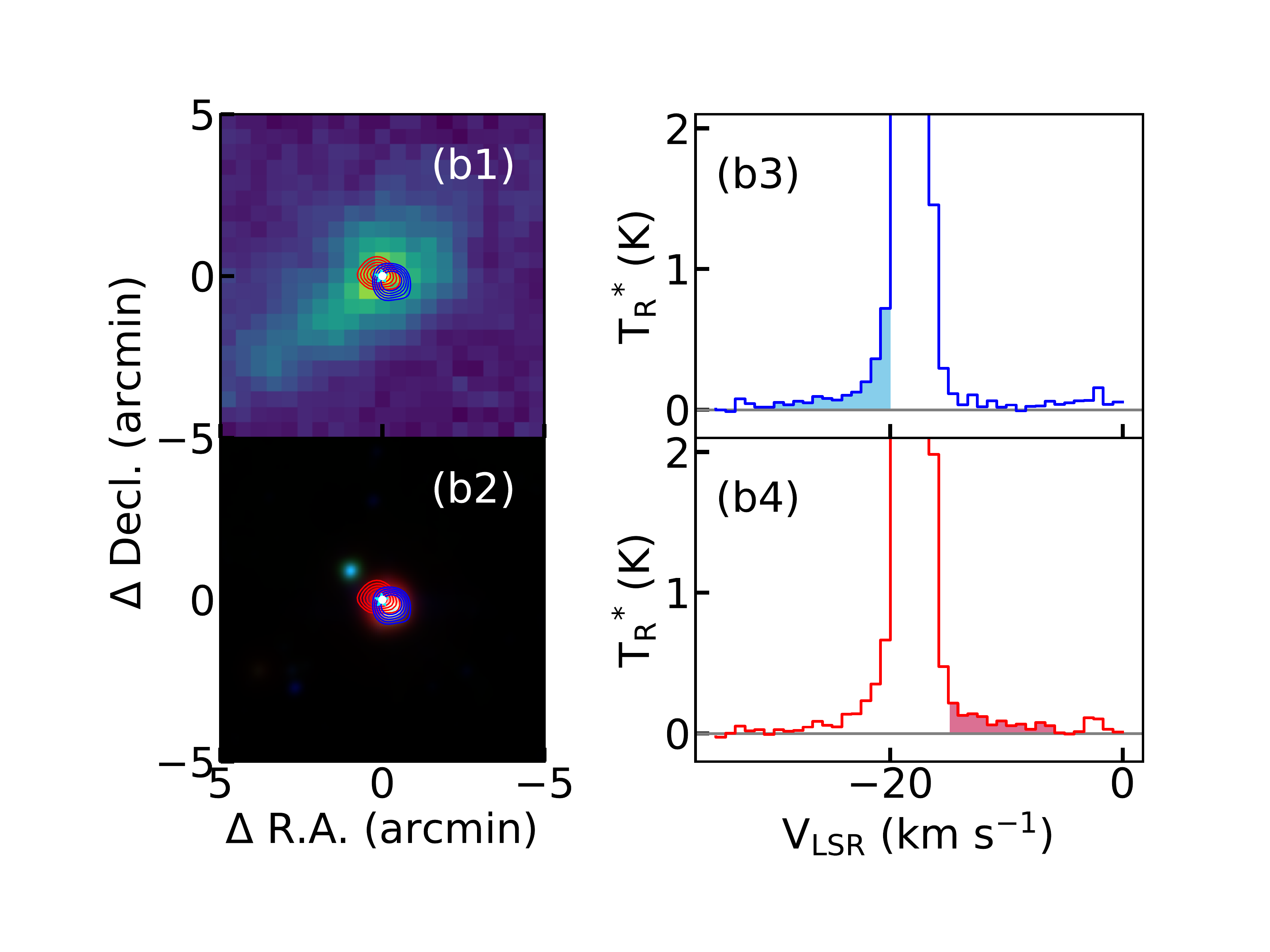}}
	\subfigure[\hco]{\includegraphics[width=8cm]{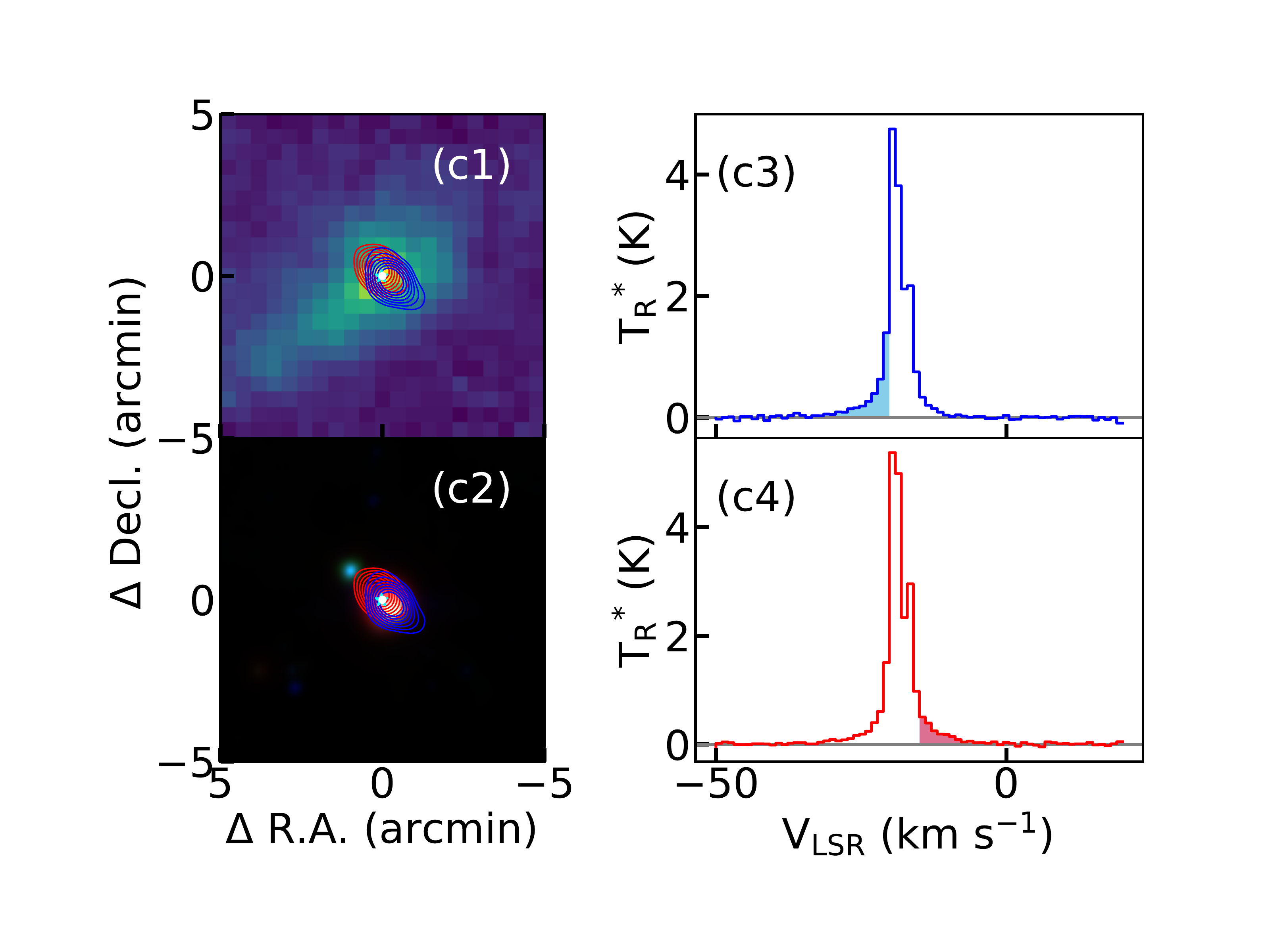}}
	\subfigure[CS]{\includegraphics[width=8cm]{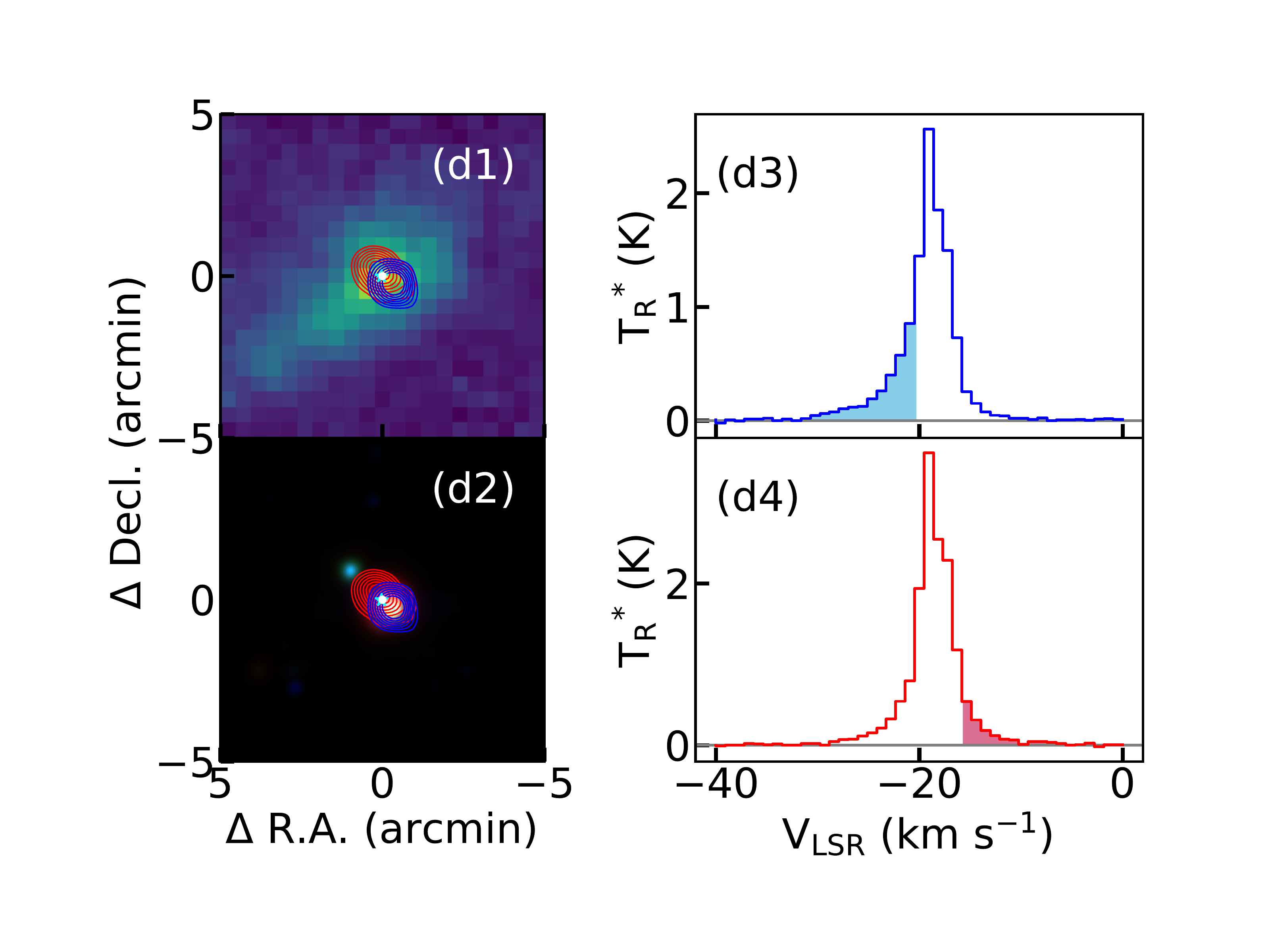}}
	\caption{Profile maps of the outflow in L1287. (a1) Integrated \xxco map with \co blue/red lobe contours (whose levels are from 30\% to 90\% of the peak intensity of each lobe). The blue star represents the IRAS source (IRAS 00038+6312). The white dot is the emission peak of \xxcoo. Position (0, 0) represents the coordinates of the source (see details in Table~\ref{tab:list}). (a2) The background shows a false-color RGB WISE image constructed using 4.6 (blue), 12 (green), and 22~$\mu$m (red) data, while the other features are the same as in panel (a1). (a3) Blue spectrum of \co at the blue emission peak position of the \co outflow. The data have been smoothed to 5$\delta v$. The blue shading of the spectrum indicates the blue line wing velocity of $^{12}$CO. (a4) Red spectrum of \co at the red emission peak position of the \co outflow. The data have been smoothed to 5$\delta v$. The red shading of the spectrum indicates the red line wing velocity of $^{12}$CO. For the other subpanels, the descriptions are the same as in panels (a1--a4), but the outflows are of (b1--b4) $^{13}$CO, (c1--c4) \hcoo, and (d1--d4) CS.}
	\label{fig:L1287}
\end{figure}

The \co bipolar outflows of L1287 were first detected by \citet{Snell+etal+1990}, and confirmed by \citet{Yang+etal+1991,Xu+etal+2006b}. Benefitting from our high-sensitivity observations (see Table~\ref{tab:noise}), the velocity range of the blue lobe of this source has been extended from $-$31~\kms \citep{Yang+etal+1991} to $-$34~\kmss. The $-$2~\kms component in the red line wing was contaminated, so the velocity range cannot be extended. After updating the velocity range, the structure of the bipolar outflow remained unchanged. It is aligned along the northeast--southwest direction (see panel (a1) of Figure~\ref{fig:L1287}), which is similar to the results of \citet{Snell+etal+1990,Yang+etal+1991,Xu+etal+2006b}.

L1287's \hco outflow was mapped by \citet{Yang+etal+1991}. The velocity ranges have been extended with our data from $-$23.8~\kms to $-$33.0~\kms for the blue lobe, and from $-$13.2~\kms to $-$6.0~\kms for the red lobe. Our results are similar to those of \citet{Yang+etal+1991}. 

The \xco and CS outflows of L1287 were also detected (see panels (c1) and (d1) of Figure~\ref{fig:L1287}, respectively), which are also aligned along the northeast--southwest direction. However, the distances between the emission peaks of the red and blue lobes of the \xcoo, \hco and CS outflows are smaller than those of \coo. Compared with the \co outflow, the locations of the other three outflows are shifted a little to the southeast direction, so that the midpoint of the red and blue lobes is closer to the position of the IRAS source. Meanwhile, both the \co and \hco outflows have larger blue lobes than red ones, while the sizes of the two lobes for \xco and CS are similar. 

IRAS 00338+6312 (see the star in Figure~\ref{fig:L1287}) is located at the center of these bipolar outflows. There is also WISE emission located at the center of the four groups of bipolar outflows. The core traced by CS, HCN, \hcoo, and NH$_{3}$ \citep{Walker+Masheder+1997,Zinchenko+etal+1997} are all associated with the IRAS source. Meanwhile, the emission peak of \xxco is associated with the IRAS source. All of these tracers seem to indicate that the same source drives all the bipolar outflows.

\subsubsection{G176.51+00.20}
\begin{figure}
	\centering
	\vspace{-0.3cm}
	\subfigure[\co]{\includegraphics[width=8cm]{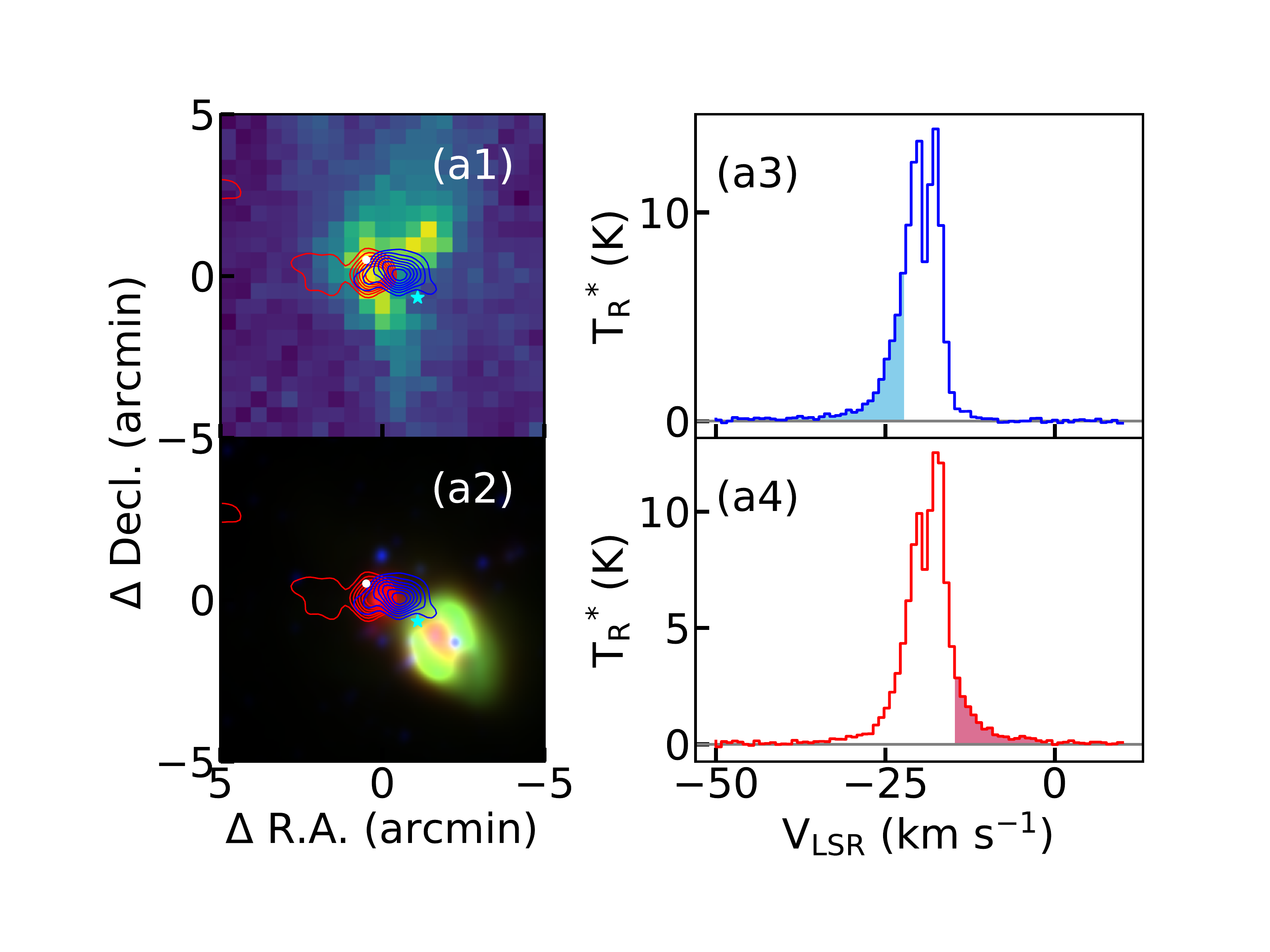}}
	\subfigure[\xco]{\includegraphics[width=8cm]{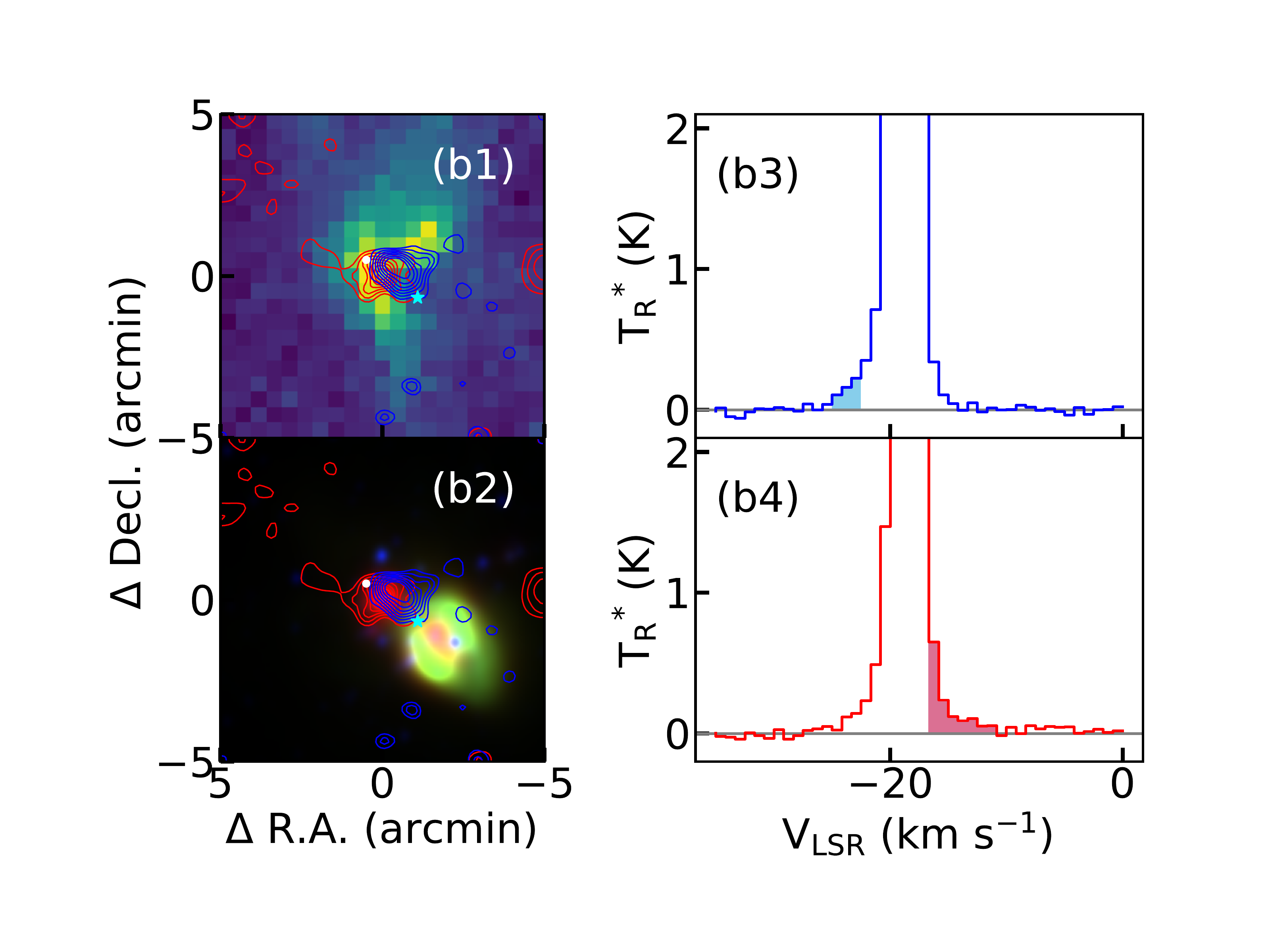}}
	\subfigure[\hco]{\includegraphics[width=8cm]{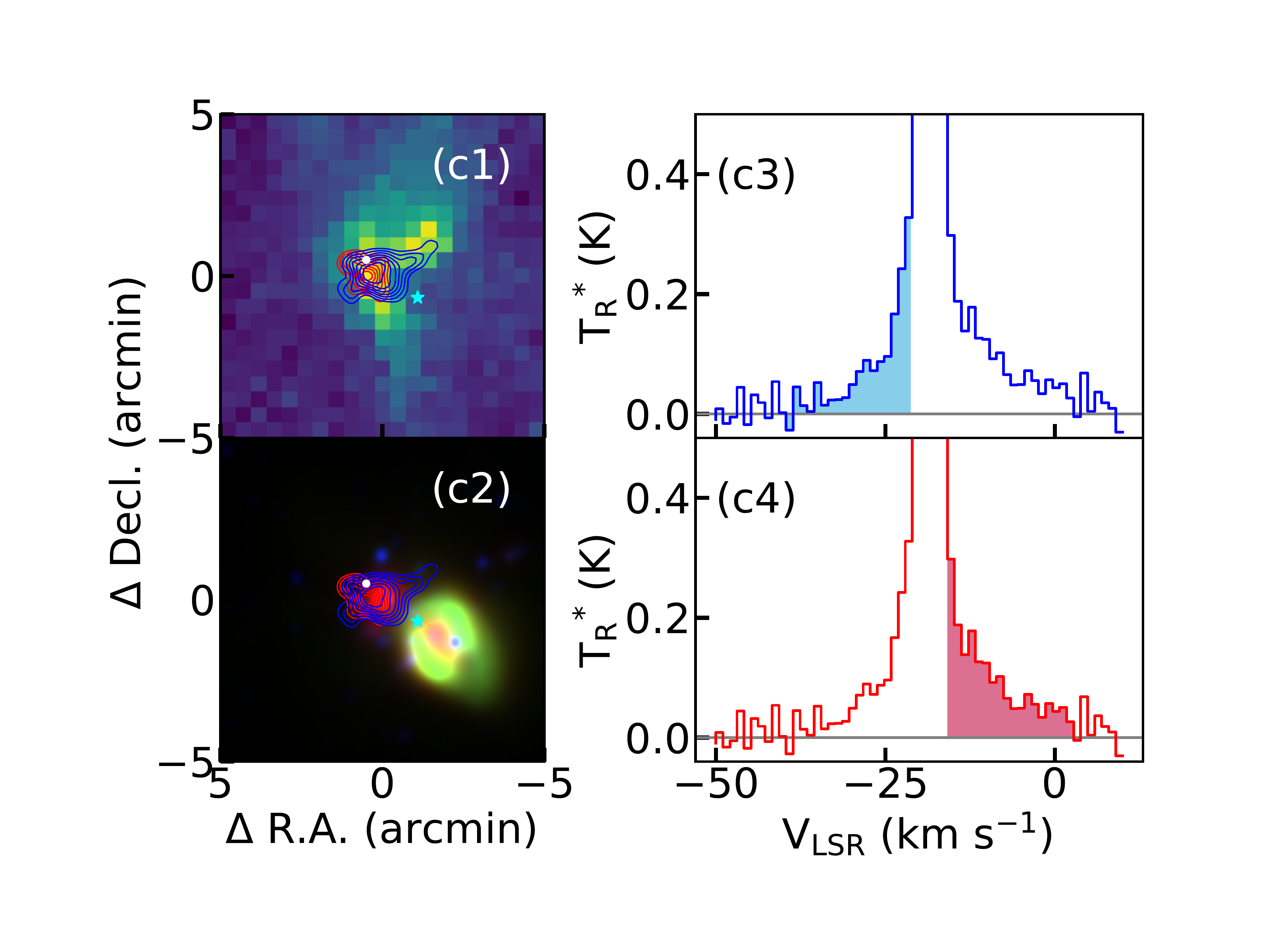}}
	\subfigure[CS]{\includegraphics[width=8cm]{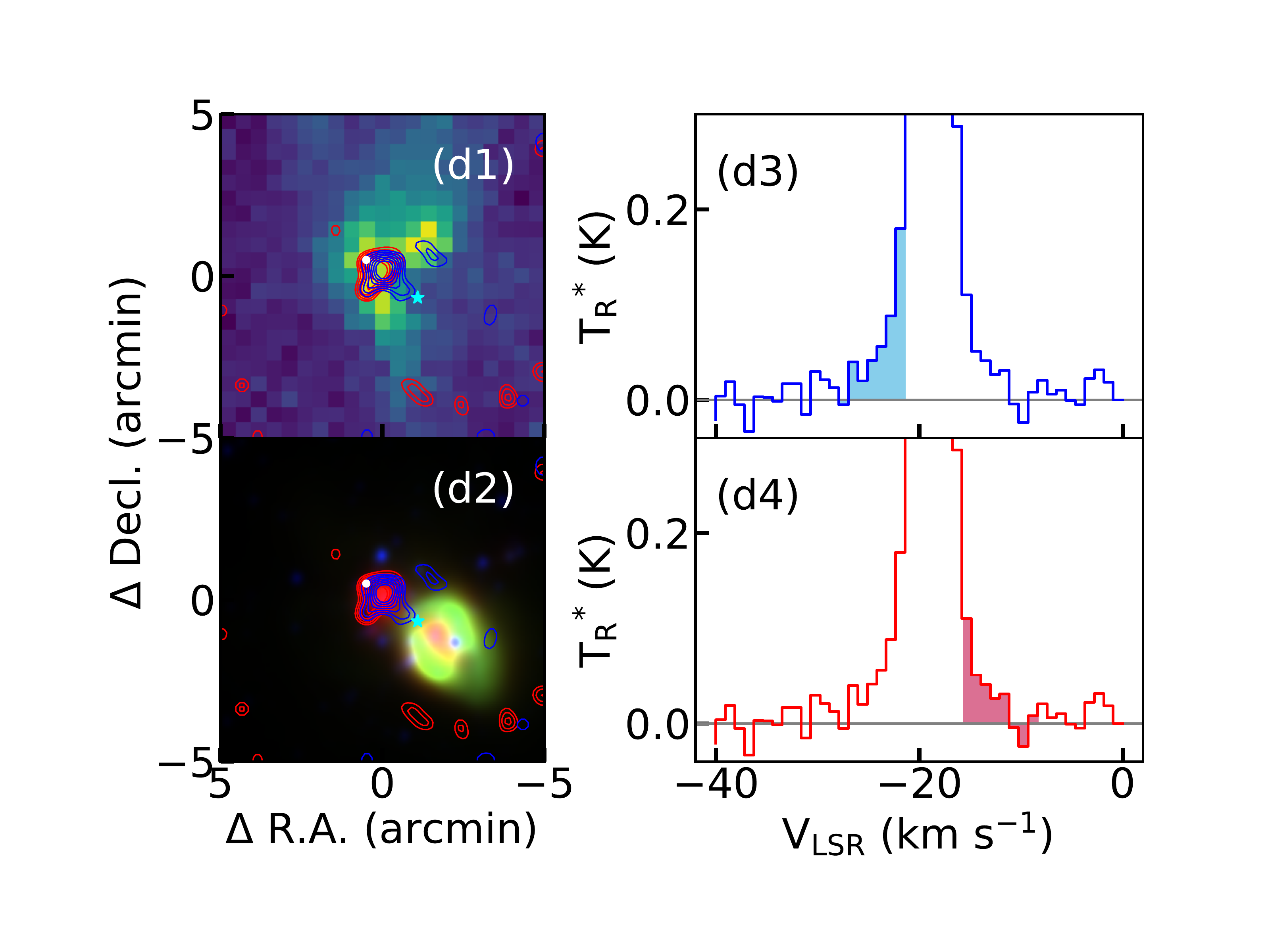}}
	\caption{Profile maps of the outflows from G176.51+00.20. The description of each map is the same as that in Figure~\ref{fig:L1287}, except the blue star is IRAS 05345+3157.}
	\label{fig:G17651}
\end{figure}

The velocity range has been extended from $-$30~\kms to $-$47~\kms in the blue wing and from $-$5~\kms to $-$1~\kms in the red wing of the \co outflow from G176.51+00.20 \cite[which is also named AFGL~5157, and NGC~1985, ][see panel (a1) of Figure~\ref{fig:G17651}]{Snell+etal+1988}. The \co bipolar outflows are along the east--west direction, which is coincident with the results of the study of \citet{Snell+etal+1988}.

The bipolar outflows of \hcoo, \xcoo, and CS have been mapped for the first time in this work (see panels (b1), (c1), and (d1) of Figure~\ref{fig:G17651}, respectively). The directions of the \hcoo, \xcoo, and CS outflows are similar to those of the \co outflows. The distances between the emission peaks of the red and blue lobes of the \hcoo, \xcoo, and CS outflows are also smaller than those of \coo. Different from the other outflows, however, the red and blue lobes of the CS outflow are very close to each other, and its structure is not as extended as the other outflows. We also found that the velocity range of the red lobe of the \hco outflow is even broader than that of the \co outflow (see Table~\ref{tab:parameters} and Figure~\ref{fig:G17651}).

There is WISE emission located at the center of bipolar outflows, which could be the source of excitation. Although the emission is faint, it is associated with the emission peak of \xxcoo. The nearest IRAS source, IRAS 05345+3157 (see the star in Figure~\ref{fig:G17651}), is located near the emission peak of the blue lobe, and it might be the source of excitation of the outflows.

\subsubsection{Mon R2}
\begin{figure}
	\centering
	\vspace{-0.3cm}
	\subfigure[\co]{\includegraphics[width=8cm]{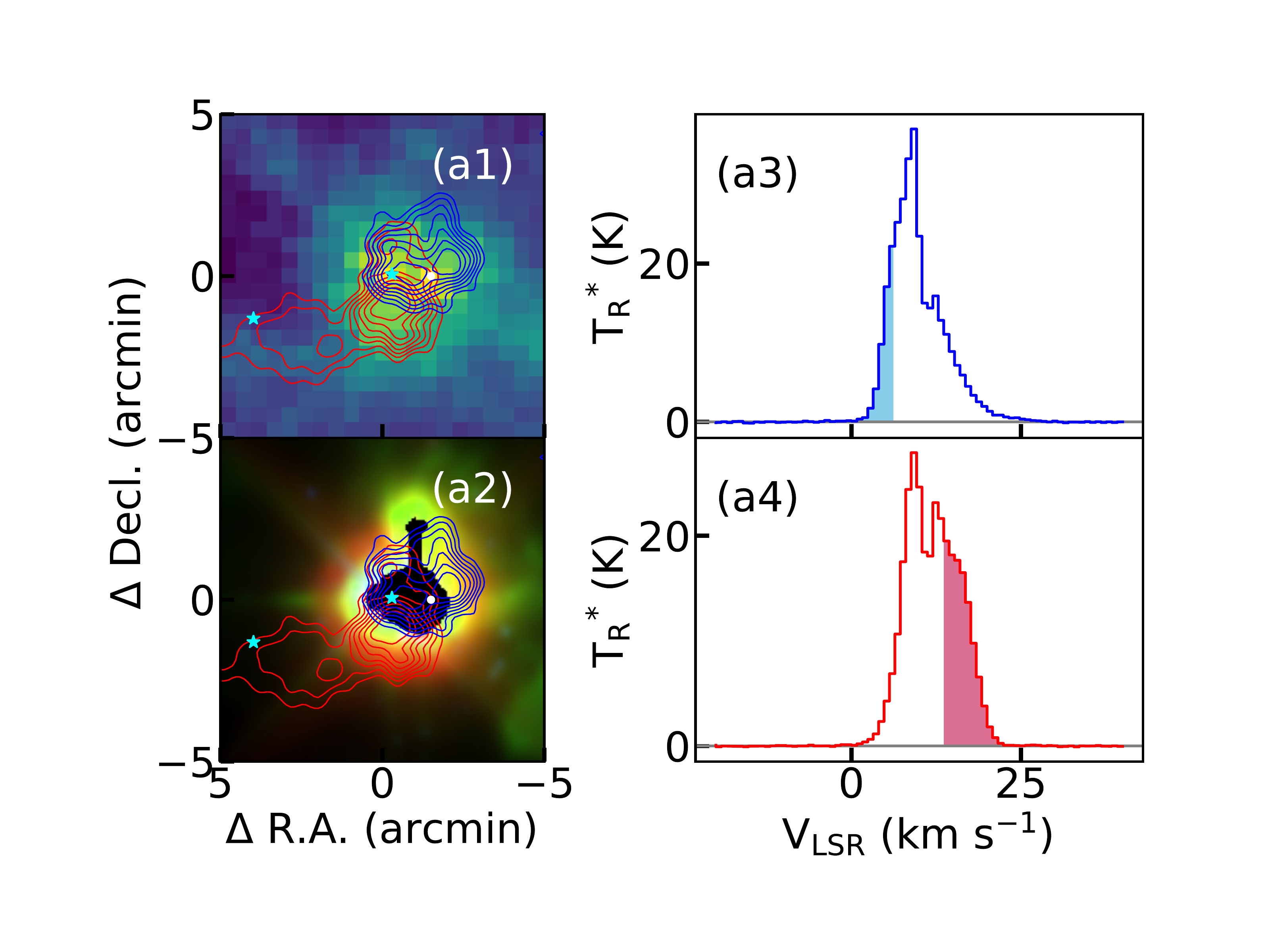}}
	\subfigure[\xco]{\includegraphics[width=8cm]{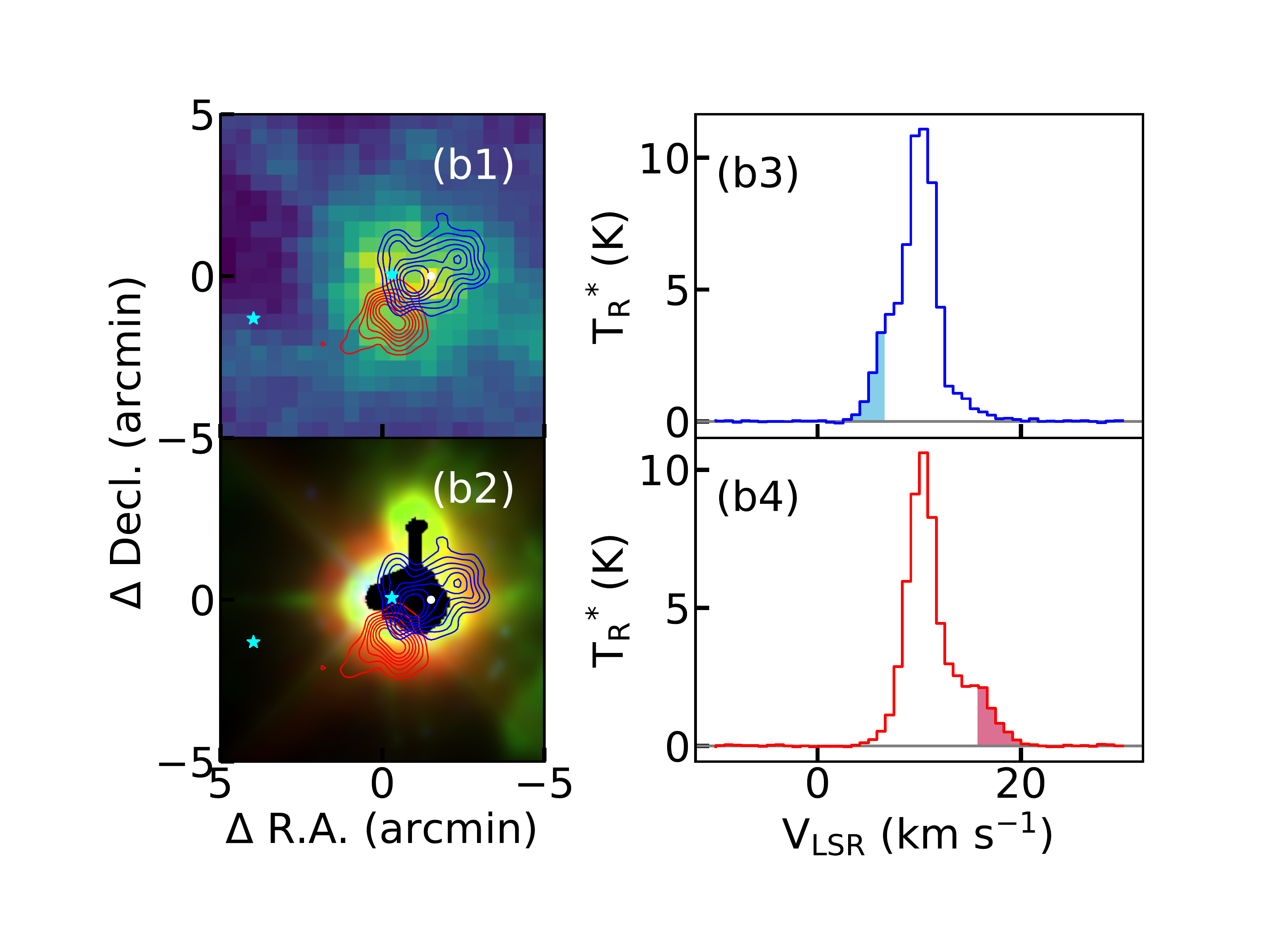}}
	\subfigure[\hco]{\includegraphics[width=8cm]{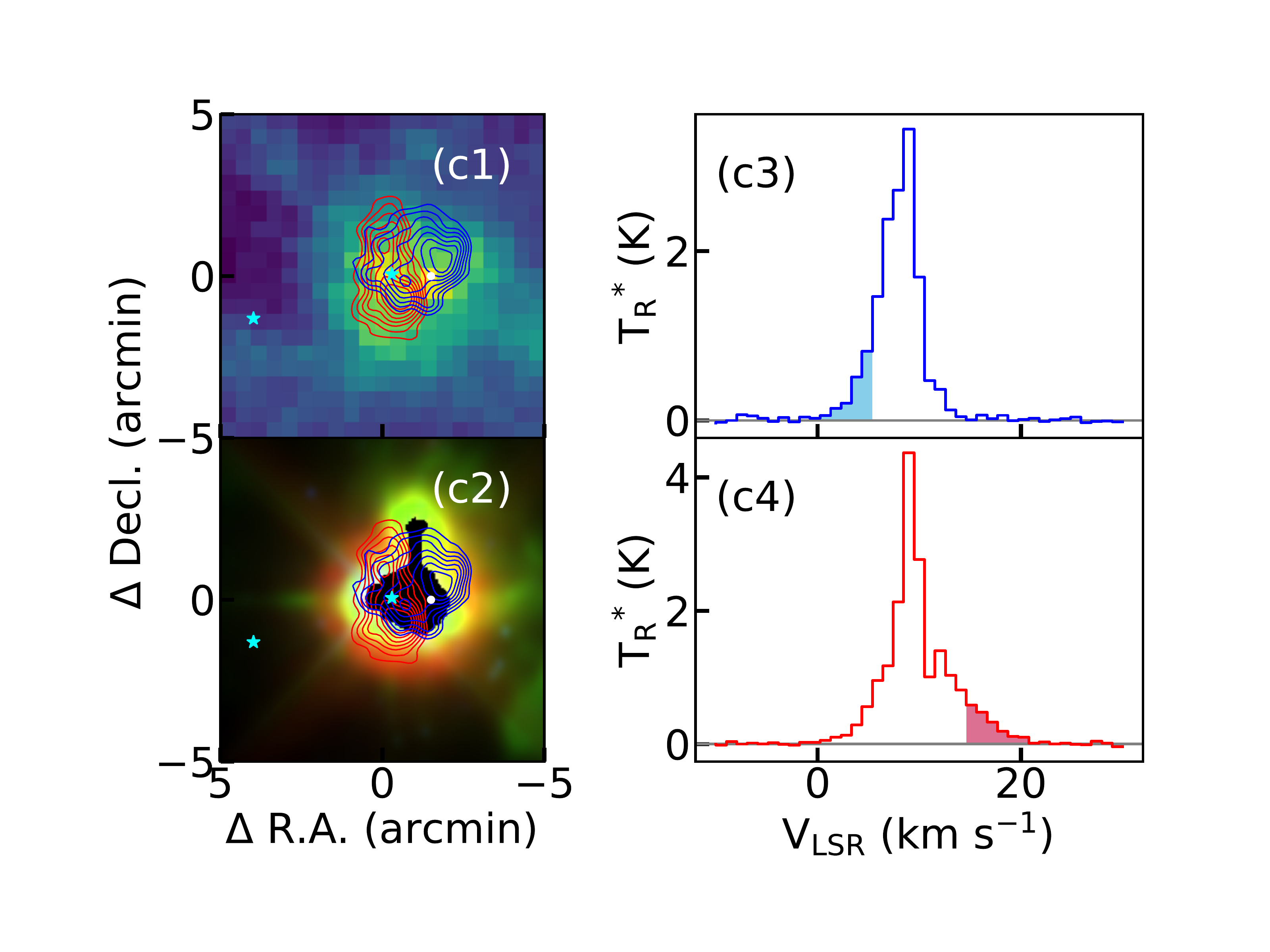}}
	\subfigure[CS]{\includegraphics[width=8cm]{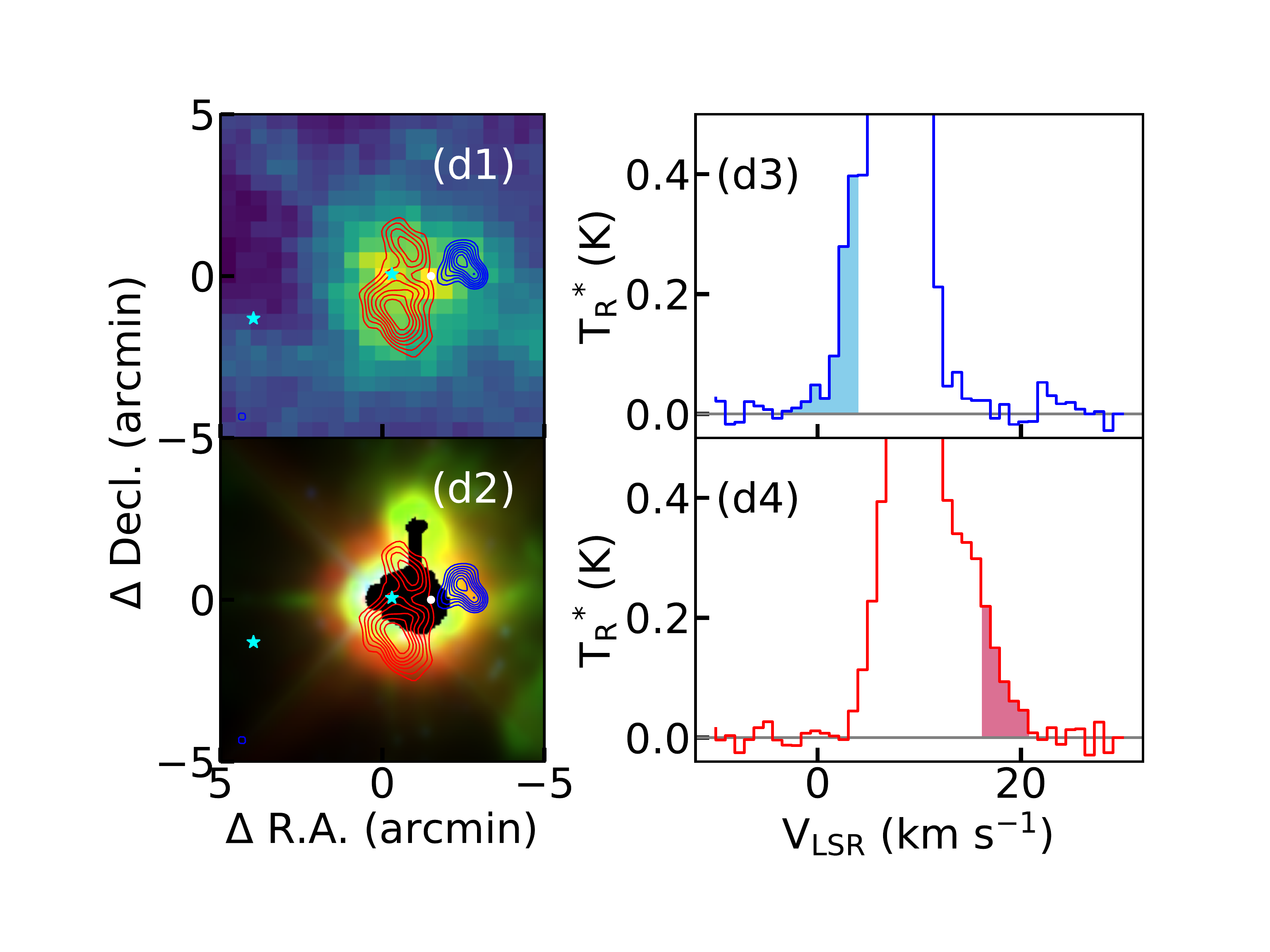}}
	\caption{Profile maps of the outflows of Mon R2. The description of each map is the same as that in Figure~\ref{fig:L1287}. The blue star at the center is IRAS 06035-0622, and that in the southeast is IRAS 06056-0621.}
	\label{fig:MonR2}
\end{figure}

We detected \coo, \hcoo, \xcoo, and CS outflows in Mon R2. The \co bipolar outflow of Mon R2 has been widely studied by many researchers \citep{Bally+Lada+1983,Meyers-Rice+Lada+1991,Xu+etal+2006b}. The velocity range of the red wing of the \co outflow has been extended from 22~\kms \citep{Meyers-Rice+Lada+1991} to 30~\kmss. This source is so complex that observations with different resolutions may result in different structural details of its outflows. With beam size used here (i.e., $\sim 50 ''$), the pair of bipolar outflow in the north \cite[see][]{Xu+etal+2006b} partly overlap with the southern bipolar outflow and perhaps also with some surrounding gas. Therefore, the outflow appears relatively extended (see panel (a1) of Figure~\ref{fig:MonR2}). This morphology is similar to the outflow shown in figure 1 in \citet{Meyers-Rice+Lada+1991}, where their beam size ($\sim 60 ''$) was similar to ours. Thus, the structure  ascertained in this work is similar to that of \citet{Meyers-Rice+Lada+1991}, the second pair of bipolar outflows is hard to be separated under current resolution, which might be interferenced by other components. 

The \hcoo, \xcoo, and CS outflows of this source were mapped for the first time here. Comparing the four groups of outflows, both the \co and \hco outflows present two red and blue emission peaks, although these peaks are hard to separate from each other. However, the two peaks in the red lobe of the CS outflow could be easily separated. Furthermore, from the \xco outflow, only one emission peak can be seen in the red lobe (see Figure~\ref{fig:MonR2}).

IRAS 06053-0622 is located at the center of the bipolar outflows and near the emission peak of \xxcoo. Meanwhile, the WISE emission of this region is so strong that the WISE data are saturated. All of these tracers seem to indicate that the IRAS source is the source of excitation of these outflows.

\
\subsubsection{NGC2264}
\begin{figure}
 	\centering
 	\vspace{-0.3cm}
 	\subfigure[\co]{\includegraphics[width=8cm]{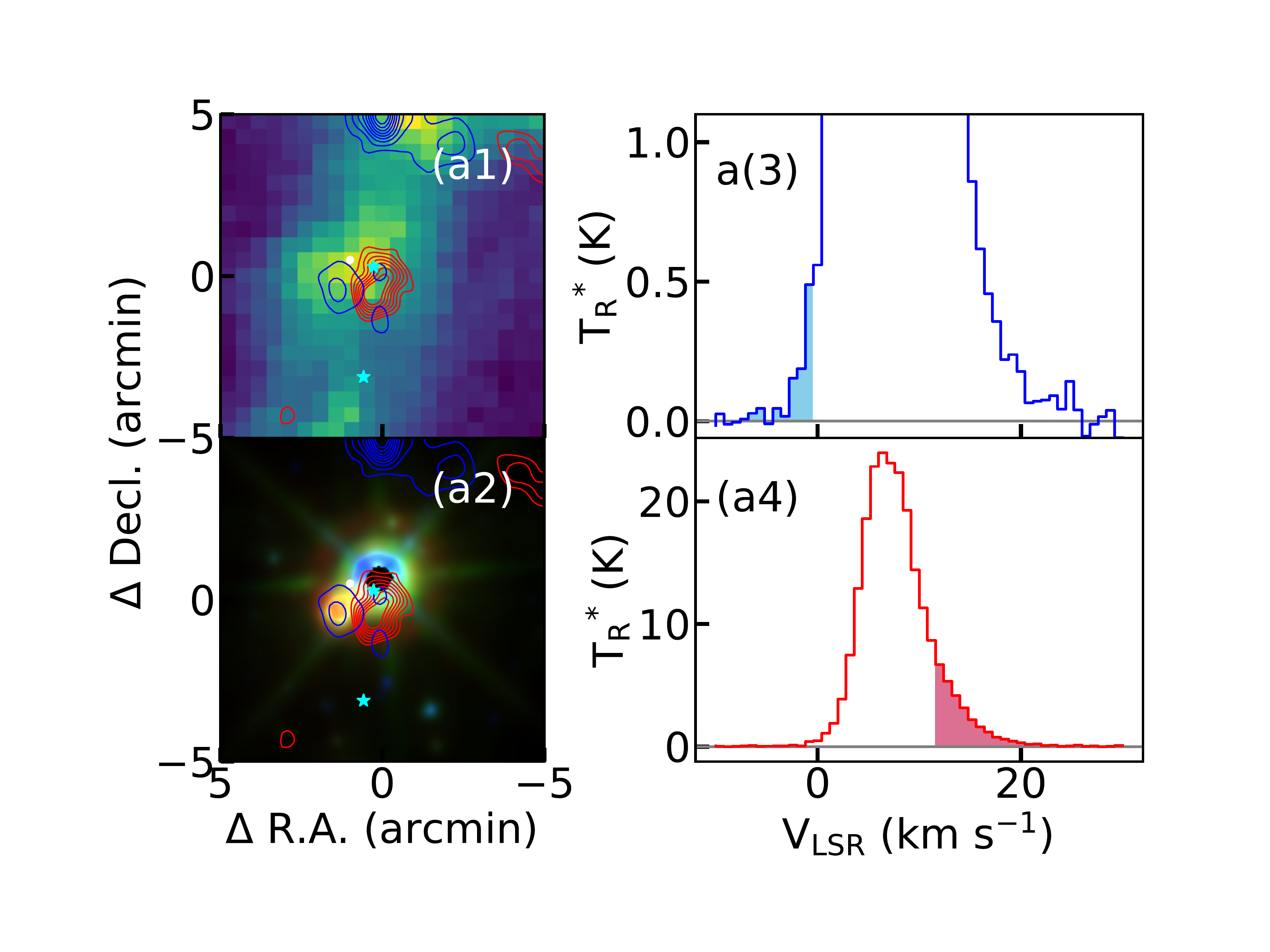}}
 	\subfigure[\xco]{\includegraphics[width=8cm]{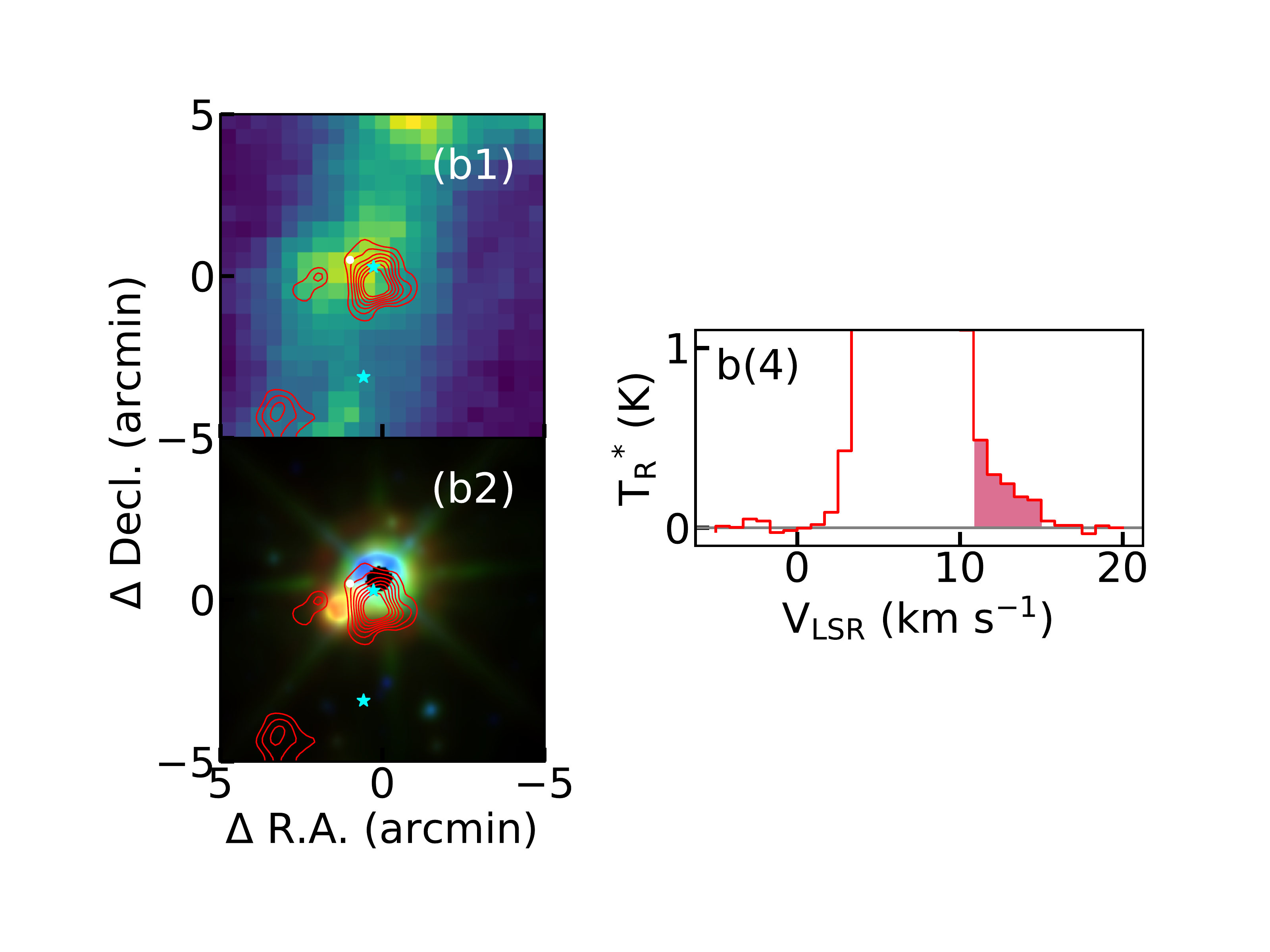}}
 	\subfigure[\hco]{\includegraphics[width=8cm]{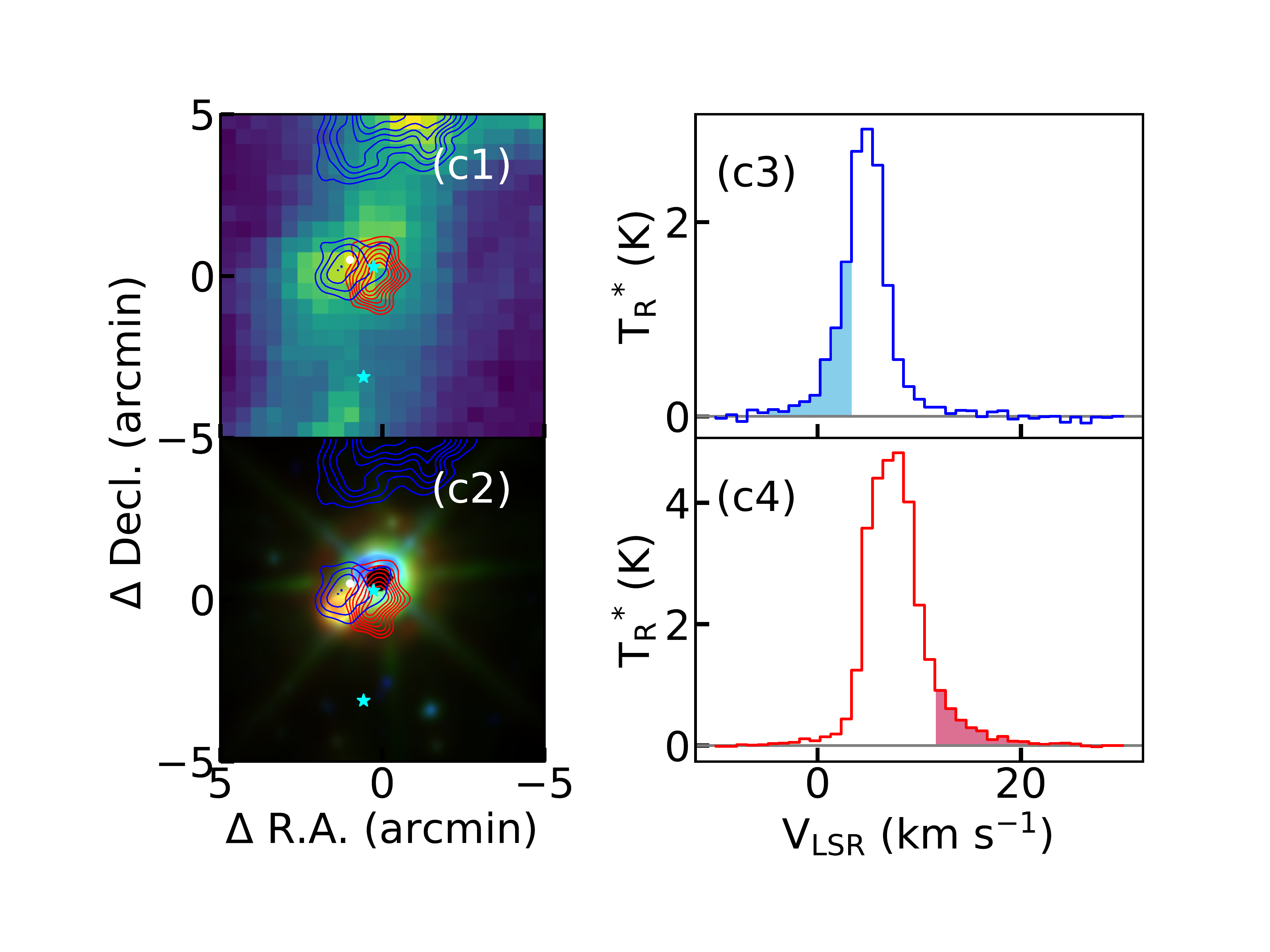}}
 	\subfigure[CS]{\includegraphics[width=8cm]{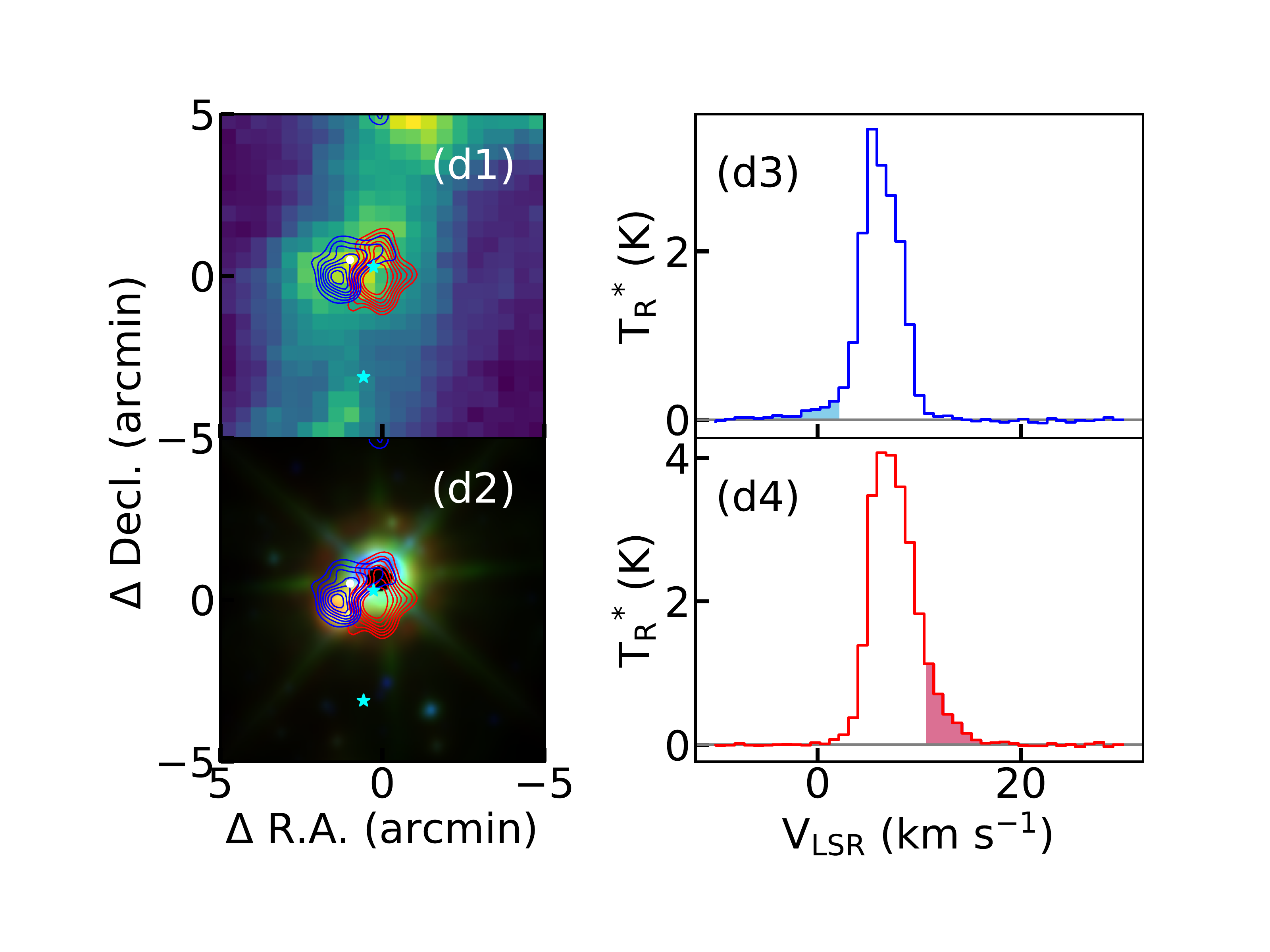}}
 	\caption{Profile maps of the outflow in NGC2264. The contour levels of the \co outflow are from 20\% to 90\% of the peak intensity of each outflow lobe. The blue star at the center is IRAS 06384+0932, and that at the south is IRAS 06384+0929. The description of each map is the same as that in Figure~\ref{fig:L1287}}
 	\label{fig:NGC2264}
\end{figure}

Previously, only a \co red lobe was detected toward NGC2264 \citep{Bally+Lada+1983,Margulis+etal+1988}; however, we have also detected a blue lobe (see panel (a1) of Figure~\ref{fig:NGC2264}). Therefore, the \co outflow of NGC2264 is not, as once believed, a one-side red lobe, but instead is a bipolar outflow. The emission in the north is so strong (see blue contour in panel (a1) of Figure~\ref{fig:NGC2264}) that it has a disastrous effect on the mapping process of its blue lobe outflow; hence, the blue lobe is easily to be ignored. Thanks to the high-sensitivity observations obtained with PMODLH, we successfully mapped the blue lobe of the outflow.  

The \xcoo, \hcoo, and CS outflows of this source were detected for the first time. Except for \xcoo, the other outflows of this source are bipolar and distributed along the east--west direction. In fact, the blue line wings of \xco is higher than the Gaussian fitting, but the structures of the blue lobe in the contour map are too chaotic. Hence, the blue lobe of \xcoo, if present, was not detected. The intensities of the blue lobes of \hco and CS are stronger than that of \co (see Figure~\ref{fig:NGC2264}).

IRAS 06384+0932 is located at the center of the bipolar outflows. Meanwhile, the WISE emission near the outflow is saturated in this region. Furthermore, the emission peak of \xxco is also near the IRAS source and WISE emission. Therefore, IRAS 06384+0932 may be the source of excitation of the outflows.

\subsubsection{G090.21+02.32}
\begin{figure}
	\centering
	\vspace{-0.3cm}
	\subfigure[\co]{\includegraphics[width=8cm]{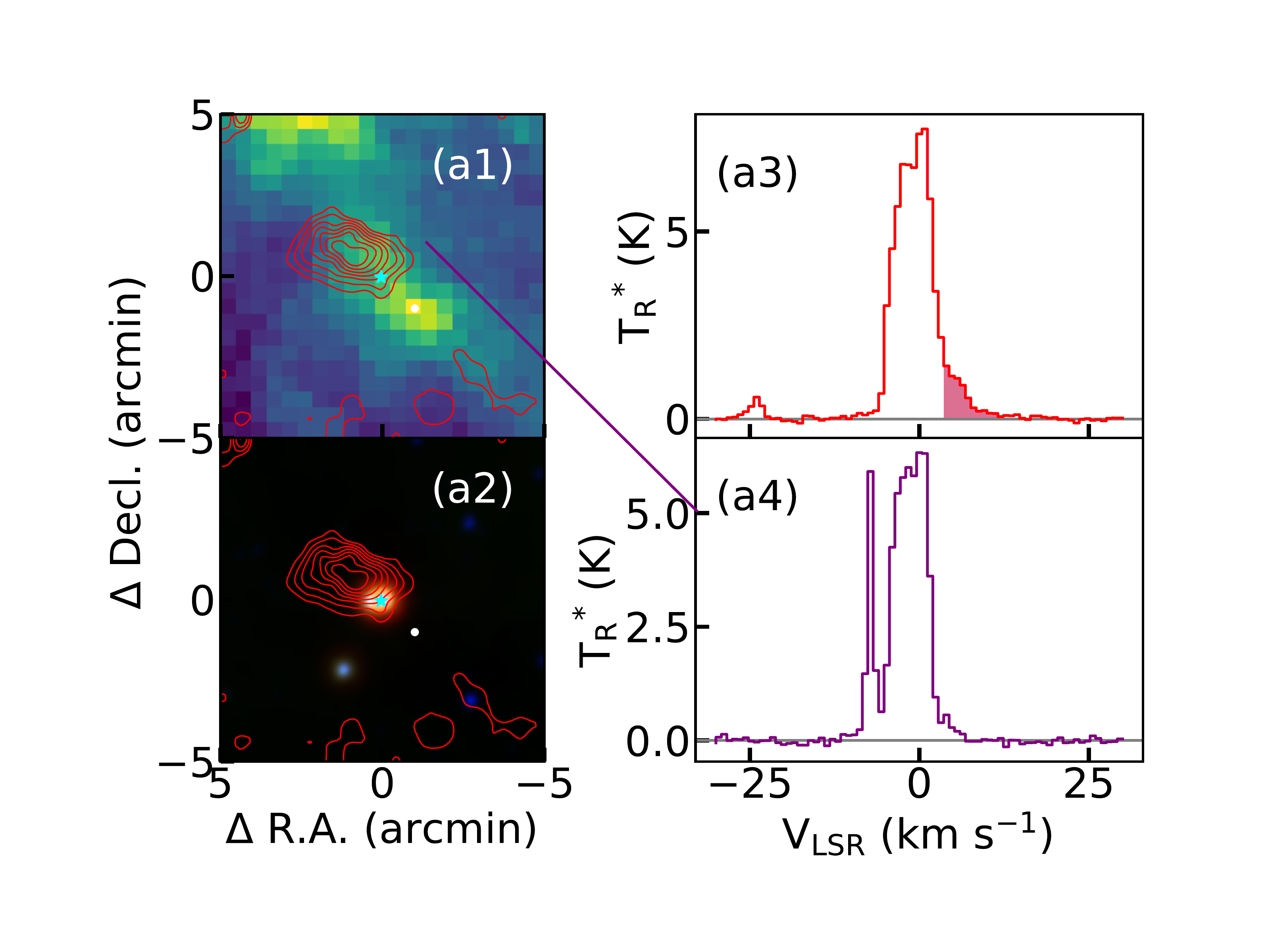}}
	\subfigure[\xco]{\includegraphics[width=8cm]{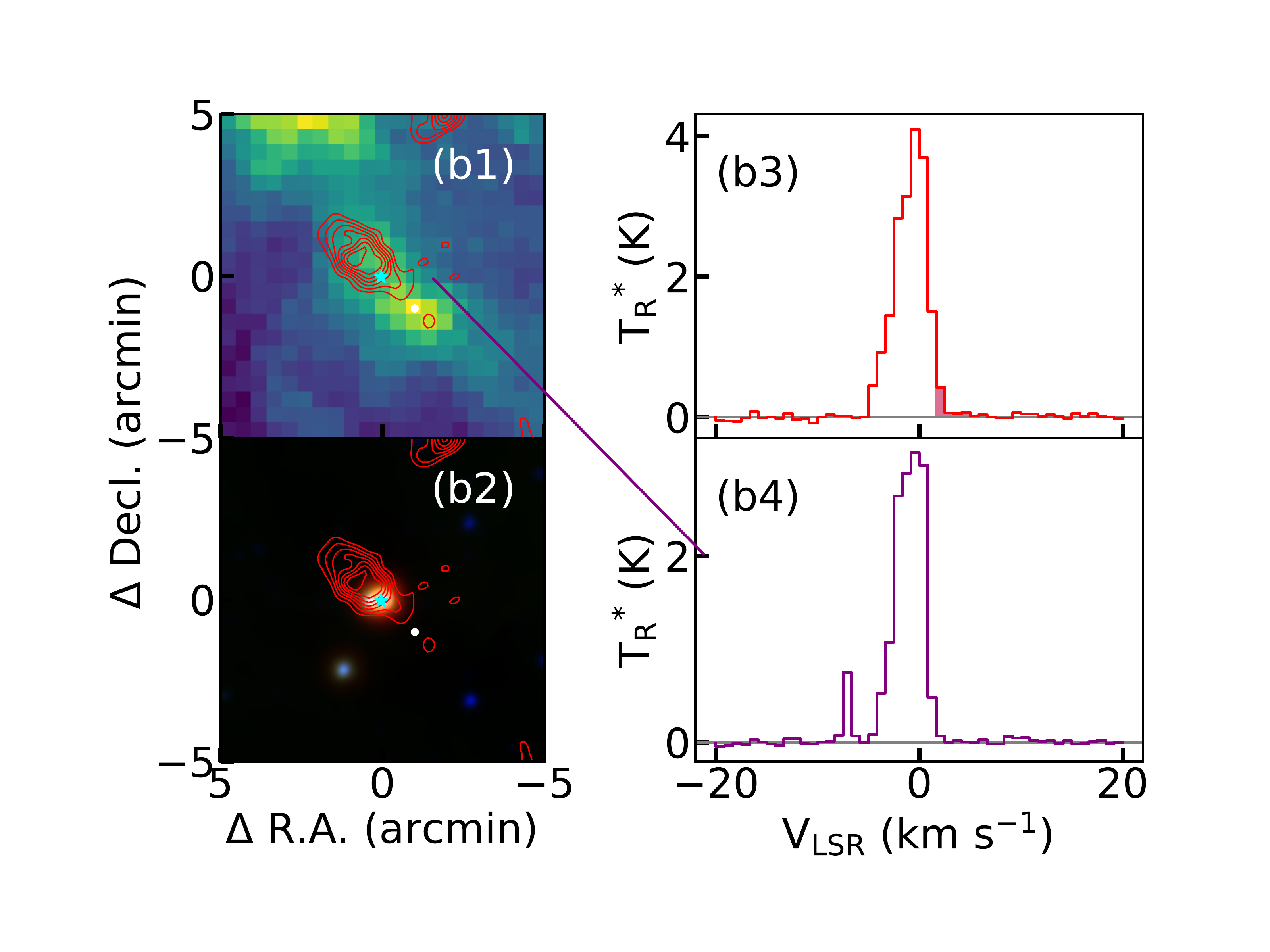}}
	\subfigure[\hco]{\includegraphics[width=8cm]{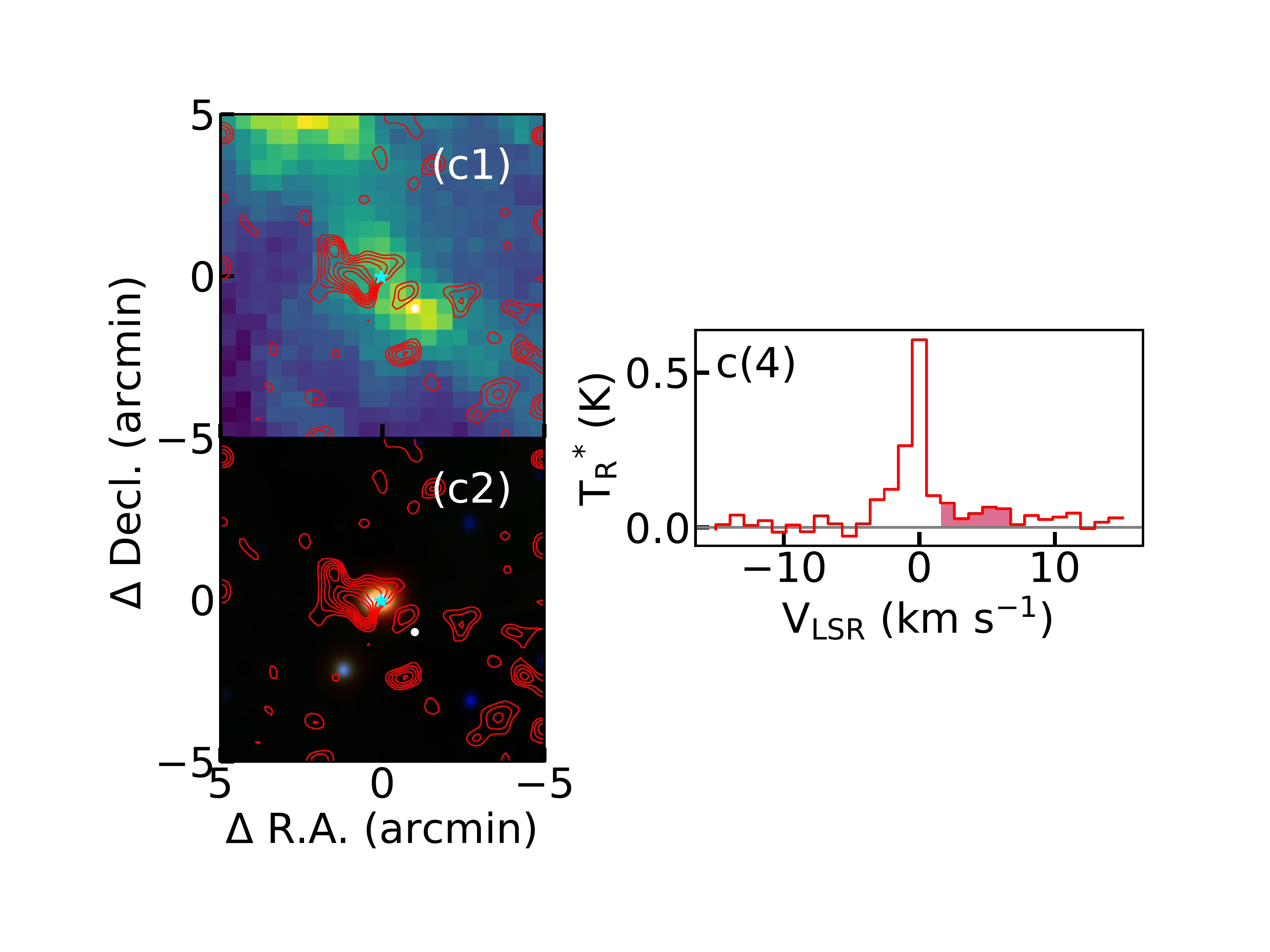}}
	\subfigure[CS]{\includegraphics[width=8cm]{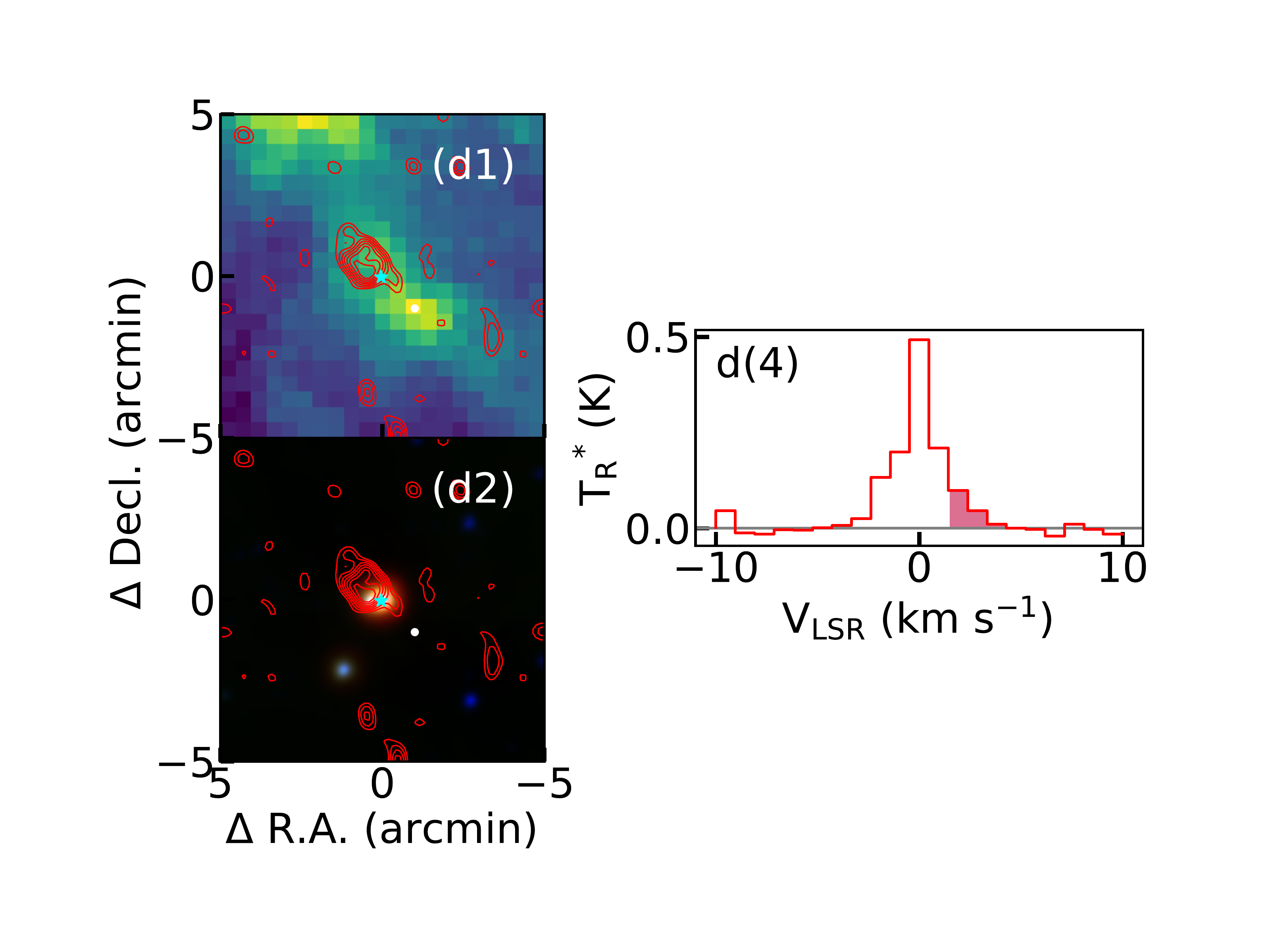}}
	\caption{Profile maps of the outflows in G090.21+02.32. Panel (a4) shows the \co spectrum at the position which is marked by the endpoint of the straight purple line in (a1), and the position is the peak position of the blue emission found by Clark (1986). Panels (b1--b4) are similar to panels (a1--a4), while the spectrum is from \xcoo. The description of each map is the same as that in Figure~\ref{fig:L1287}.}
	\label{fig:G09021}
\end{figure}

The \co bipolar outflow of G090.21+02.32 was first detected by \citet{Clark+1986} (which is near the dark cloud L988a). However, there is strong emission in the blue wing at the peak of the blue lobe in \citet{Clark+1986} (see panel (a4) of Figure~\ref{fig:G09021}), but this emission cannot be the blue lobe of the \co outflow. Instead, we suggest that G090.21+02.32 only possesses a red lobe rather than a bipolar \co outflows.

We also detected \xcoo, \hcoo, and CS outflows for the first time for this source. There are multi-velocity components at $\sim$1.0~\kms in the CS and \xco line profiles. Therefore, we considered that there might be two features along the line of sight (see panel (e) of Figure~\ref{fig:spectra}), and G090.21+02.32 only has red lobes of \xcoo, \hco and CS. The emission intensities of \hco and CS of this source, especially the red lobes of the outflows, are weak.

IRAS 21007+4951 is located near the outflow, which is associated with the H$_2$O maser, UC H$_{\rm II}$ and WISE emission \citep{Xu+etal+2013,Wood+Churchwell+1989}. All of these dense tracers indicate that the IRAS source might be a possible excitation source of the outflows.

\subsubsection{G105.41+09.87}
\begin{figure}
	\centering
	\vspace{-0.3cm}
	\subfigure[\co]{\includegraphics[width=8cm]{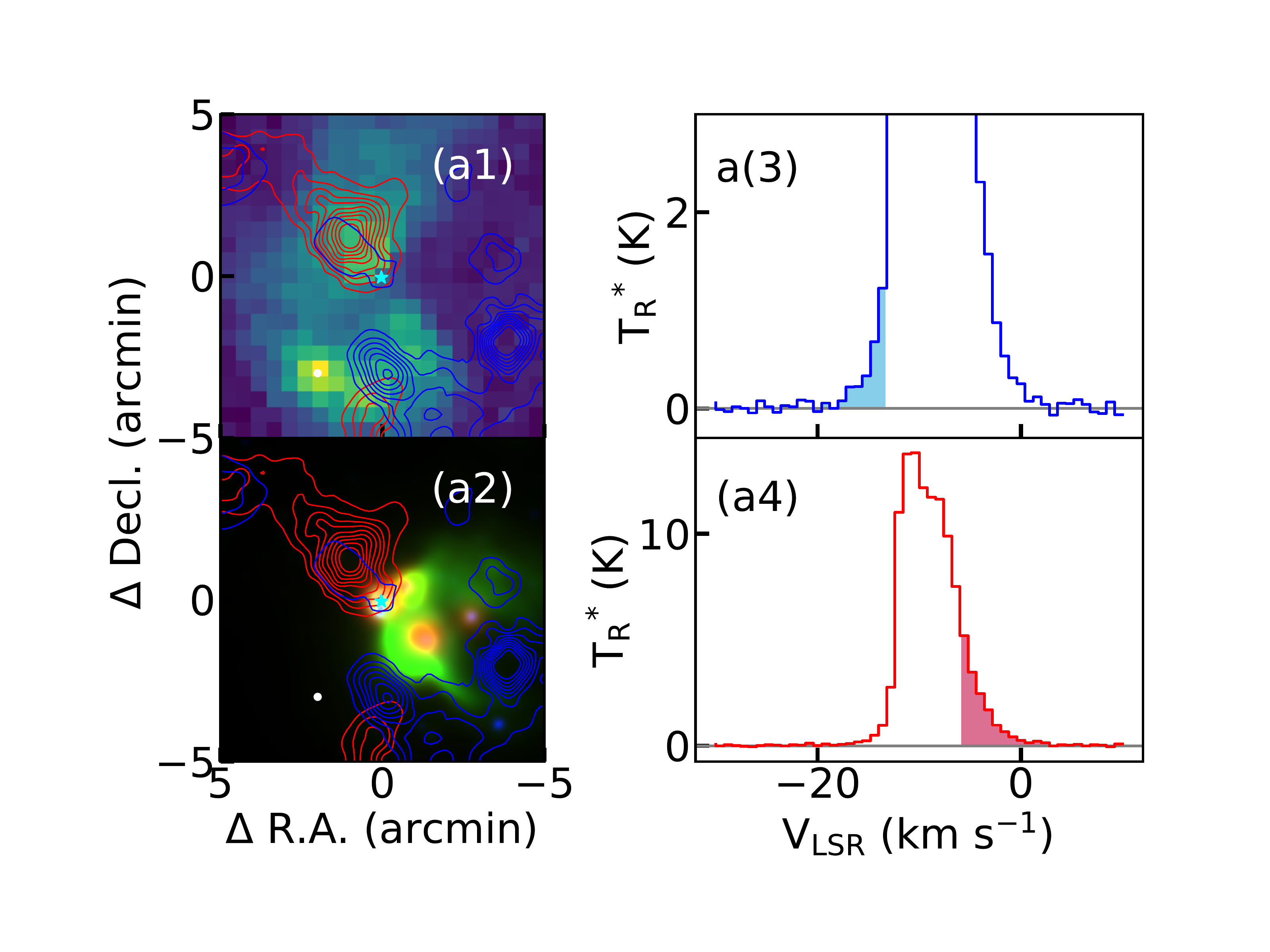}}
	\subfigure[\xco]{\includegraphics[width=8cm]{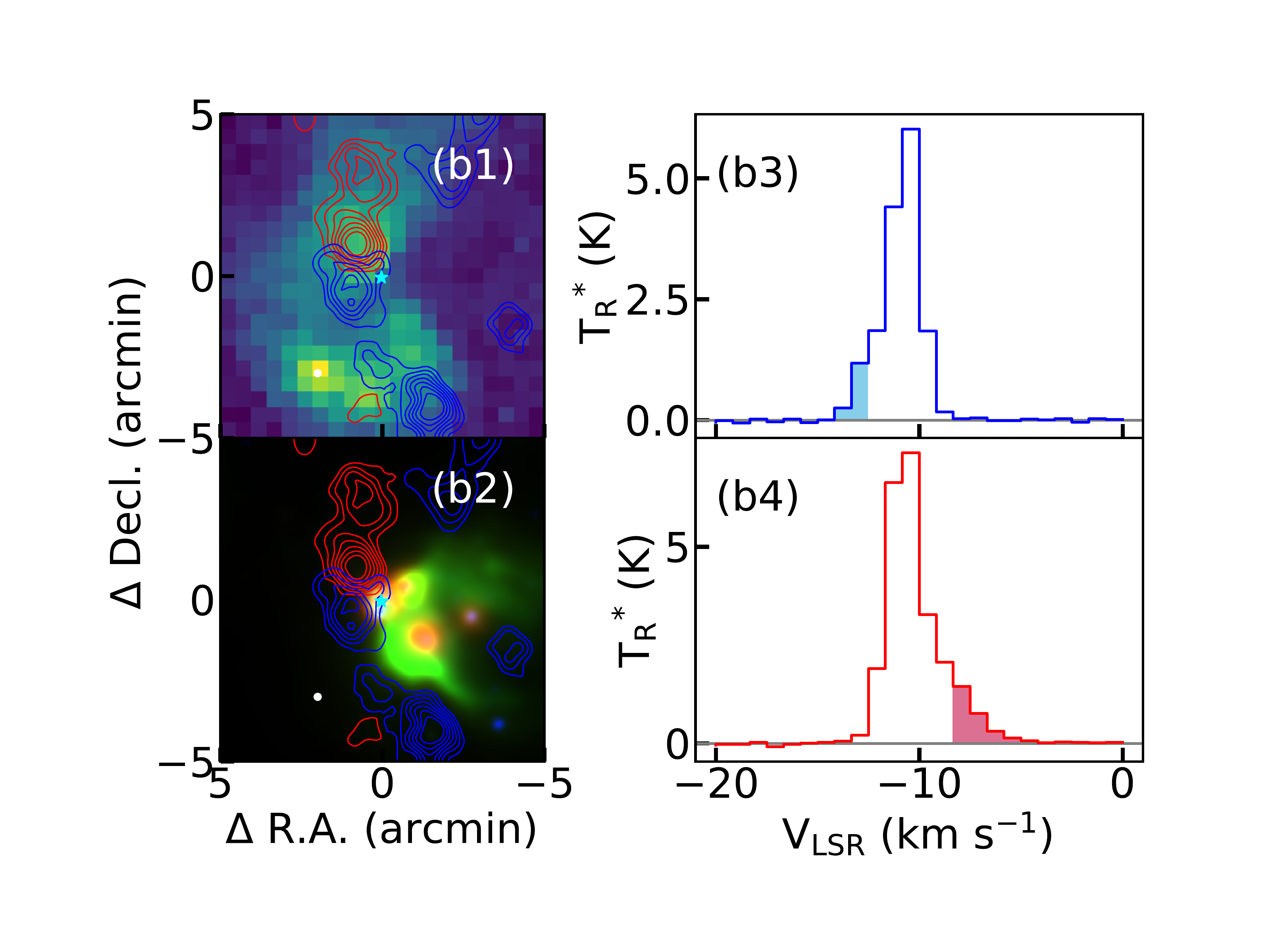}}
	\subfigure[\hco]{\includegraphics[width=8cm]{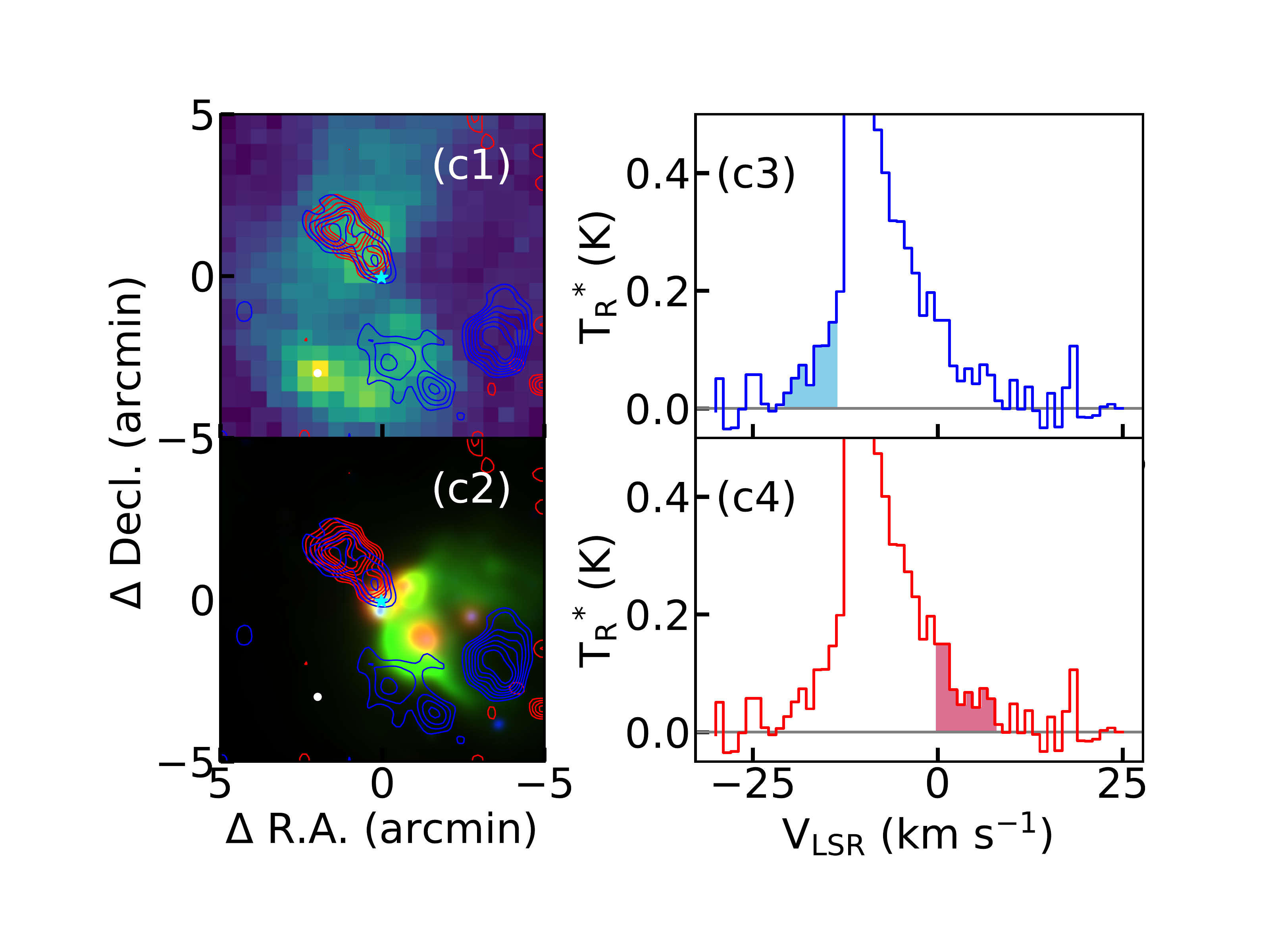}}
	\subfigure[CS]{\includegraphics[width=8cm]{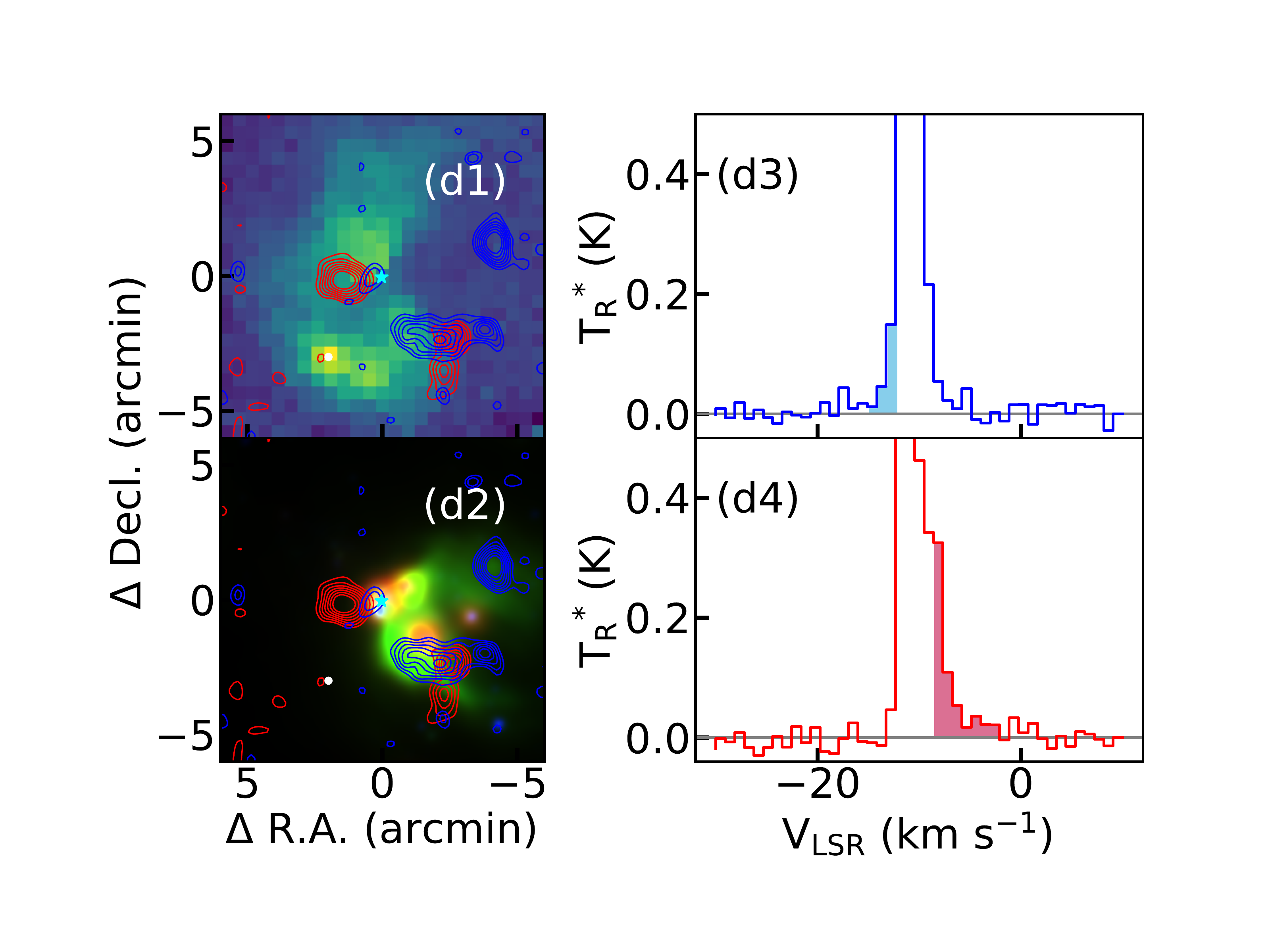}}
	\caption{Profile maps of the outflows in G105.41+09.87. The blue star at the center is IRAS 21418+6552, and the contour levels of the \co outflow are from 10\% to 90\% of the peak intensity of each outflow lobe. The other descriptions of each map are the same as those in Figure~\ref{fig:L1287}.}
	\label{fig:G10541}
\end{figure}

G105.41+09.87 is near the open cluster NGC~7129 \citep{Trinidad+etal+2004}, and this region is extremely complex. It includes two far-infrared sources \cite[i.e., NGC~7129 FIRS1 and NGC~7129 FIRS2, ][]{Bechis+etal+1978} and two Herbig-Haro objects \cite[i.e., HH~103 and HH~105][]{Edwards+Snell+1983}. \citet{Fuente+etal+2001} detected multiple \co outflows in this region. We also detected three \co bipolar outflows, where two are located close to the edge of the figure and the other one is located at the center (see panel (a1) of Figure~\ref{fig:G10541}). In this work, we have mainly analyzed the outflows at the center. 

The \xcoo, \hco and CS outflows were detected for the first time. We detect \hcoo, \xco and CS bipolar outflows in this region. Similar to the \co outflow, we mainly analyzed the bipolar outflow at the center. The \co outflow of G105.41+09.87 is extended, while the \xcoo, \hco and CS outflows are more concentrated, especially the CS outflows (see Figure~\ref{fig:G10541}).

There are two emission peaks of \xxco in this region. One is located at the center, which is associated with the IRAS source (IRAS 21418+6552) and the WISE emission, and the other one is located in the south. An H$_2$O maser is associated with the IRAS source \citep{Xu+etal+2013}, so the outflow in the center might be associated with it. Meanwhile, the emission of \xxco in the south might be associated with the outflow in the south.
\subsubsection{IRAS 22198+6336}
\begin{figure}
	\centering
	\vspace{-0.3cm}
	\subfigure[\co]{\includegraphics[width=8cm]{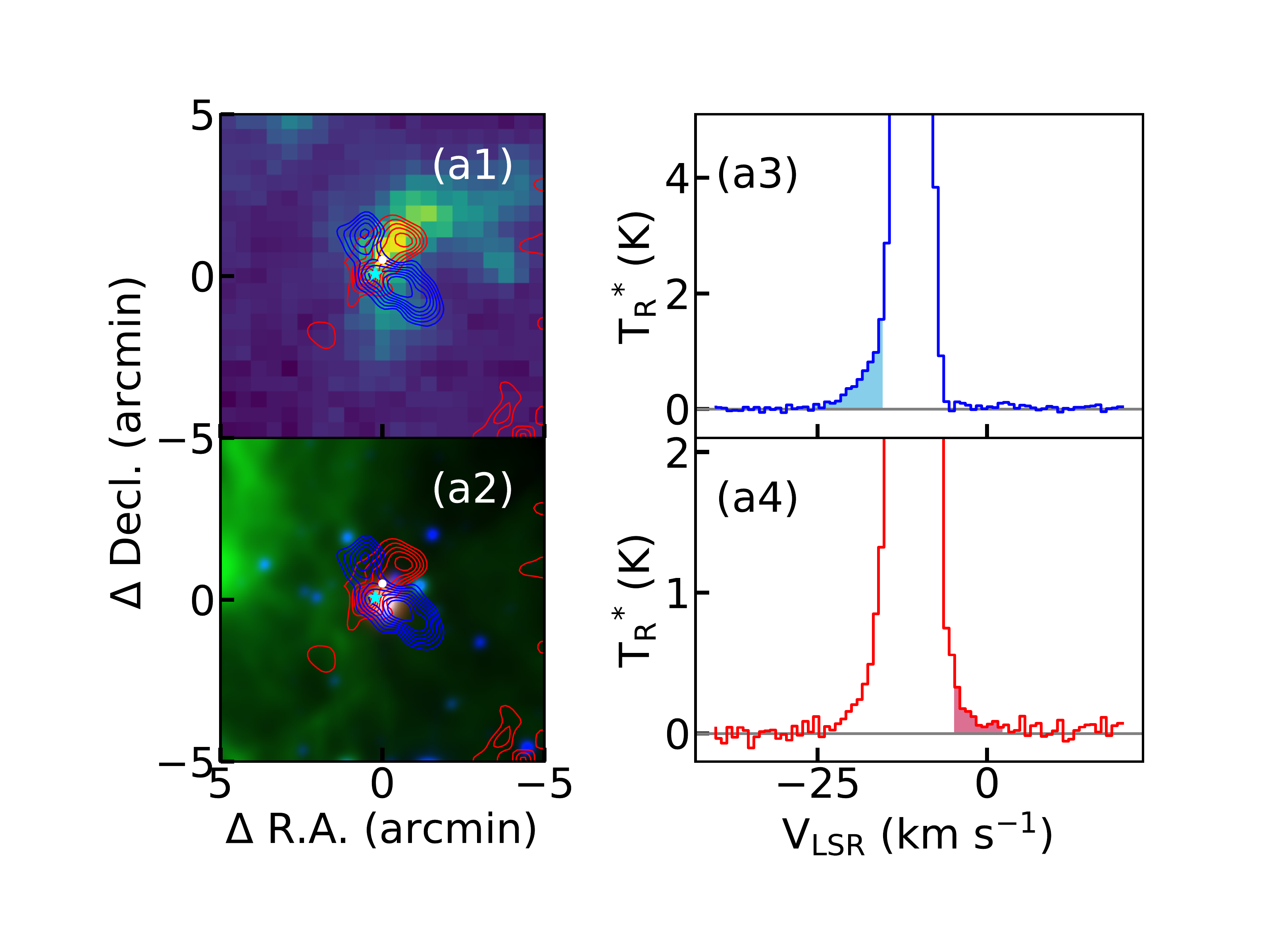}}
	\subfigure[\xco]{\includegraphics[width=8cm]{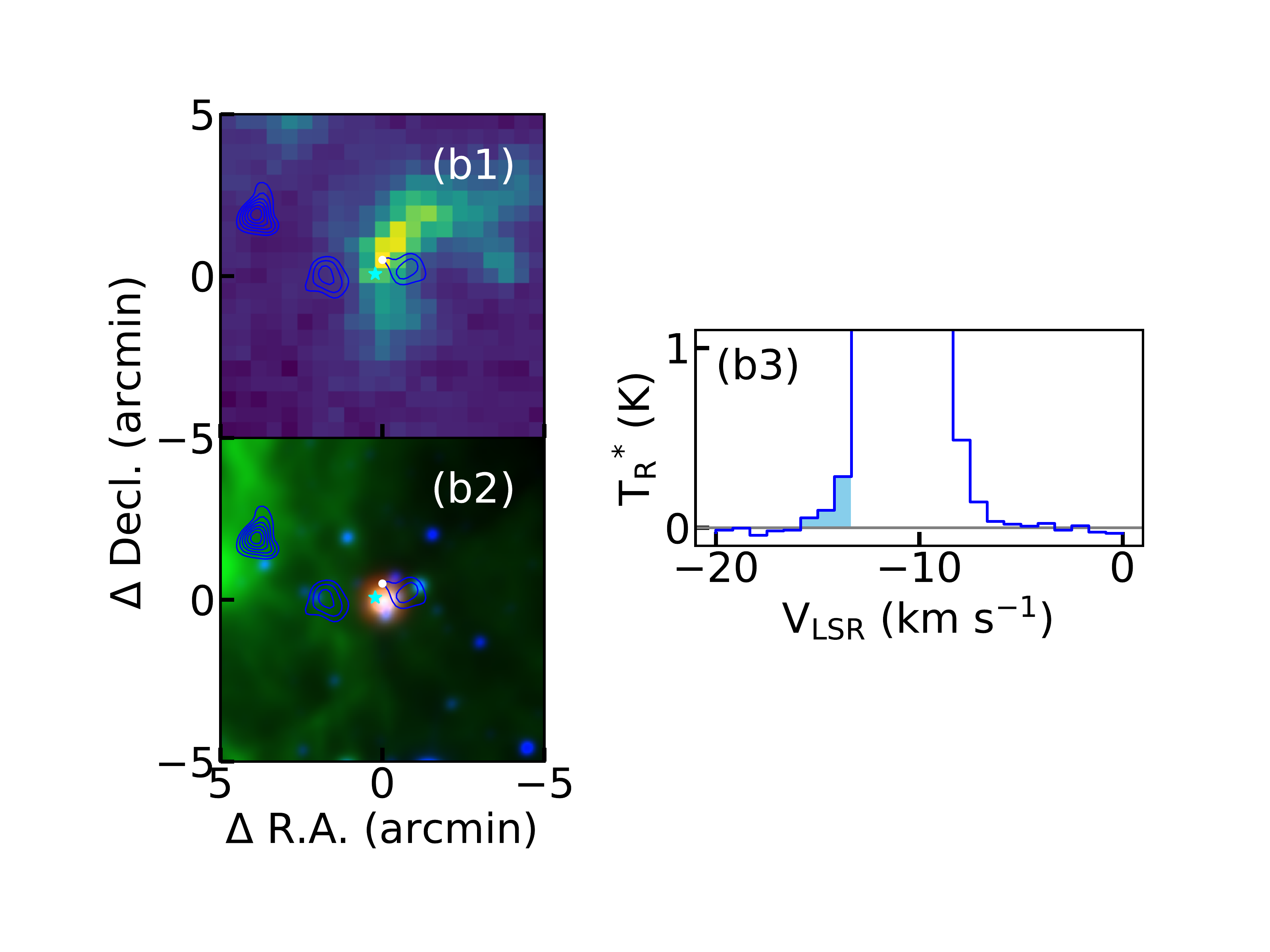}}
	\subfigure[\hco]{\includegraphics[width=8cm]{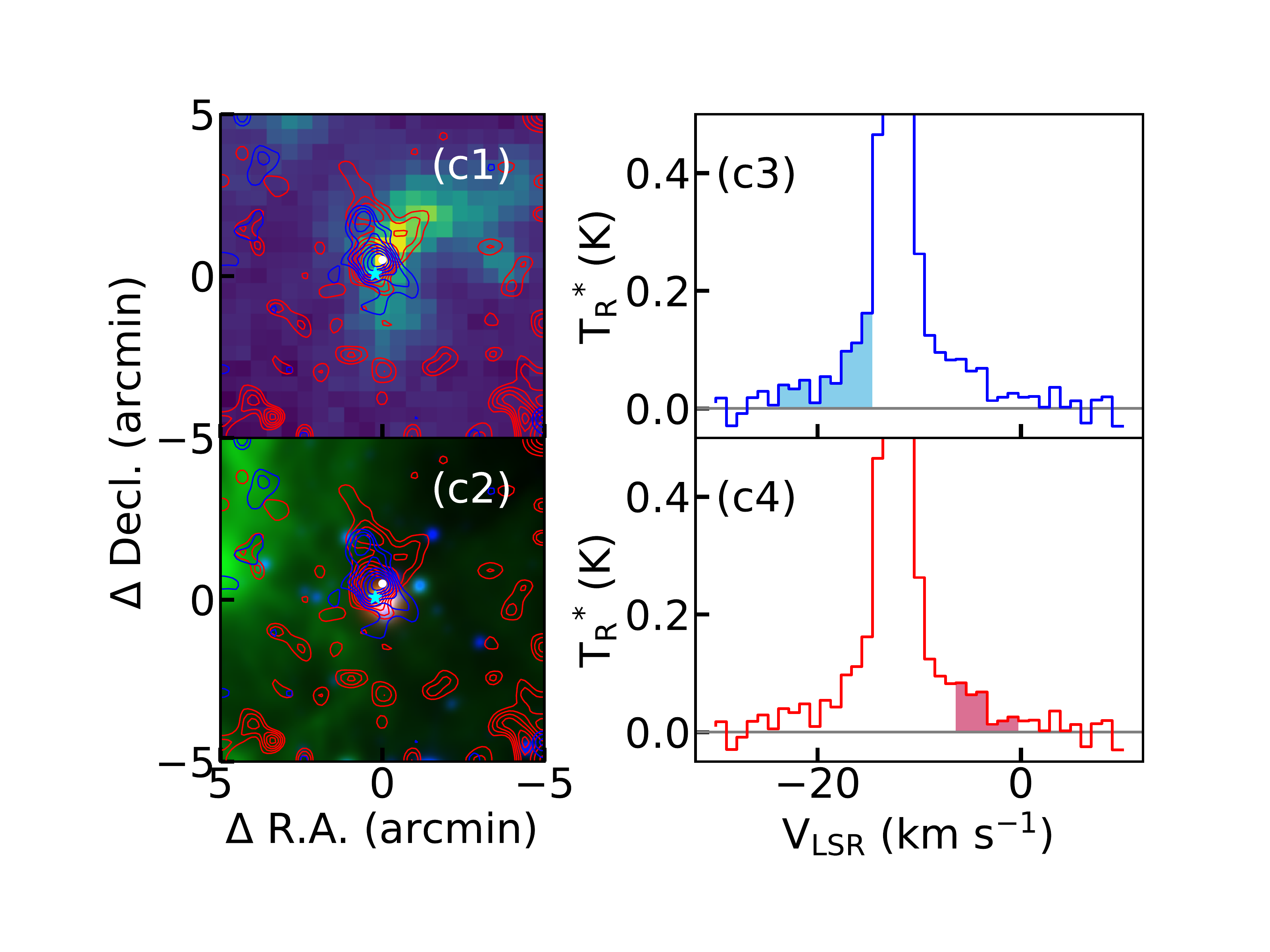}}
	\subfigure[CS]{\includegraphics[width=8cm]{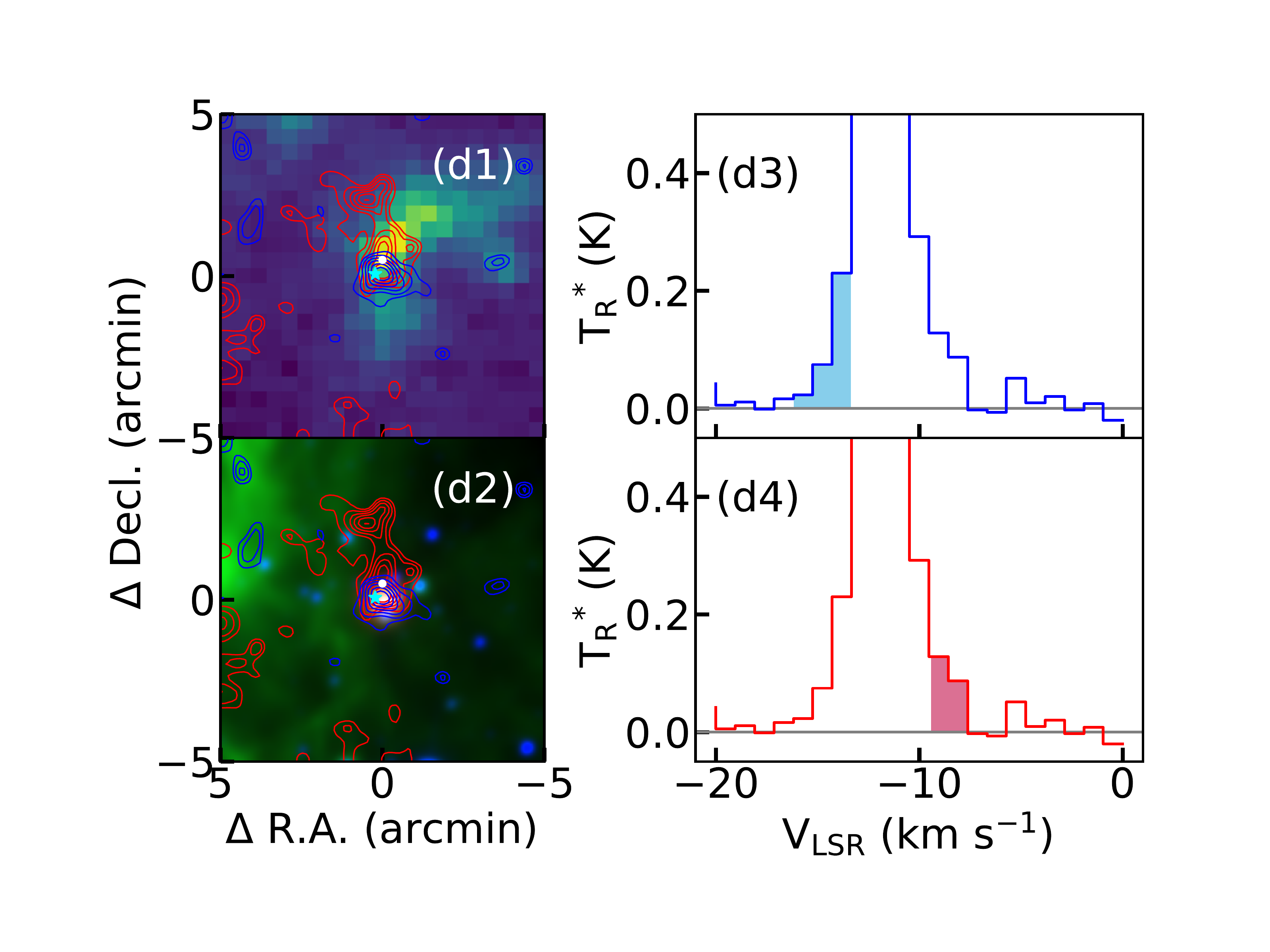}}
	\caption{Profile maps of the outflows in IRAS 22198+6336. The description of each map is the same as that in Figure~\ref{fig:L1287}. The blue star is IRAS 22198+6336, and the contour levels of the CS outflow are from 40\% to 90\% of the peak intensity of each outflow lobe.}
	\label{fig:22198}
\end{figure}

IRAS 22198+6336 is located near the dark cloud L1204A. The \co bipolar outflows of the source was first studied by \citet{Fukui+1989}, although detailed description was not provided. Thus, our work is the first to provide a detailed description of the outflows of this source. IRAS 22198+6336 presents not bipolar, but multiple \co outflows (see panel (a1) of Figure~\ref{fig:22198}), where the two red lobes are aligned the northwest--southeast direction, and the two blue lobes are aligned the northeast--southwest direction. Meanwhile, these blue  and red lobes intersect. As both the red and blue lobes have two peaks, we infer that there are probably two pairs of bipolar outflows in this region. The southern one is along the northeast--southwest direction, and the northern one is along the east--west direction.

We also detected \hco and CS bipolar outflows in this region, as well as the blue lobe of an \xco outflow. However, different to the structures seen in the \co outflows, both the \hco and CS outflows seem to have one pair of bipolar outflow. The blue lobes of the \hco and CS outflows are more concentrated than that of the \co outflows. Although the red lobes of the \hco and CS outflows are extended and their structures are different from those of the \co outflows, they only have one emission peak. As the brightness temperatures of \hco and CS are relatively lower than those of the other sources, the outflow maps of \hco and CS are weak.

IRAS 22198+6336 is located at the center of the bipolar outflows, and the WISE emission is also near the IRAS source. Hence, IRAS 22198+6336 might be the excitation source of the outflows.
\subsubsection{L1206}
\begin{figure}
	\centering
	\vspace{-0.3cm}
	\subfigure[\co]{\includegraphics[width=8cm]{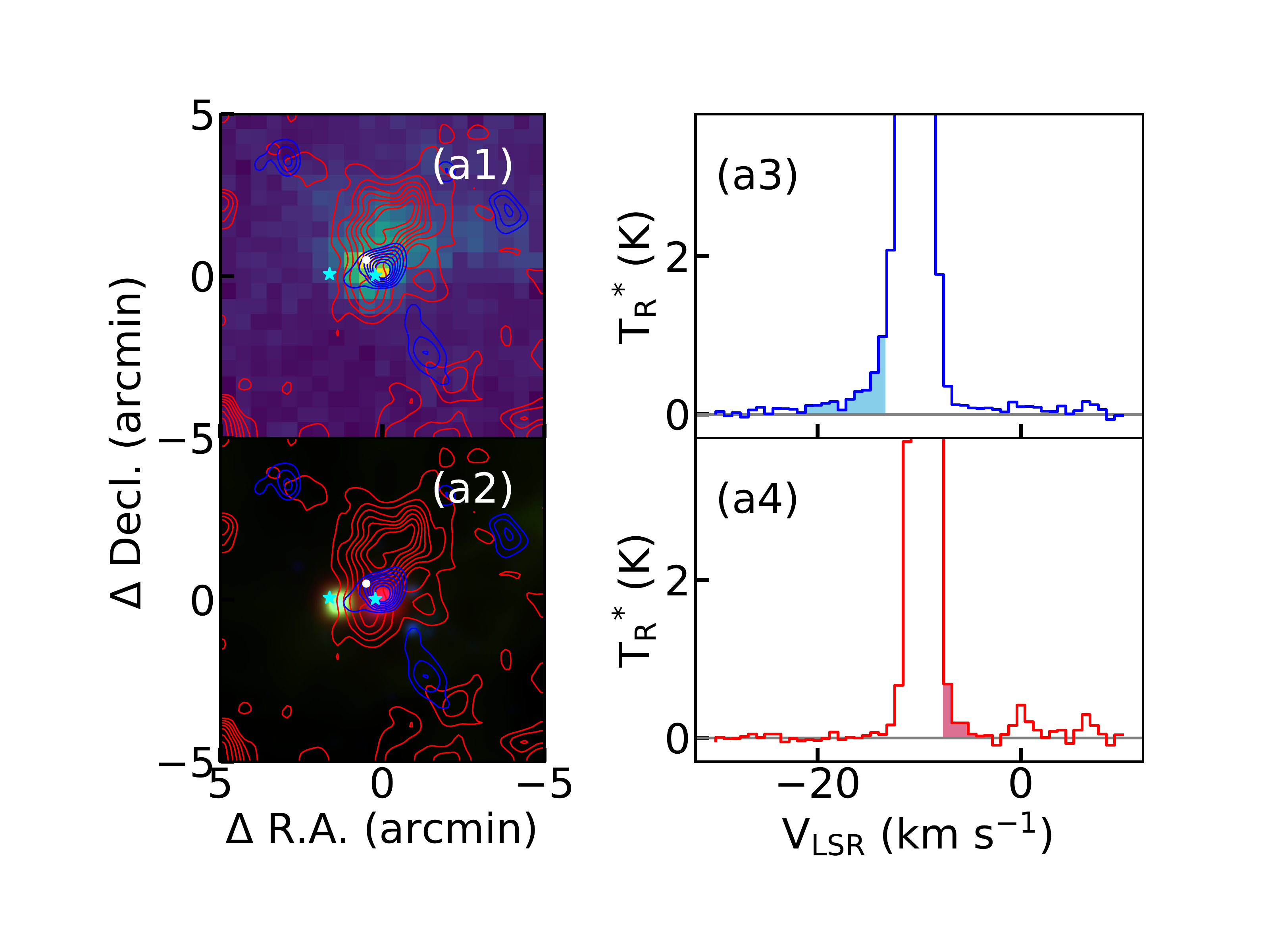}}
	\subfigure[\xco]{\includegraphics[width=8cm]{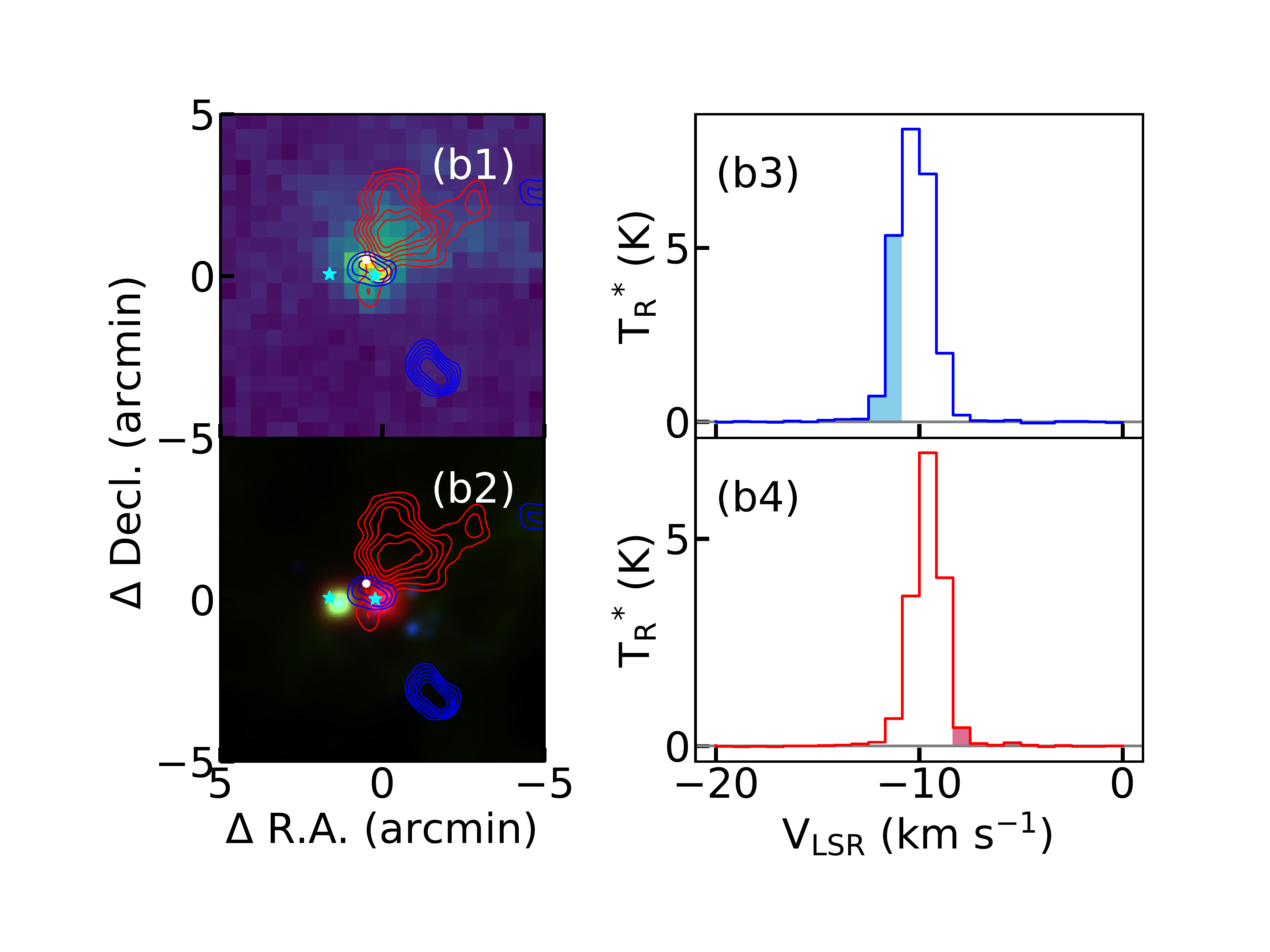}}
	\subfigure[\hco]{\includegraphics[width=8cm]{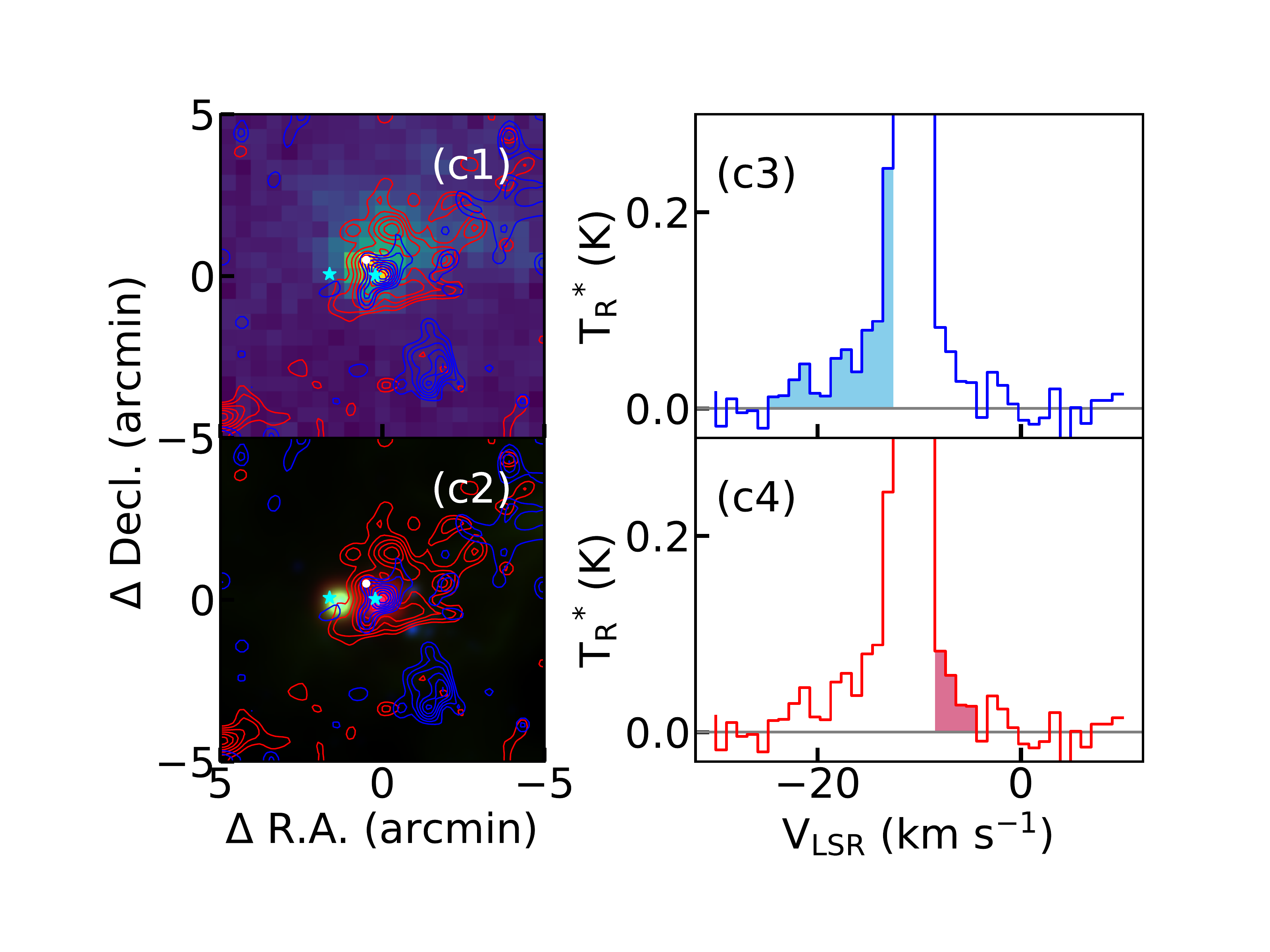}}
	\subfigure[CS]{\includegraphics[width=8cm]{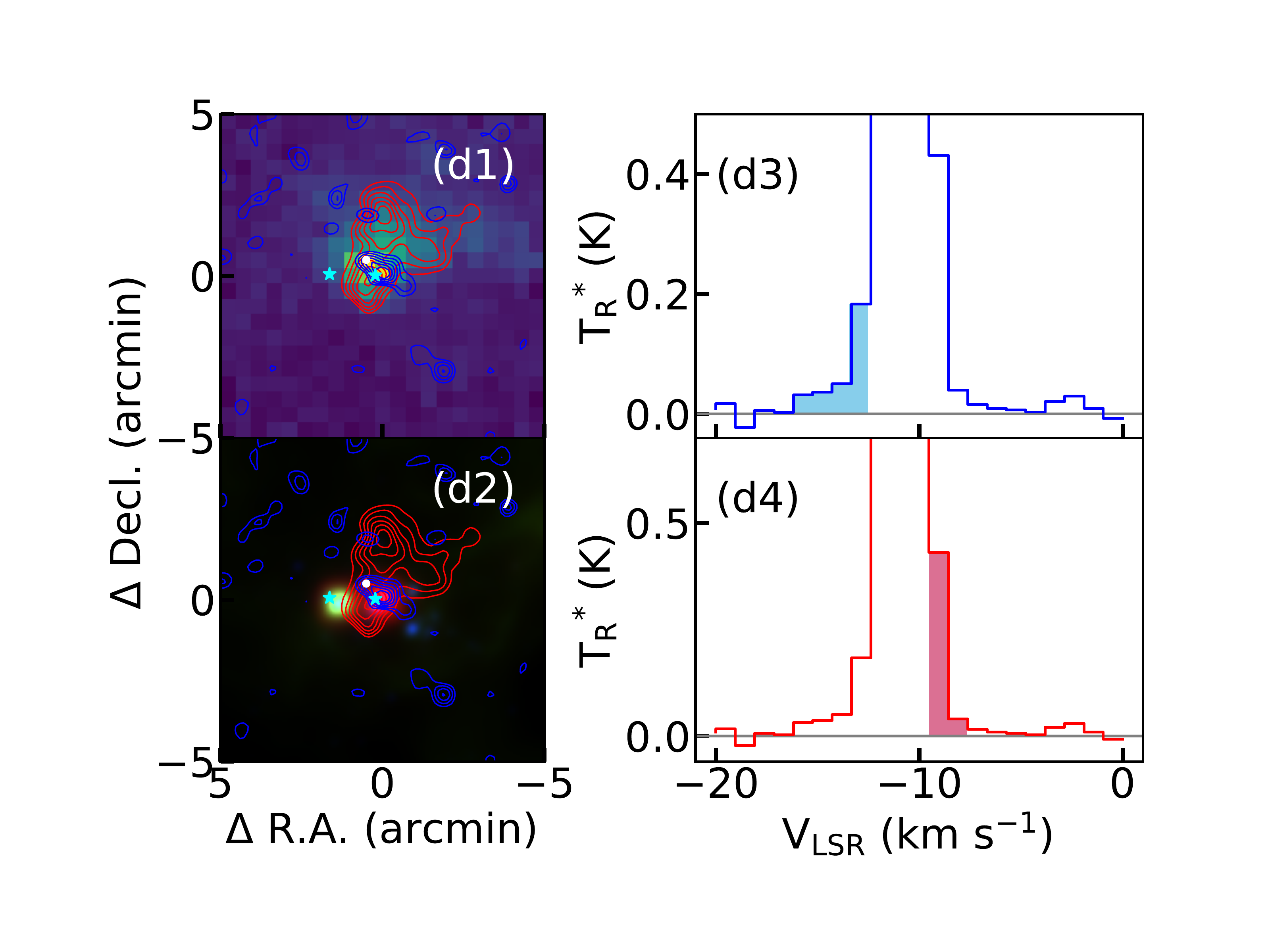}}
	\caption{Profile maps of the outflows in L1206. The description of each map is the same as that in Figure~\ref{fig:L1287}. The blue star at the center is IRAS 22272+6358A, and that in the east is IRAS 22272+6358B.}
	\label{fig:L1206}
\end{figure}

Only the blue lobe of the \co outflow was detected in L1206 by \citet{Sugitani+etal+1989,Xu+etal+2006b}, while \citet{Beltran+etal+2006} and \citet{Liu+etal+2020} successfully detected the blue and red lobes of the \co outflow. We also detected a pair of bipolar outflow of this source. Hence, there is a pair of bipolar \co outflows in L1206 (see details in panel (a1) of Figure~\ref{fig:L1206}). Apart from \coo, we detected bipolar \xcoo, \hcoo, and CS outflows from this source. The structures of these four groups of outflows are similar, i.e., the extended red lobes are located to the northwest and the compact blue lobes are located at the center. In general, the bipolar outflows are aligned along the northwest--southeast direction. Except for the CS outflow, the red lobes of the \coo, \xcoo, and \hco outflows are complex, where the different components seen in the red lobes are difficult to be separated with each other under the present resolution.

There are two IRAS sources in this region (i.e., IRAS 22272+6358A and IRAS 22272+6358B). IRAS 22272+6358A is located near the center of the bipolar outflows. Meanwhile, there is WISE emission with red color, which implies an earlier object relative to the eastern yellow-green one. The emission peak of \xxco is also near IRAS 22272+6358A and the red WISE emission. Therefore, IRAS 22272+6358A might be the source of excitation of the outflows.
\subsubsection{Cep A}
\begin{figure}
	\centering
	\vspace{-0.3cm}
	\subfigure[\co]{\includegraphics[width=8cm]{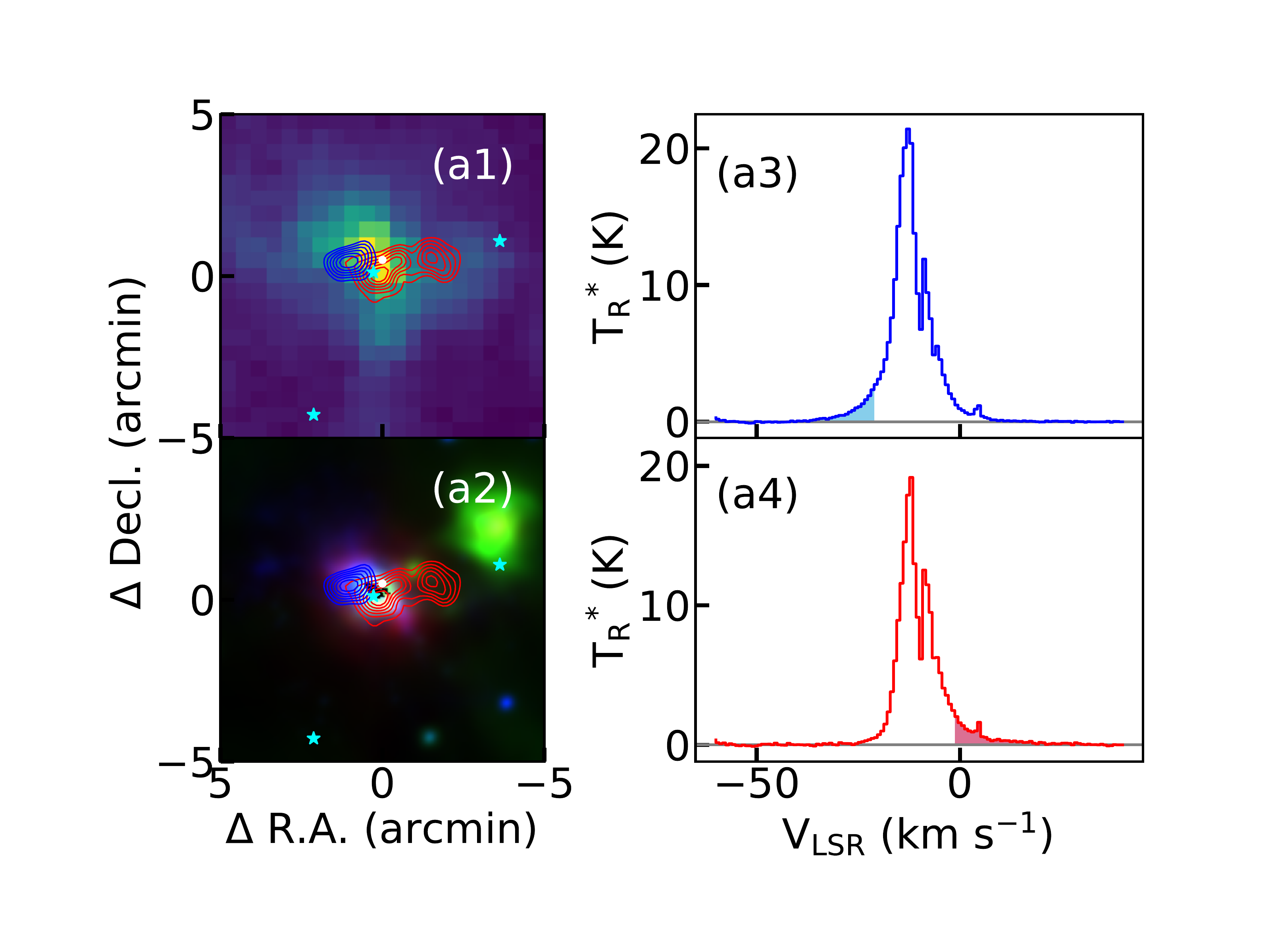}}
	\subfigure[\xco]{\includegraphics[width=8cm]{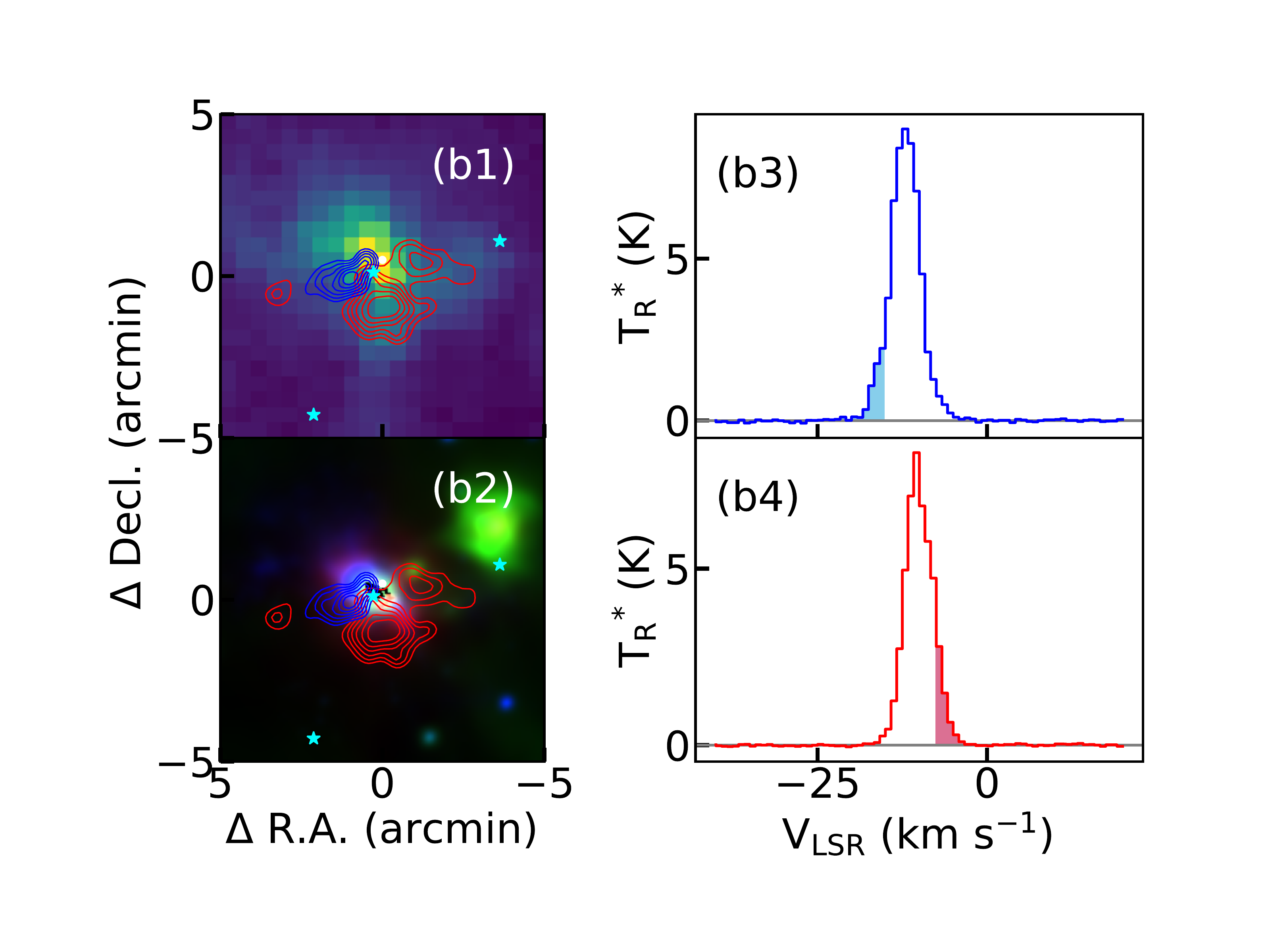}}
	\subfigure[\hco]{\includegraphics[width=8cm]{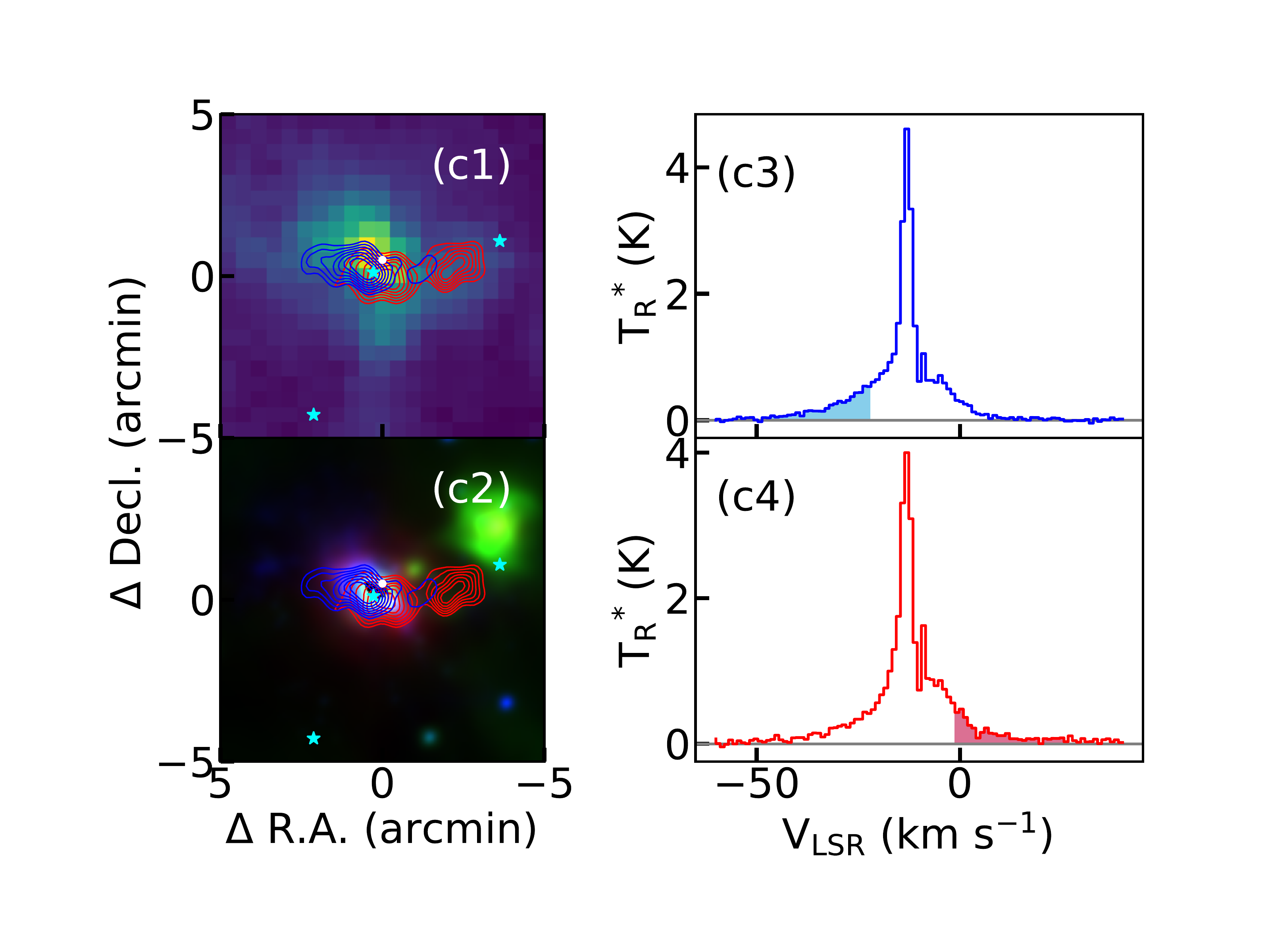}}
	\subfigure[CS]{\includegraphics[width=8cm]{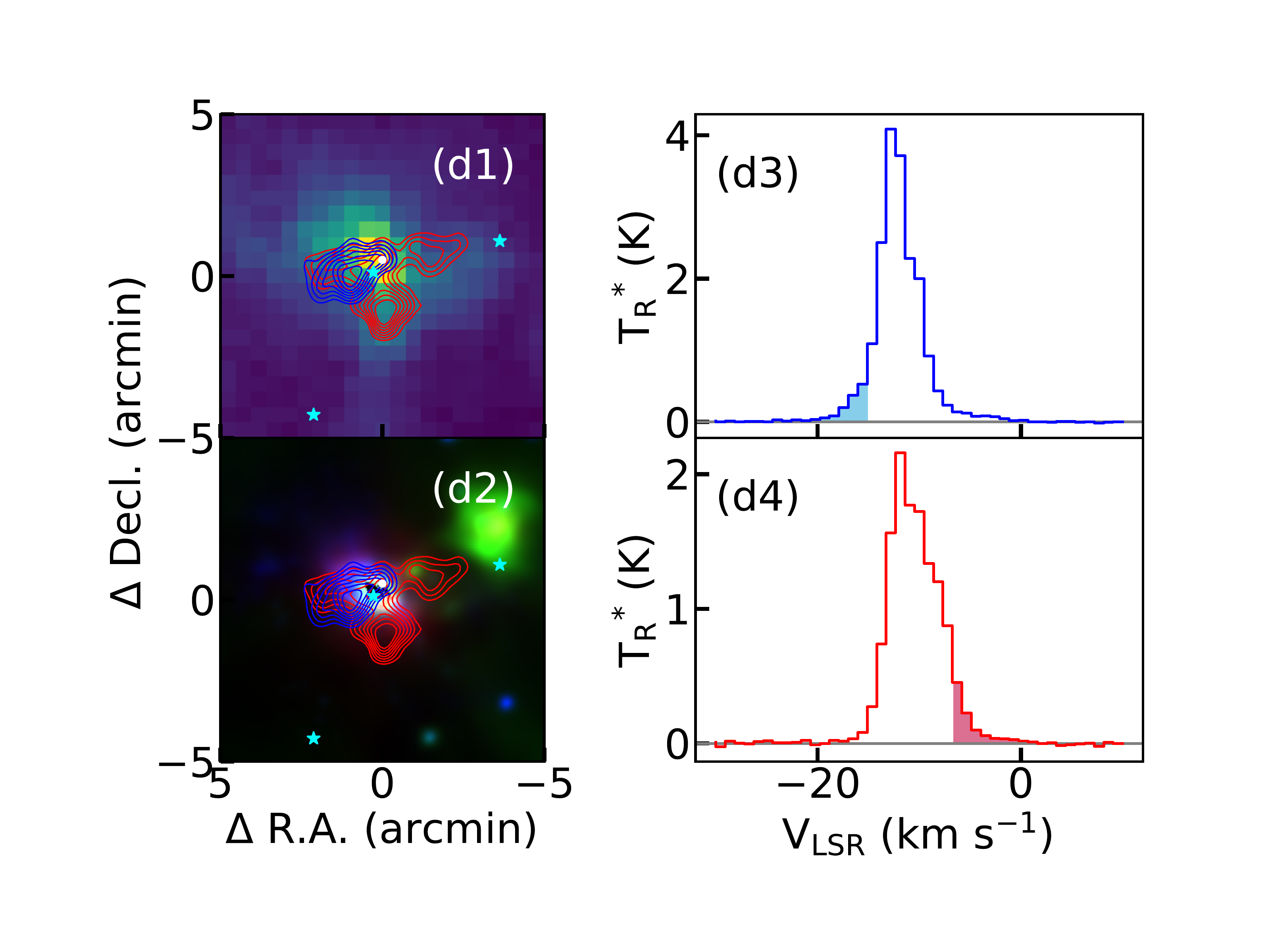}}
	\caption{Profile maps of the outflows in Cep A. The description of each map is the same as that in Figure~\ref{fig:L1287}. The blue star at the center is IRAS 22543+6145, while that in the southeast is IRAS 22544+6141, and that in the northwest is IRAS 22540+6146.}
	\label{fig:CepA}
\end{figure}

We have detected \coo, \xcoo, \hcoo, and CS outflows from this source (see Figure~\ref{fig:CepA}). The configurations of the \co bipolar outflow (see panel (a1) Figure~\ref{fig:CepA}) is similar to the results of \citet{Rodriguez+etal+1980,Ho+etal+1982,Narayanan+Walker+1996,Xu+etal+2006b}. The velocity ranges of the \co outflows have been extended from $-$36~\kms to $-$42~\kms in the blue lobe, and from 14~\kms to 33~\kms in the red lobe \citep{Rodriguez+etal+1980}. After updating the velocity ranges, the structures of the outflows are also similar to the morphologies detected by \citet{Ho+etal+1982,Xu+etal+2006b}. 

The red lobes of the \coo, \xcoo, and \hco outflows appear to have two peaks. For the CS outflow, there are even three components in the red lobe. The sizes of the blue lobes are similar for the four groups of outflows, but they differ greatly from those of the corresponding red lobes. The red lobes of the \xco and CS outflows are significantly more extended than those of \co and \hcoo. Different to the other outflows, the \xco outflow has an additional red lobe in the east.

There are three IRAS sources (IRAS 22543+6145, IRAS 22540+6146, and IRAS 22544+6141) in this region. IRAS 22543+6143, near the white region with strong WISE emission, is located at the center of the bipolar outflows (see Figure~\ref{fig:CepA}), which indicates that IRAS 22543+6413 is a possible excitation source of the outflows.

\clearpage

\begin{longrotatetable}
	\begin{deluxetable*}{ccccccccccc}
		\renewcommand\arraystretch{1}
		\centering
		\setlength\tabcolsep{6pt}
		\tablecolumns{4}
		\tabletypesize{\normalsize}
		\tablewidth{15cm}
		\tablecaption{Physical properties of the \co outflows. \label{tab:12CO}}
		\tablehead{
			\colhead{Source} & \colhead{Index} & \colhead{Lobe} & \colhead{$N$} & \colhead{$M_{\rm lobe}$} & \colhead{$\langle \Delta v_{\rm lobe} \rangle$} & \colhead{$l_{\rm lobe}$} & \colhead{$P_{\rm lobe}$} & \colhead{$E_{\rm lobe}$} & \colhead{$t_{\rm lobe}$} & \colhead{$L_{\rm m (lobe)}$}\\
			\colhead{} & \colhead{} & \colhead{} &  \colhead{($10^{20}\ \rm cm^{-2}$)} & \colhead{($M_{\odot}$)} & \colhead{($\rm km\ s^{-1}$)} & \colhead{($\rm pc$)} & \colhead{($M_{\odot}\ \rm km\ s^{-1}$)} & \colhead{($10^{44}\ \rm erg$)} & \colhead{($10^{4}\ \rm yr$)} & \colhead{($10^{-2}\ L_{\odot}$)} \\
			\colhead{(1)}   & \colhead{(2)} & \colhead{(3)} & \colhead{(4)} & \colhead{(5)} & \colhead{(6)} & \colhead{(7)} & \colhead{(8)} & \colhead{(9)} & \colhead{(10)} & \colhead{(11)}
		}
		\startdata
		\multirow{2}[0]{*}{L1287} & 1     & blue  & 4.1   & 2.0   & 7.7   & 0.7   & 15.2  & 15.9  & 2.3   & 53.9  \\
		& 2     & red   & 3.4   & 1.1   & 12.4  & 0.5   & 13.6  & 18.3  & 1.7   & 81.7  \\ \hline
		\multirow{2}[0]{*}{G176.51+00.20} & 3     & blue  & 3.1   & 1.3   & 15.9  & 0.6   & 20.9  & 42.4  & 1.0   & 331.9  \\
		& 4     & red   & 2.1   & 0.7   & 11.3  & 0.5   & 8.0   & 13.4  & 0.9   & 119.0  \\ \hline
		\multirow{2}[0]{*}{Mon R2} & 5     & blue  & 3.9   & 7.3   & 8.6   & 1.2   & 63.3  & 54.7  & 5.2   & 82.0  \\
		& 6     & red   & 13.4  & 17.2  & 12.2  & 1.0   & 209.2  & 257.0  & 2.3   & 848.4  \\ \hline
		\multirow{2}[0]{*}{NGC2264} & 7     & blue  & 0.3   & 0.1   & 18.2  & 0.5   & 1.2   & 2.1   & 1.9   & 8.3  \\
		& 8     & red   & 3.5   & 1.4   & 13.9  & 0.6   & 19.6  & 27.6  & 1.9   & 110.5  \\ \hline
		G090.21+02.32 & 9     & red   & 0.8   & 0.5   & 10.7  & 0.7   & 5.9   & 6.6   & 3.1   & 16.3  \\ \hline
		\multirow{2}[0]{*}{G105.41+09.87} & 10    & blue  & 0.4   & 0.4   & 9.9   & 1.0   & 3.8   & 5.1   & 2.4   & 16.7  \\
		& 11    & red   & 2.5   & 2.2   & 11.4  & 0.7   & 24.6  & 28.6  & 2.7   & 82.0  \\ \hline
		\multirow{3}[0]{*}{IRAS 22198+6336} & 12    & blue  & 0.6   & 0.4   & 11.4  & 0.8   & 4.1   & 4.9   & 3.0   & 12.6  \\
		& 13    & ...   & ...   & ...   & ...   & ...   & ...   & ...   & ...   & ... \\
		& 14    & red   & 0.2   & 0.1   & 15.3  & 0.7   & 1.0   & 1.5   & 2.7   & 4.2  \\ \hline
		\multirow{2}[0]{*}{L1206} & 15    & blue  & 0.3   & 0.1   & 11.4  & 0.5   & 0.8   & 0.9   & 2.0   & 3.6  \\
		& 16    & red   & 0.1   & 0.1   & 6.1   & 0.7   & 0.5   & 0.3   & 6.6   & 0.4  \\ \hline
		\multirow{3}[0]{*}{CepA} & 17    & blue  & 2.4   & 0.5   & 27.7  & 0.5   & 15.1  & 43.3  & 0.8   & 431.6  \\
		& 18    & red   & 2.3   & 1.2   & 31.5  & 0.7   & 38.9  & 129.1  & 0.8   & 1264.2  \\
		& 19    & ...   & ...   & ...   & ...   & ...   & ...   & ...   & ...   & ... \\ \hline
		\multirow{4}[0]{*}{Statistic} & min   &       & 0.1   & 0.1   & 6.1   & 0.5   & 0.5   & 0.3   & 0.8   & 0.4  \\
		& max   &       & 13.4  & 17.2  & 31.5  & 1.2   & 209.2  & 257.0  & 6.6   & 1264.2  \\
		& mean  &       & 2.5   & 2.2   & 13.9  & 0.7   & 26.2  & 38.3  & 2.4   & 204.0  \\
		& median &       & 2.3   & 0.7   & 11.4  & 0.7   & 13.6  & 15.9  & 2.3   & 81.7  \\
		\enddata
		\tablewidth{15cm}
		\tablecomments{ (1) Source name.  (2) Index. The index of the outflow lobe of each source is unique, and a vacant index corresponds to an outflow of other molecules. (3) Red/blue lobe. (4) H$_{2}$ density of the red/blue lobe. (5) Mass of the red/blue lobe.  (6) Velocity of the red/blue lobe. (7) Length of the red/blue lobe. (8) Momentum of the red/blue lobe. (9) Kinetic energy of the red/blue lobe. (10) Dynamical timescale of the red/blue lobe. (11) Mechanical luminosity of the red/blue lobe. The last four lines are the mean, median, min, and max values of these parameters. }		
	\end{deluxetable*}
\end{longrotatetable}

\begin{longrotatetable}
	\begin{deluxetable*}{ccccccccccc}
		\renewcommand\arraystretch{1}
		\centering
		\setlength\tabcolsep{6pt}
		\tablecolumns{4}
		\tabletypesize{\normalsize}
		\tablewidth{15cm}
		\tablecaption{Physical proterties of the \xco outflows. \label{tab:13CO}}
		\tablehead{
			\colhead{Source} & \colhead{Index} & \colhead{lobe} & \colhead{$N$} & \colhead{$M_{\rm lobe}$} & \colhead{$\langle \Delta v_{\rm lobe} \rangle$} & \colhead{$l_{\rm lobe}$} & \colhead{$P_{\rm lobe}$} & \colhead{$E_{\rm lobe}$} & \colhead{$t_{\rm lobe}$} & \colhead{$L_{\rm m (lobe)}$} \\
			\colhead{} & \colhead{} & \colhead{} &  \colhead{($10^{20}\ \rm cm^{-2}$)} & \colhead{($M_{\odot}$)} & \colhead{($\rm km\ s^{-1}$)} & \colhead{($\rm pc$)} & \colhead{($M_{\odot}\ \rm km\ s^{-1}$)} & \colhead{($10^{45}\ \rm erg$)} & \colhead{($10^{4}\ \rm yr$)} & \colhead{($10^{-1}\ L_{\odot}$)} \\
			\colhead{(1)}   & \colhead{(2)} & \colhead{(3)} & \colhead{(4)} & \colhead{(5)} & \colhead{(6)} & \colhead{(7)} & \colhead{(8)} & \colhead{(9)} & \colhead{(10)} & \colhead{(11)}
		}
		
		\startdata
		\multirow{2}[0]{*}{L1287} & 1     & blue  & 9.9   & 3.2   & 9.7   & 0.5   & 30.7  & 3.5   & 1.9   & 14.4  \\
		& 2     & red   & 9.4   & 3.8   & 9.7   & 0.5   & 36.5  & 4.2   & 2.1   & 15.8  \\ \hline
		\multirow{2}[0]{*}{G176.51+00.20} & 3     & blue  & 3.3   & 2.3   & 9.6   & 0.7   & 21.8  & 2.0   & 5.0   & 3.1  \\
		& 4     & red   & 9.8   & 3.4   & 5.7   & 0.5   & 19.4  & 1.2   & 3.4   & 2.8  \\ \hline
		\multirow{2}[0]{*}{Mon R2} & 5     & blue  & 39.7  & 48.5  & 7.8   & 1.1   & 376.4  & 28.2  & 8.4   & 26.0  \\
		& 6     & red   & 42.2  & 17.2  & 12.0  & 0.5   & 206.4  & 23.9  & 2.3   & 79.2  \\ \hline
		\multirow{2}[0]{*}{NGC2264} & 7    & ...   & ...   & ...   & ...   & ...   & ...   & ...   & ...   & ... \\
		& 8     & red   & 12.0  & 4.8   & 7.2   & 0.6   & 34.7  & 2.6   & 4.5   & 4.4  \\ \hline
		G090.21+02.32 & 9     & red   & 6.4   & 2.4   & 3.7   & 0.6   & 8.9   & 0.3   & 6.8   & 0.4  \\ \hline
		\multirow{2}[0]{*}{G105.41+09.87} & 10    & blue  & 6.5   & 6.1   & 3.7   & 0.8   & 22.8  & 0.8   & 11.2  & 0.6  \\
		& 11    & red   & 23.9  & 17.3  & 5.5   & 0.7   & 95.6  & 5.3   & 7.1   & 5.8  \\ \hline
		\multirow{3}[0]{*}{IRAS 22198+6336} & 12    & blue1 & 4.1   & 2.1   & 5.1   & 0.6   & 10.4  & 0.5   & 5.9   & 0.7  \\
		& 13    & blue2 & 4.9   & 2.7   & 4.6   & 0.6   & 12.6  & 0.6   & 5.9   & 0.7  \\
		& 14    & ...   & ...   & ...   & ...   & ...   & ...   & ...   & ...   & ... \\ \hline
		\multirow{2}[0]{*}{L1206} & 15    & blue  & 8.8   & 2.9   & 3.5   & 0.6   & 10.1  & 0.3   & 8.4   & 0.3  \\
		& 16    & red   & 8.7   & 3.8   & 3.3   & 0.7   & 12.6  & 0.4   & 8.2   & 0.4  \\ \hline
		\multirow{3}[0]{*}{CepA} & 17    & blue  & 28.1  & 8.9   & 9.9   & 0.6   & 87.8  & 8.5   & 2.9   & 22.8  \\
		& 18    & red1  & 44.4  & 32.1  & 7.5   & 0.7   & 239.9  & 17.7  & 5.1   & 27.0  \\
		& 19    & red2  & 31.7  & 15.7  & 7.7   & 0.6   & 121.7  & 9.2   & 4.1   & 17.3  \\ \hline
		\multirow{4}[0]{*}{Statistic} & min   &       & 3.3   & 2.1   & 3.3   & 0.5   & 8.9   & 0.3   & 1.9   & 0.3  \\
		& max   &       & 44.4  & 48.5  & 12.0  & 1.1   & 376.4  & 28.2  & 11.2  & 79.2  \\
		& mean  &       & 17.3  & 10.4  & 6.8   & 0.6   & 79.3  & 6.4   & 5.5   & 13.0  \\
		& median &       & 9.8   & 3.8   & 7.2   & 0.6   & 30.7  & 2.6   & 5.1   & 4.4  \\
		\enddata
		\tablewidth{15cm}
		\tablecomments{The description of each column is the same as that of Table~\ref{tab:12CO}.}		
	\end{deluxetable*}
\end{longrotatetable}

\begin{longrotatetable}
	\begin{deluxetable*}{ccccccccccc}
		\renewcommand\arraystretch{1}
		\centering
		\setlength\tabcolsep{5pt}
		\tablecolumns{4}
		\tabletypesize{\normalsize}
		\tablewidth{15cm}
		\tablecaption{Physical properties of the \hco outflows. \label{tab:HCO}}
		\tablehead{
			\colhead{Source} & \colhead{Index} & \colhead{lobe} & \colhead{$N$} & \colhead{$M_{\rm lobe}$} & \colhead{$\langle \Delta v_{\rm lobe} \rangle$} & \colhead{$l_{\rm lobe}$} & \colhead{$P_{\rm lobe}$} & \colhead{$E_{\rm lobe}$} & \colhead{$t_{\rm lobe}$} & \colhead{$L_{\rm m (lobe)}$} \\
			\colhead{} & \colhead{} & \colhead{} & \colhead{($10^{21}\ \rm cm^{-2}$)} & \colhead{($M_{\odot}$)} & \colhead{($\rm km\ s^{-1}$)} & \colhead{($\rm pc$)} & \colhead{($10\ M_{\odot}\ \rm km\ s^{-1}$)} & \colhead{($10^{45}\ \rm erg$)} & \colhead{($10^{4}\ \rm yr$)} & \colhead{($L_{\odot}$)} \\
			\colhead{(1)}   & \colhead{(2)} & \colhead{(3)} & \colhead{(4)} & \colhead{(5)} & \colhead{(6)} & \colhead{(7)} & \colhead{(8)} & \colhead{(9)} & \colhead{(10)} &\colhead{(11)}
		}
		\startdata
		\multirow{2}[0]{*}{L1287} & 1     & blue  & 3.0   & 14.4  & 8.5   & 0.5   & 122.8  & 14.0  & 1.6   & 6.7  \\
		& 2     & red   & 2.8   & 11.0  & 11.1  & 0.5   & 122.7  & 14.7  & 1.9   & 5.9  \\ \hline
		\multirow{2}[0]{*}{G176.51+00.20} & 3     & blue  & 1.4   & 11.8  & 16.1  & 0.8   & 189.1  & 38.7  & 2.0   & 15.2  \\
		& 4     & red   & 2.4   & 4.1   & 14.3  & 0.5   & 58.8  & 11.3  & 1.1   & 7.8  \\ \hline
		\multirow{2}[0]{*}{Mon R2} & 5     & blue  & 1.9   & 29.8  & 9.6   & 1.0   & 287.1  & 27.6  & 4.7   & 4.5  \\
		& 6     & red   & 2.5   & 28.3  & 14.2  & 1.0   & 400.7  & 56.4  & 3.9   & 11.2  \\ \hline
		\multirow{2}[0]{*}{NGC2264} & 7     & blue  & 2.3   & 17.5  & 7.3   & 0.7   & 128.1  & 11.0  & 3.5   & 2.4  \\
		& 8     & red   & 4.0   & 16.1  & 13.2  & 0.6   & 213.0  & 29.1  & 2.2   & 10.4  \\ \hline
		G090.21+02.32 & 9     & red   & 0.3   & 1.1   & 6.7   & 0.6   & 7.2   & 0.5   & 4.7   & 0.1  \\ \hline
		\multirow{2}[0]{*}{G105.41+09.87} & 10    & blue  & 0.6   & 6.4   & 9.2   & 1.0   & 58.4  & 5.8   & 4.9   & 0.9  \\
		& 11    & red   & 0.9   & 7.5   & 26.7  & 0.8   & 200.3  & 51.8  & 2.1   & 19.0  \\ \hline
		\multirow{3}[0]{*}{IRAS 22198+6336} & 12    & blue  & 0.7   & 1.8   & 9.4   & 0.6   & 17.3  & 1.9   & 2.6   & 0.6  \\
		& 13    & ...   & ...   & ...   & ...   & ...   & ...   & ...   & ...   & ... \\
		& 14    & red   & 0.3   & 1.6   & 13.6  & 0.8   & 22.1  & 2.9   & 4.3   & 0.5  \\ \hline
		\multirow{2}[0]{*}{L1206} & 15    & blue  & 0.6   & 1.3   & 5.7   & 0.5   & 7.4   & 0.5   & 3.3   & 0.1  \\
		& 16    & red   & 0.4   & 2.3   & 5.0   & 0.7   & 11.6  & 0.6   & 5.8   & 0.1  \\ \hline
		\multirow{3}[0]{*}{CepA} & 17    & blue  & 7.5   & 23.6  & 32.5  & 0.6   & 765.8  & 272.9  & 0.8   & 263.0  \\
		& 18    & red1  & 4.1   & 9.1   & 34.1  & 0.6   & 311.8  & 114.5  & 0.8   & 115.7  \\
		& 19    & red2  & 4.6   & 18.7  & 37.4  & 0.6   & 700.6  & 282.7  & 0.7   & 294.9  \\ \hline
		\multirow{4}[0]{*}{Statistic} & min   &       & 0.3   & 1.1   & 5.0   & 0.5   & 7.2   & 0.5   & 0.7   & 0.1  \\
		& max   &       & 7.5   & 29.8  & 37.4  & 1.0   & 765.8  & 282.7  & 5.8   & 294.9  \\
		& mean  &       & 2.2   & 11.5  & 15.3  & 0.7   & 201.4  & 52.1  & 2.8   & 42.2  \\
		& median &       & 2.1   & 10.1  & 12.2  & 0.6   & 125.4  & 14.3  & 2.4   & 6.3  \\
		\enddata
		\tablewidth{15cm}
		\tablecomments{The description of each column is the same as that of Table~\ref{tab:12CO}.}		
	\end{deluxetable*}
\end{longrotatetable}

\begin{longrotatetable}
	\begin{deluxetable*}{ccccccccccc}
		\renewcommand\arraystretch{1}
		\centering
		\setlength\tabcolsep{8pt}
		\tablecolumns{4}
		\tabletypesize{\normalsize}
		\tablewidth{15cm}
		\tablecaption{Physical properties of the CS outflows. \label{tab:CS}}
		\tablehead{
			\colhead{Source} & \colhead{Index} & \colhead{Lobe} & \colhead{$N$} & \colhead{$M_{\rm lobe}$} & \colhead{$\langle \Delta v_{\rm lobe} \rangle$} & \colhead{$l_{\rm lobe}$} & \colhead{$P(\rm lobe)$} & \colhead{$E(\rm lobe)$} & \colhead{$t_{\rm lobe}$} & \colhead{$L_{\rm m (lobe)}$} \\
			\colhead{} & \colhead{} & \colhead{} & \colhead{($10^{20}\ \rm cm^{-2}$)} & \colhead{($M_{\odot}$)} & \colhead{($\rm km\ s^{-1}$)} & \colhead{($\rm pc$)} & \colhead{($M_{\odot}\ \rm km\ s^{-1}$)} & \colhead{($10^{44}\ \rm erg$)} & \colhead{($10^{4}\ \rm yr$)} & \colhead{($10^{-2}\ L_{\odot}$)} \\
			\colhead{(1)}   & \colhead{(2)} & \colhead{(3)} & \colhead{(4)} & \colhead{(5)} & \colhead{(6)} & \colhead{(7)} & \colhead{(8)} & \colhead{(9)} & \colhead{(10)} &\colhead{(11)}
		}
		\startdata
		\multirow{2}[0]{*}{L1287} & 1     & blue  & 12.8  & 4.1   & 8.0   & 0.5   & 32.4  & 32.4  & 1.8   & 136.3  \\
		& 2     & red   & 7.1   & 3.4   & 9.8   & 0.5   & 33.0  & 35.5  & 1.9   & 143.1  \\ \hline
		\multirow{2}[0]{*}{G176.51+00.20} & 3     & blue  & 1.6   & 0.6   & 8.2   & 0.5   & 4.6   & 3.8   & 2.6   & 11.2  \\
		& 4     & red   & 1.2   & 0.6   & 9.8   & 0.6   & 6.1   & 6.4   & 2.7   & 18.2  \\ \hline
		\multirow{2}[0]{*}{Mon R2} & 5     & blue  & 2.8   & 0.4   & 14.1  & 0.5   & 5.3   & 7.3   & 1.9   & 29.2  \\
		& 6     & red   & 2.6   & 2.1   & 14.7  & 0.7   & 31.5  & 44.3  & 3.3   & 103.1  \\ \hline
		\multirow{2}[0]{*}{NGC2264} & 7     & blue  & 3.6   & 1.1   & 10.1  & 0.5   & 10.9  & 11.5  & 2.1   & 41.6  \\
		& 8     & red   & 16.8  & 8.4   & 9.5   & 0.7   & 79.8  & 76.4  & 3.7   & 158.9  \\ \hline
		G090.21+02.32 & 9     & red   & 0.8   & 0.2   & 4.3   & 0.5   & 0.9   & 0.4   & 5.1   & 0.6  \\ \hline
		\multirow{2}[0]{*}{G105.41+09.87} & 10    & blue  & 0.8   & 0.3   & 4.8   & 0.6   & 1.6   & 0.8   & 6.1   & 1.0  \\
		& 11    & red   & 2.5   & 1.1   & 5.7   & 0.5   & 6.1   & 3.4   & 5.5   & 4.7  \\ \hline
		\multirow{3}[0]{*}{IRAS 22198+6336} & 12    & blue  & 0.7   & 0.3   & 3.9   & 0.6   & 1.1   & 0.4   & 11.1  & 0.3  \\
		& 13    & ...   & ...   & ...   & ...   & ...   & ...   & ...   & ...   & ... \\
		& 14    & red   & 0.7   & 0.5   & 4.9   & 0.7   & 2.2   & 1.0   & 11.4  & 0.7  \\ \hline
		\multirow{2}[0]{*}{L1206} & 15    & blue  & 1.1   & 0.2   & 5.8   & 0.5   & 1.4   & 0.8   & 3.9   & 1.6  \\
		& 16    & red   & 3.9   & 2.2   & 2.4   & 0.6   & 5.2   & 1.3   & 10.1  & 1.0  \\ \hline
		\multirow{3}[0]{*}{CepA} & 17    & blue  & 5.5   & 2.7   & 7.6   & 0.7   & 20.7  & 15.4  & 5.0   & 24.1  \\
		& 18    & red1  & 4.8   & 3.2   & 13.1  & 0.8   & 42.0  & 55.0  & 3.5   & 123.0  \\
		& 19    & red2  & 2.8   & 1.0   & 11.9  & 0.6   & 12.0  & 14.4  & 2.7   & 41.6  \\ \hline
		\multirow{4}[0]{*}{Statistic} & min   &       & 0.7   & 0.2   & 2.4   & 0.5   & 0.9   & 0.4   & 1.8   & 0.3  \\
		& max   &       & 16.8  & 8.4   & 14.7  & 0.8   & 79.8  & 76.4  & 11.4  & 158.9  \\
		& mean  &       & 3.5   & 1.7   & 8.3   & 0.6   & 15.5  & 16.4  & 4.9   & 41.4  \\
		& median &       & 2.7   & 1.0   & 8.1   & 0.6   & 6.1   & 6.8   & 3.6   & 21.2  \\
		\enddata
		\tablewidth{15cm}
		\tablecomments{The description of each column is the same as that of Table~\ref{tab:12CO}.}		
	\end{deluxetable*}
\end{longrotatetable}

\begin{figure}
	\vspace{-0.3cm}
	\subfigure[Masses]{\includegraphics[width=8.5cm]{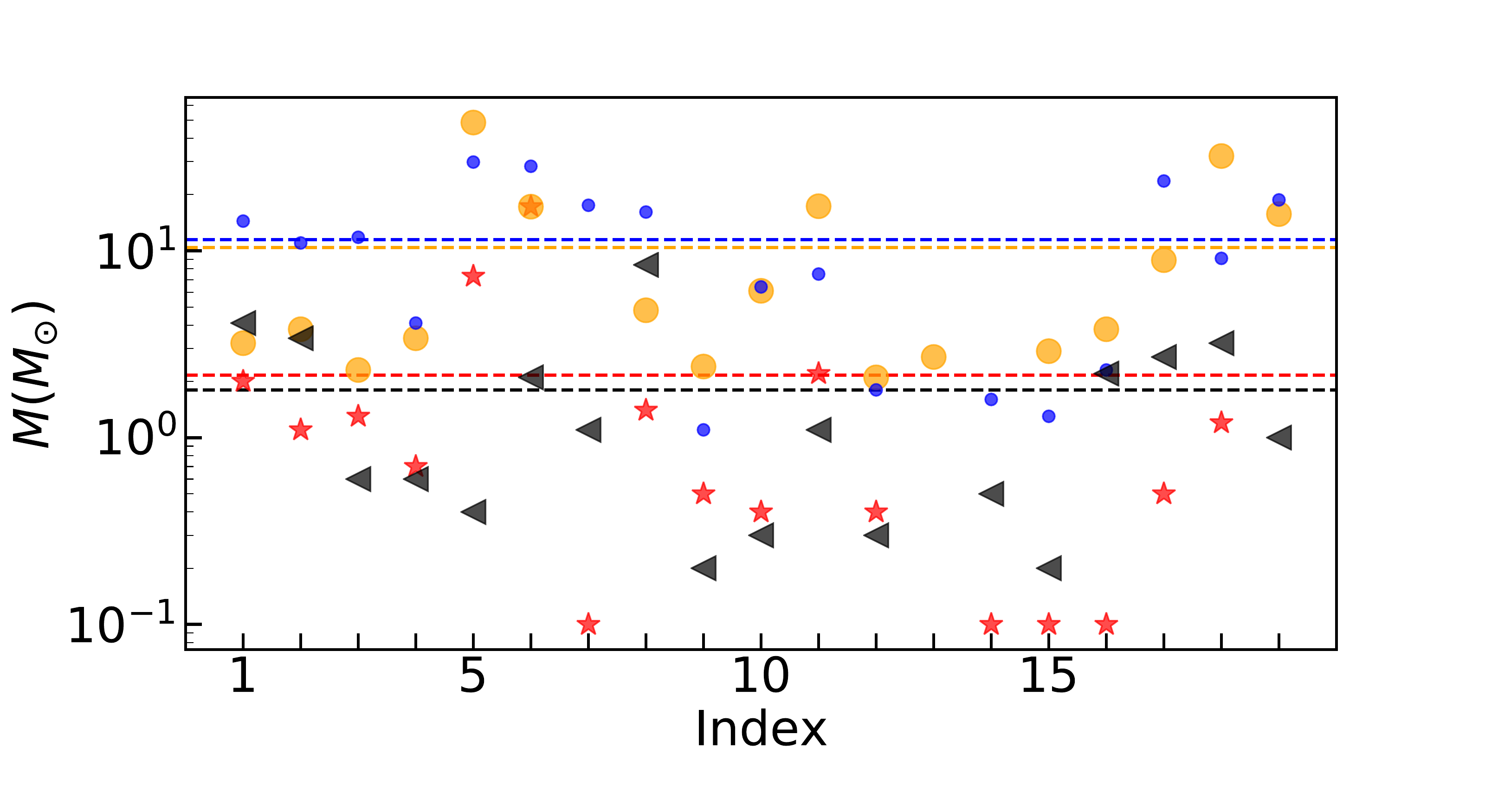}}
	\subfigure[Momenta]{\includegraphics[width=8.5cm]{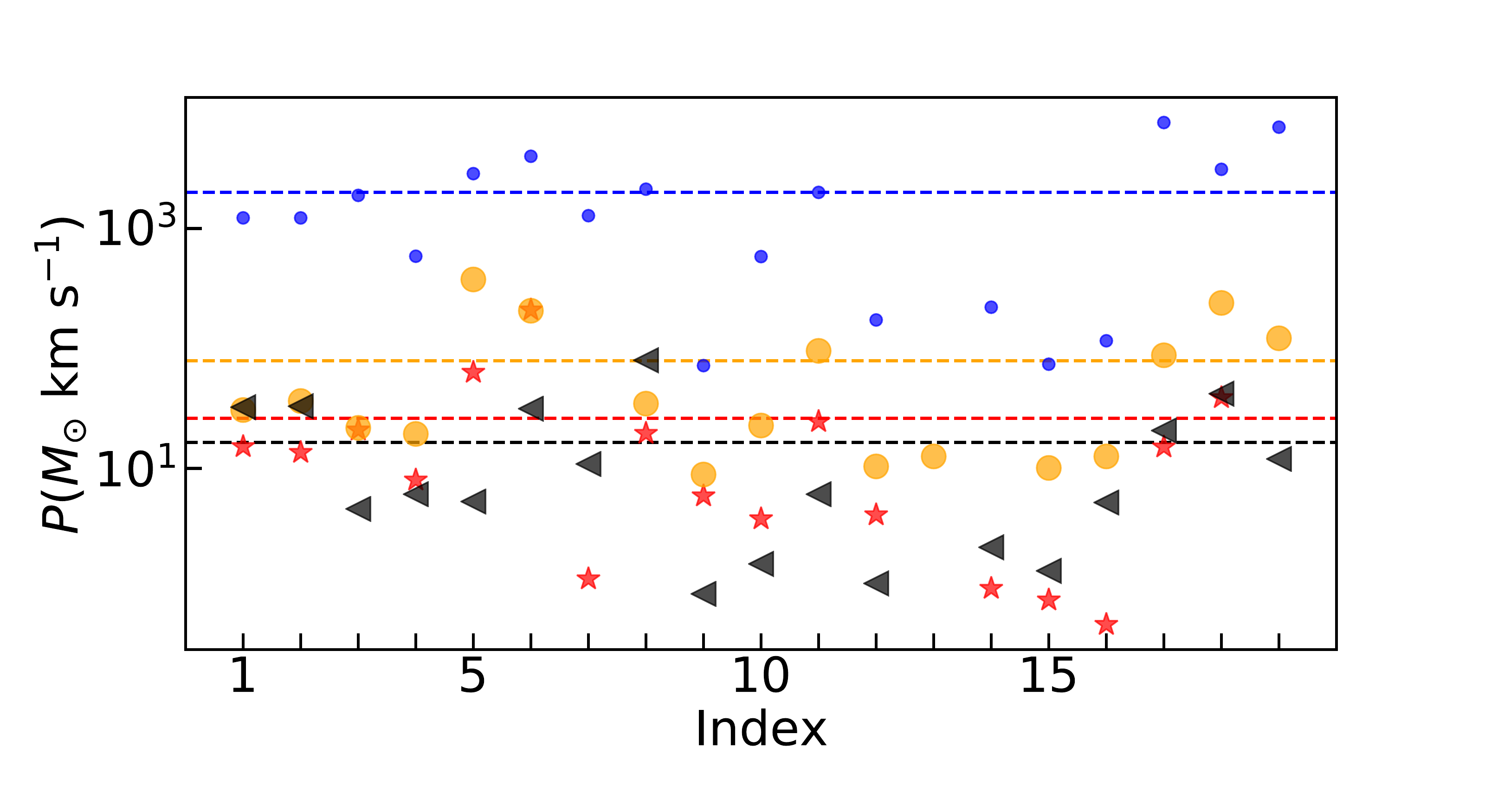}}
	\subfigure[Dynamical timescales]{\includegraphics[width=8.5cm]{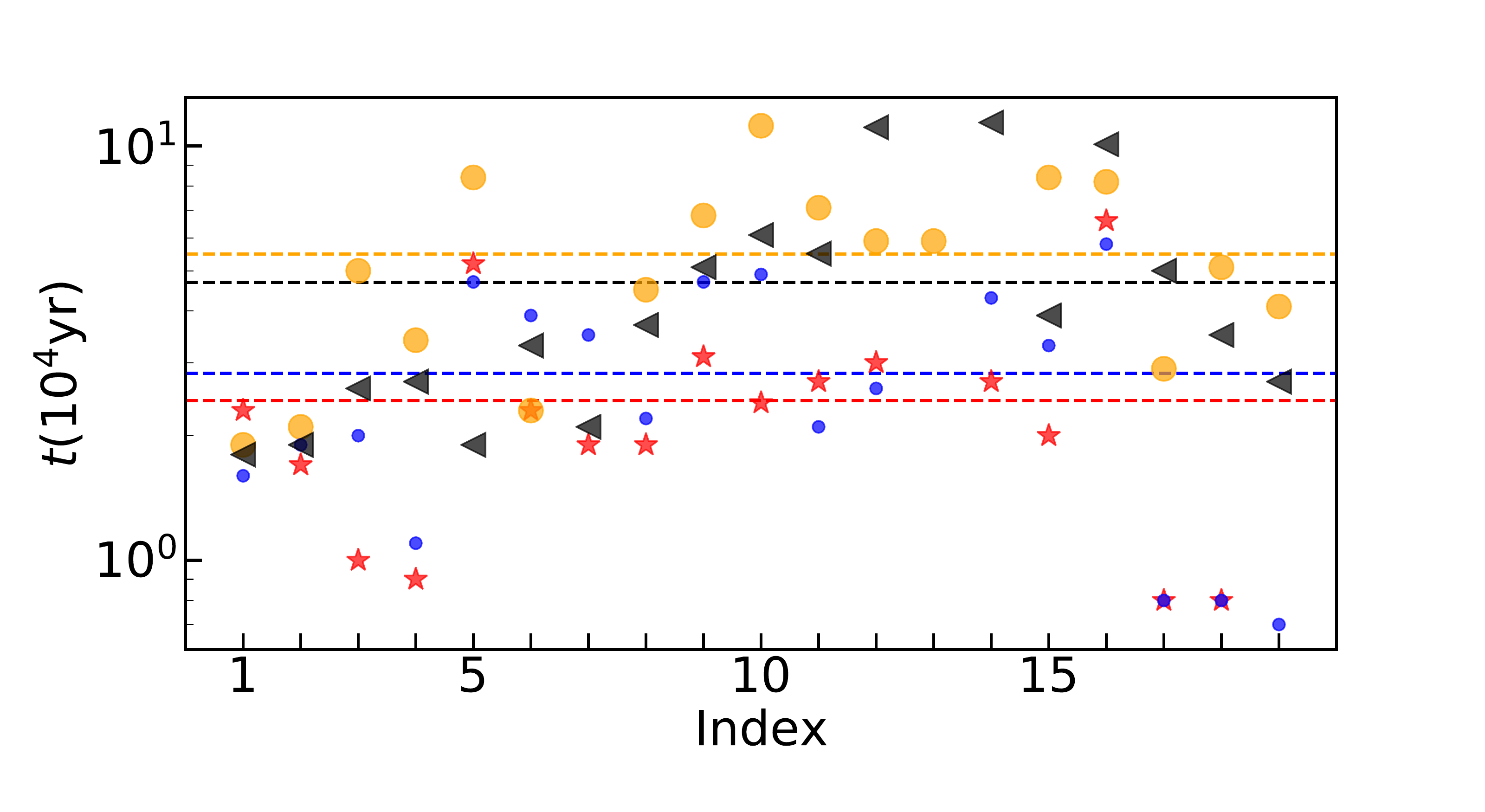}}
	\subfigure[Kinetic energies]{\includegraphics[width=8.5cm]{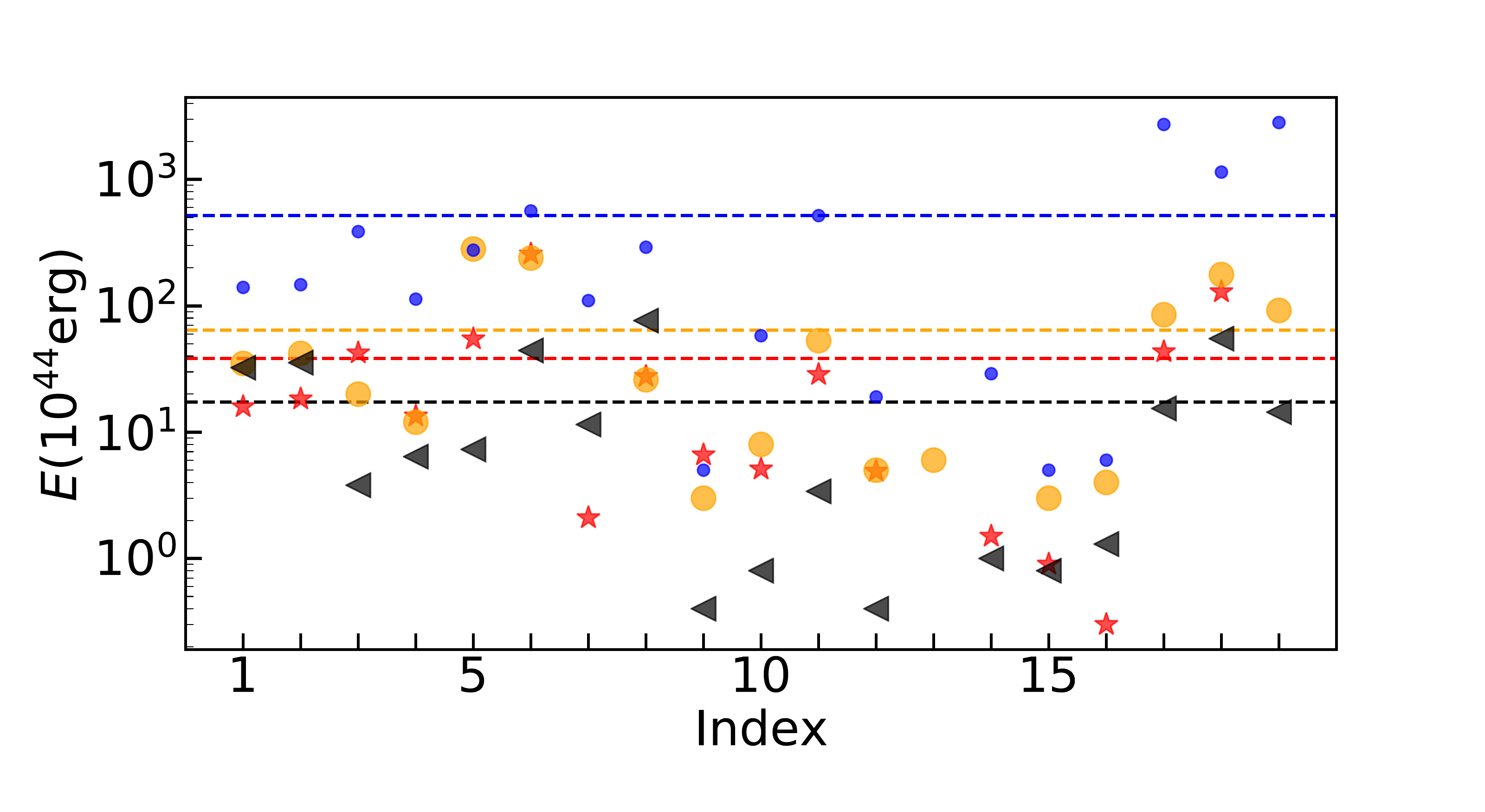}}
	\subfigure[Mechanical luminosities]{\includegraphics[width=8.5cm]{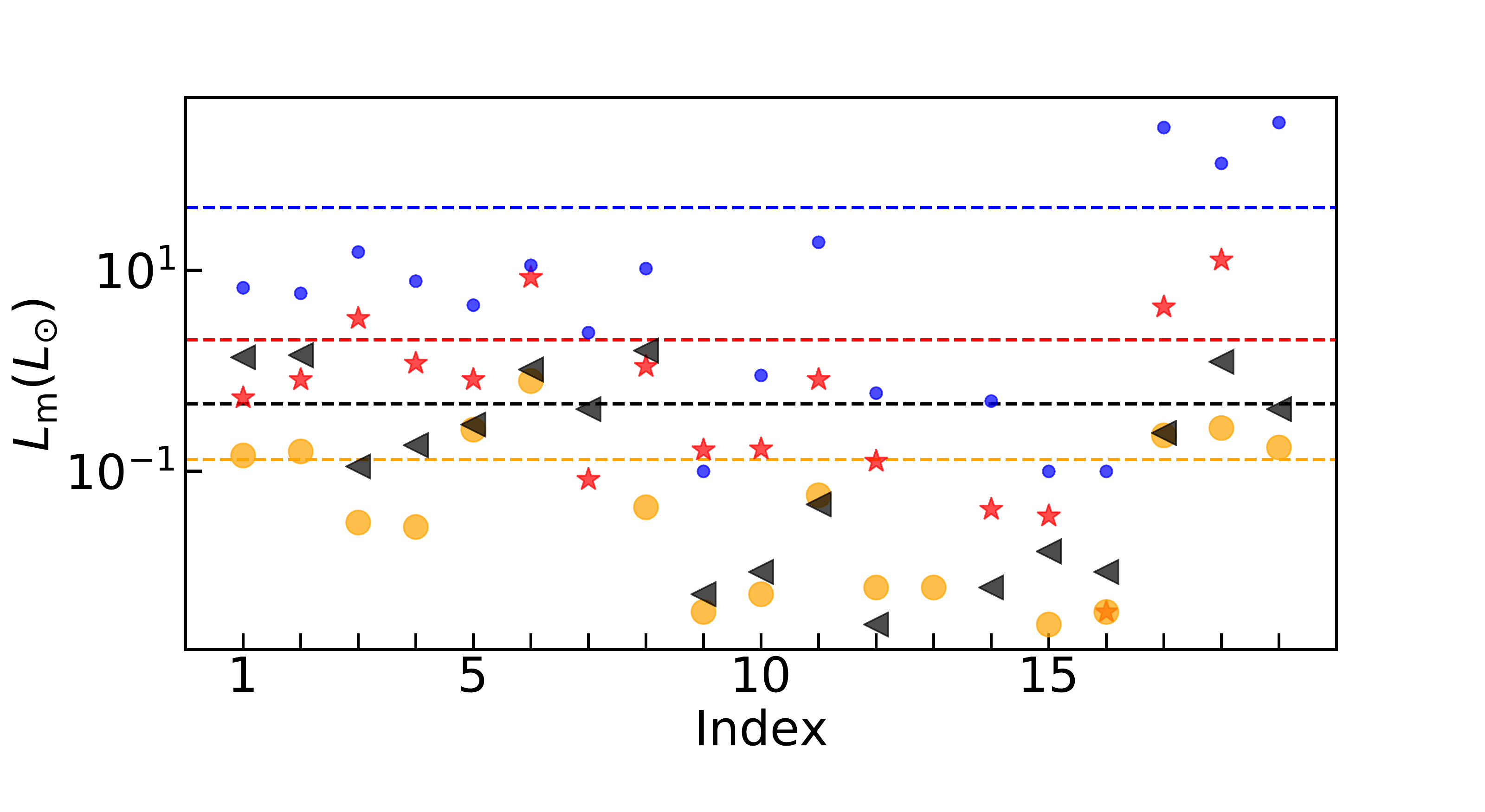}}
	\subfigure[Lengths]{\includegraphics[width=8.5cm]{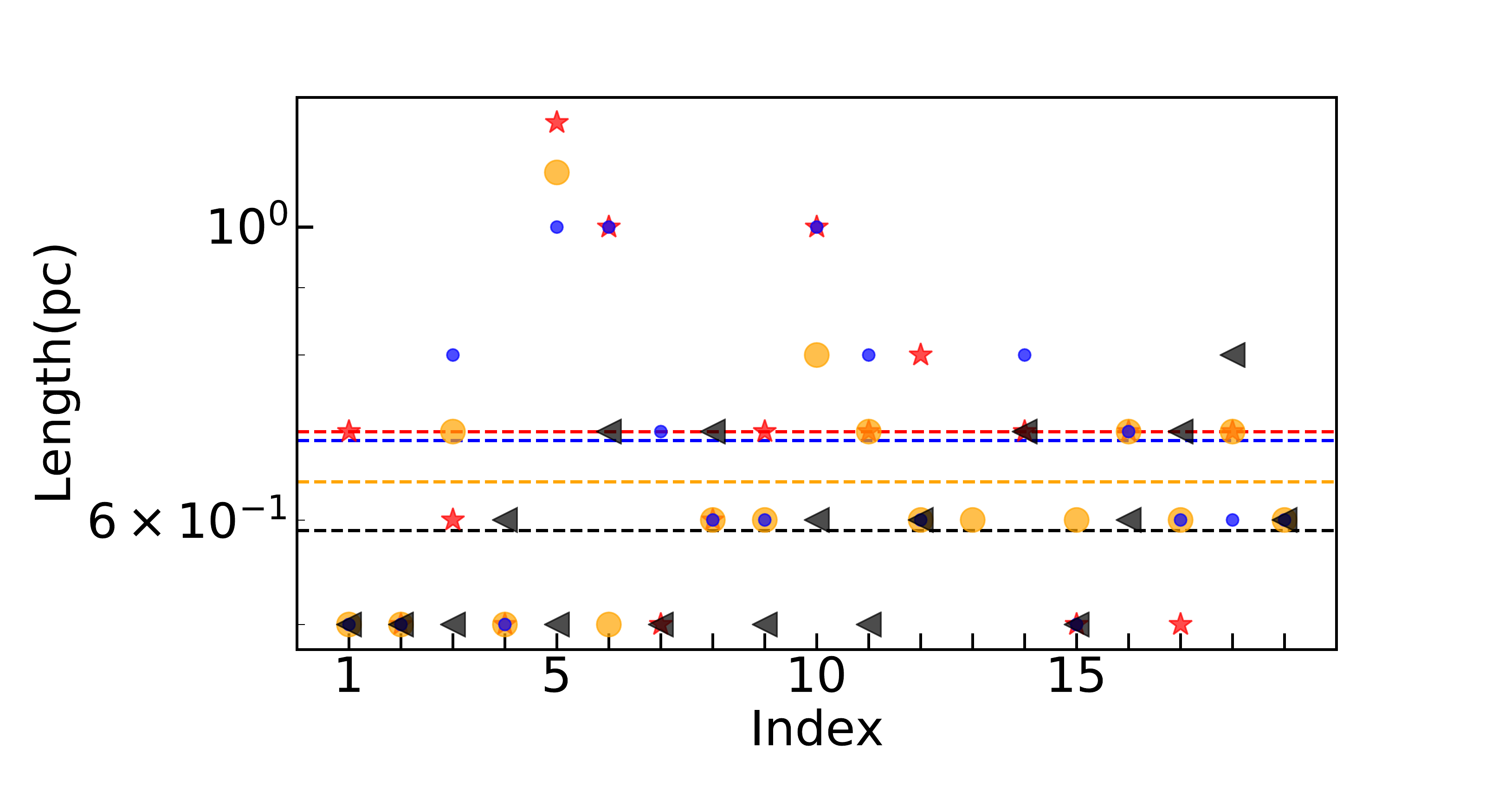}}
	\subfigure[Velocities]{\includegraphics[width=8.5cm]{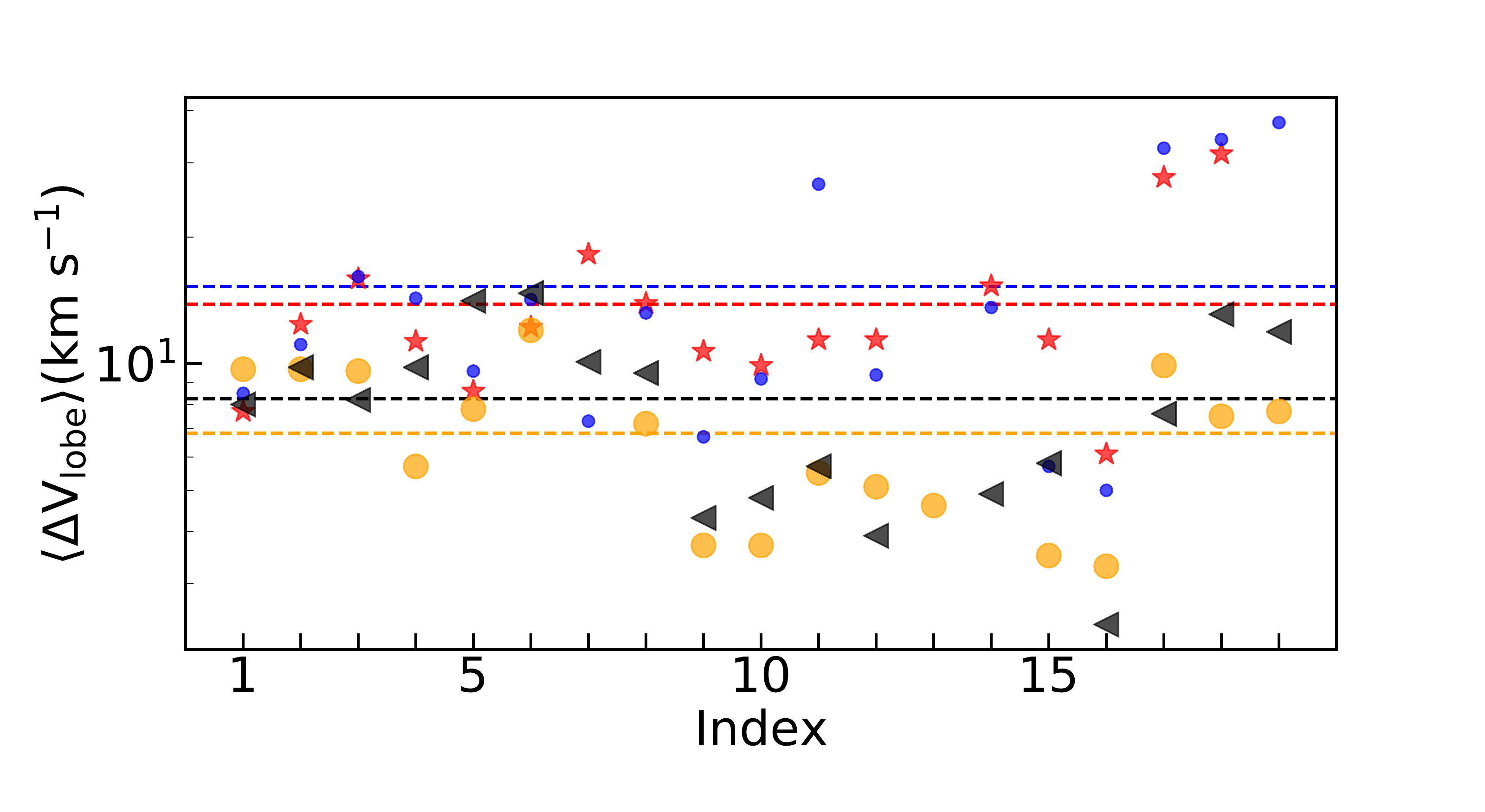}}
	\subfigure[Legend]{\includegraphics[width=8.5cm]{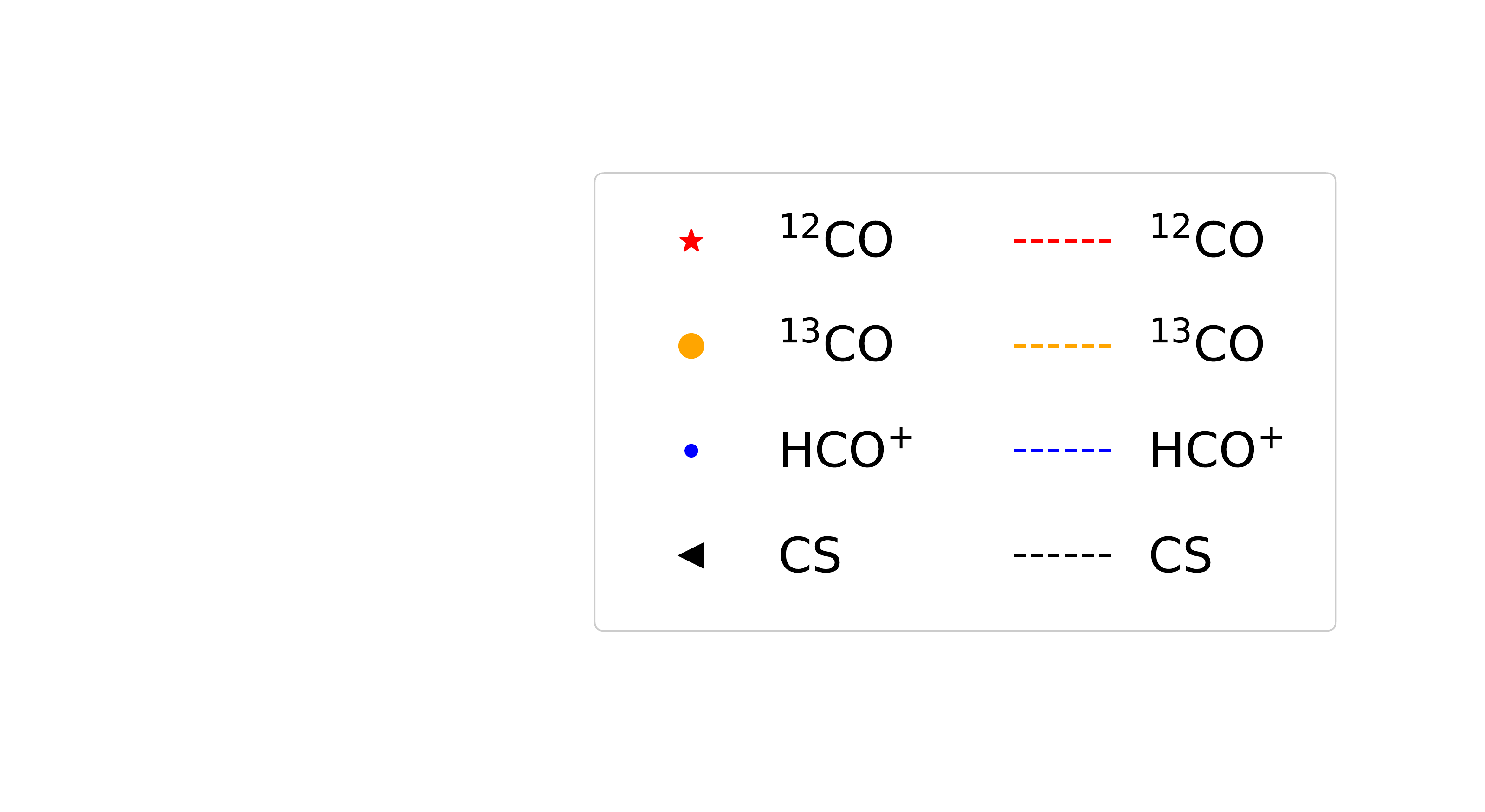}}
	\caption{The distributions of the physical properties of the outflows. The dashed lines are the mean value of the physical properties. The meaning of each index is given in Tables~\ref{tab:12CO} --~\ref{tab:CS}.}
	\label{fig:phys}
\end{figure} 
\clearpage

\subsection{Physical properties of the outflows}
Due to the accurate distances of these nearby sources and the high-sensitivity observations, we have obtained more accurate physical properties than found previously for these objects. The lengths, masses, momenta, kinetic energies, mechanical luminosities, and dynamical timescales were calculated (see Appendix~\ref{appendix}). The physical quantities of these molecular lines are summarized in Tables~\ref{tab:12CO} --~\ref{tab:CS}, and the distributions of these quantities are displayed in Figure~\ref{fig:phys}. For \coo, \xcoo, and CS, the molecular outflow of each lobe has a typical momentum of a few 10~\msun~\kmss, a kinetic energy of a few $10^{45}$~erg, a dynamical timescale of a few $10^{4}$~yr, and a mechanical luminosity of a few $10^{-1}$~$L_{\odot}$. However, the momenta, kinetic energies and mechanical luminosities of the \hco outflows are an order of magnitude larger than those of \coo, \xco and CS. The typical outflow mass for \xco and \hco is a few 10 \msun \kmss, which is an order magnitude larger than \co and CS. Although the masses of \xco outflow is silimar to \hcoo, the momenta, kinetic energies and mechanical luminosities of \xco are lower than those of \hcoo. It might result from the outflow velocity for \hco is much larger than \xcoo. However, there is no order of magnitude difference in the dynamical timescales of these four kinds of outflows.

We also obtained more accurate outflow velocities ($\langle \Delta v \rangle$) than previously found, where our results show that $\langle \Delta v \rangle ({\rm ^{12}CO}) \approx \langle \Delta v \rangle ({\rm HCO^{+}}) > \langle \Delta v \rangle ({\rm CS}) \approx \langle \Delta v \rangle ({\rm ^{13}CO})$. This relationship might indicate that relatively high-velocity outflows can be traced by \co and \hcoo, and relatively low-velocity outflows can be traced by \xco and CS. In our survey, the typical dynamical timescales ($t$) of our samples are lower than those found by \citet{Beuther+etal+2002,Zhang+etal+2005}. Meanwhile, the results show that $t_{\rm ^{13}CO} \approx t_{\rm CS} > t_{\rm ^{12}CO} \approx t_{\rm HCO^{+}}$. Considering that different gases trace different parts of the molecular clouds, the results seem to indicate that the inner gas (traced by \xco and CS) flows relatively slower than the outer gas (traced by \co and \hcoo). Meanwhile, the outer gas might disperse earlier than the inner gas. However, due to the limited number of objects in our sample (i.e., only nine sources), a larger sample size is needed to confirm/refute this tentative conclusion.

\subsection{Correlations between the outflow mass, mechanical luminosity and bolometric luminosity of the central sources}

A positive correlation was found between the masses ($M$) of the outflows and the bolometric luminosities ($L_{\rm bol}$) of the central IRAS sources (see Eq.~\ref{equ:1} and Figure~\ref{fig:LvsM}), where the slope of the best-fitting line is similar to that found by \citet{Wu+etal+2004,Maud+etal+2015}. This correlation suggests that the mass of an outflow probably depends on the nature of the central source \citep{Bally+Lada+1983,Wu+etal+2004,Maud+etal+2015}. Furthermore, the effect of the central star on the surrounding gas seems to increase from the inside out, because the slope of the best-fitting line of the molecules tracing the outer gas is greater than that tracing the inner gas. 

Similar to the results of \citet{Bontemps+etal+1996,Churchwell+1997,Wu+etal+2004,Maud+etal+2015}, we also find that the mechanical force ($F = P / \langle \Delta v \rangle$), the mass rate of the outflow ($\dot{M} = M / t$), and mechanical luminosity ($L_{\rm m}$) correlate with the bolometric luminosity of the central IRAS source (see details in Eqs.~\ref{equ:2} -- \ref{equ:4}, and Figures~\ref{fig:LvsF} -- \ref{fig:LvsL}), where the power-law indices of the best-fitting lines among those physical properties are similar to those found by \citet{Wu+etal+2004} and \citet{Maud+etal+2015}. In addition, these relationships also imply that the bolometric luminosity of the central IRAS sources and mechanical luminosity of the outflow are probably correlated with the other physical properties (e.g., the accretion rate).  Hence, we suggest the following dependencies: bolometric luminosity $\rightarrow$ accretion rate $\rightarrow$ mass-loss rate in the outflow $\rightarrow$ mechanical luminosity of outflow \cite[see also ][]{Wu+etal+2004}. Indeed, we find that the mass-loss rates in the outflows are correlated with the mechanical luminosities of the outflows (see Eq.~\ref{equ:5} and Figure~\ref{fig:LvsM_t}). This dependency was also found by \citet{Maud+etal+2015}, who had indicated that the central source with higher bolometric luminosity might entrain more material and thus can drive more powerful and more energetic outflow.

\section{Summary}
\label{sect:Sum}
We searched for outflows using multi-molecular lines toward nine nearby HMSFRs with accurate distances with the 14-m PMODLH millimeter-wavelength telescope. The main results of our study are summarized as follows:

1. \coo, \xco, \hcoo, and CS outflows were detected toward all nine sources (i.e., the detection rate of all outflows was 100\%). Bipolar or multiple outflows of \coo, \hcoo, and CS were detected for eight sources (i.e., the detection rate was about 89\%). Bipolar outflows of \xco were detected for six sources (i.e., the detection rate was about 67\%).

2. The full line widths of different molecules may have the following relationship: $\Delta V_{\rm ^{12}CO} > \Delta V_{\rm HCO^{+}} > \Delta V_{\rm CS} \approx \Delta V_{\rm ^{13}CO} > \Delta V_{\rm ^{18}CO}$.

3. \co and \hco can be used to trace relatively high-velocity outflows, and \xco and CS can be used to trace relatively low-velocity outflows. 

4. The dynamical timescale of different molecules may have the following relationship: $t_{\rm ^{13}CO} \approx t_{\rm CS} > t_{\rm ^{12}CO} \approx t_{\rm HCO^{+}}$.

5. There was a strong correlation between the bolometric luminosity of the central IRAS source and the mechanical luminosities of the outflows (including \coo, \xcoo, \hco and CS), and between the bolometric luminosities and outflow masses. The former correlation suggests a flow dependence, i.e., bolometric luminosity $\rightarrow$ accretion rate $\rightarrow$ mass-loss rate in the outflow $\rightarrow$ mechanical luminosity of outflow. The latter relationship indicates that the mass of the outflow probably depends on its driving source.

\acknowledgments
We are grateful to all the staff of Purple Mountain Observatory Delingha, especially our observer colleagues for obtaining the excellent observations. We would like to thank the anonymous referee. This work was funded by the NSFC, under grant numbers 11933011, 11873019, and 11673066, and by the Key Laboratory for Radio Astronomy. The research work was also supported by the National Natural Science Fund Committee of the Chinese Academy of Sciences Astronomical Union Funds No. U1731124, U2031202.

\appendix
\section{Deriving the outflow parameters} 
\label{appendix}
The $\rm H_{2}$ column density, $N({\rm H_{2}})$, traced by \co is given as \citet{Snell+etal+1984}
\begin{equation}
N({\rm H_{2}})=4.2 \times 10^{17} \frac{T_{\rm ex}}{e^{-5.5 / T_{\rm ex}}} \int{T_{\rm mb}}dv.
\label{eq:N12CO}
\end{equation}
where the velocity range to be integrated is that of the wing range. In this equation we assume that the gas is in local thermodynamic equilibrium (LTE), $X(\rm ^{12}CO) = [\rm ^{12}CO] / [\rm H_{2}] = 10^{-4}$ \citep{Snell+etal+1984}, and the excitation temperature, $T_{\rm ex}$, is 30 K. 

The $\rm H_{2}$ column density, $N({\rm H_{2}})$, traced by \xco is \citet{Wilson+etal+2013}
\begin{equation}
N({\rm H_{2}}) = 2.3 \times 10^{19} \frac{T_{\rm ex}}{e^{-5.3 / T_{\rm ex}}} \int{T_{\rm{mb}}}dv
\end{equation}
where the velocity range to be integrated is that of the line's wings. In this equation we assume that the gas is in LTE,  $X(\rm ^{13}CO) = [\rm ^{13}CO] / [\rm H_{2}] = 2 \times 10^{-6}$, and $T_{\rm ex} = 30\ \rm K$ \citep{Li+etal+2016}.

The $\rm H_{2}$ column density, $N({\rm H_{2}})$, traced by \hco can be found as \citet{Yang+etal+1991}
\begin{equation}
N({\rm H_{2}}) = 1.87 \times 10^{19} \frac{T_{\rm ex}}{1 - e^{-4.3/T_{\rm ex}}}\int{T_{\rm mb}}dv.
\end{equation}
where the integration range is over the range of the wings. In this equation we assume that the gas is in LTE, $X(\rm HCO^{+}) = [\rm HCO^{+}] / [\rm H_{2}] = 10^{-8}$ \citep{Turner+etal+1997} and $T_{\rm ex} = 15\ \rm K$.

The $\rm H_{2}$ column density, $N({\rm H_{2}})$, traced by CS is
\begin{equation}
N({\rm H_{2}}) = 3 \times 10^{9}\frac{k^{2}T_{\rm ex}}{4\pi^{3}\mu_{d}^{2}hv e^{-hv / k \rm {T_{ex}}}}\int T_{\rm mb} dv
\end{equation}
where $k$ is the Boltzmann constant ($\rm 1.38 \times 10^{-16}\ erg\ K^{-1}$), $h$ is the Planck constant ($\rm 6.626 \times 10^{-27}\ erg\ s$), $\mu_{d}$ is the dipole moment (1.96 D), and $v$ is the transition frequency (97.981 GHz). The velocity range of integration is that of the wing's range. In this equation we assume that the gas is in LTE, $X(\rm CS) = [\rm CS] / [\rm H_{2}] = 10^{-9}$, and $T_{\rm ex} = 20\ \rm K$ \citep{Tatematsu+etal+1998}.

The mass of the outflow lobe, $M_{\rm lobe}$, can be found as: 
\begin{equation}
M_{\rm lobe} =  N_{\rm lobe} A_{\rm lobe}\mu m_{\rm H}
\end{equation}
where $A_{\rm lobe}$ represents the area of the blue/red lobes of outflows, $\mu = 2.72$ is the mean molecular weight, and $m_{\rm H}$ is the hydrogen molecule \citep{Garden+etal+1991}. The area was estimated by the region covered by 50\% of the outflow peak.

The momentum ($P_{\rm lobe}$) and kinetic energy ($E_{\rm lobe}$) of an outflow lobe are, respectively:
\begin{equation}
P_{\rm lobe}=\sum_{A_{\rm lobe}}^{}{M_{\rm lobe}\langle \Delta v_{\rm lobe} \rangle},
\end{equation}

and

\begin{equation}
E_{\rm lobe}=\frac{1}{2}\sum_{A_{\rm lobe}}^{}{M_{\rm lobe}\langle \Delta v_{\rm lobe}^{2} \rangle}.
\end{equation}
where $\langle \Delta v_{\rm lobe} \rangle$ and $\langle \Delta v_{\rm lobe}^{2} \rangle$ are the velocity (i.e., the relative velocity with respect to the central velocity), and the square of the velocity of an outflow lobe \cite[see detail in ][]{Li+etal+2018}. The mechanical luminosity ($L_{\rm lobe}$) of an outflow lobe is
\begin{equation}
L_{\rm lobe} = E_{\rm lobe} / t_{\rm lobe}
\end{equation}
where $t_{\rm lobe}=l_{\rm lobe} / \Delta v_{\rm max}$, $\Delta v_{\rm max}$ is the maximum outflow lobe velocity, and $l_{\rm lobe}$ is the length of the outflow.

As it is hard to determine the inclination angle of an outflow, we have adopted a mean inclination angle of $57.3^{\circ}$ to conform with similar studies \citep{Bontemps+etal+1996}. The inclination and blending correction factors are cited \cite[][see table~5 of that paper]{Li+etal+2019}.

\section{Relationship between the outflow mass, mechanical luminosity and bolometric luminosity of the central sources}

\begin{figure}
	\centering
	\subfigure[\co]{\includegraphics[width=6cm]{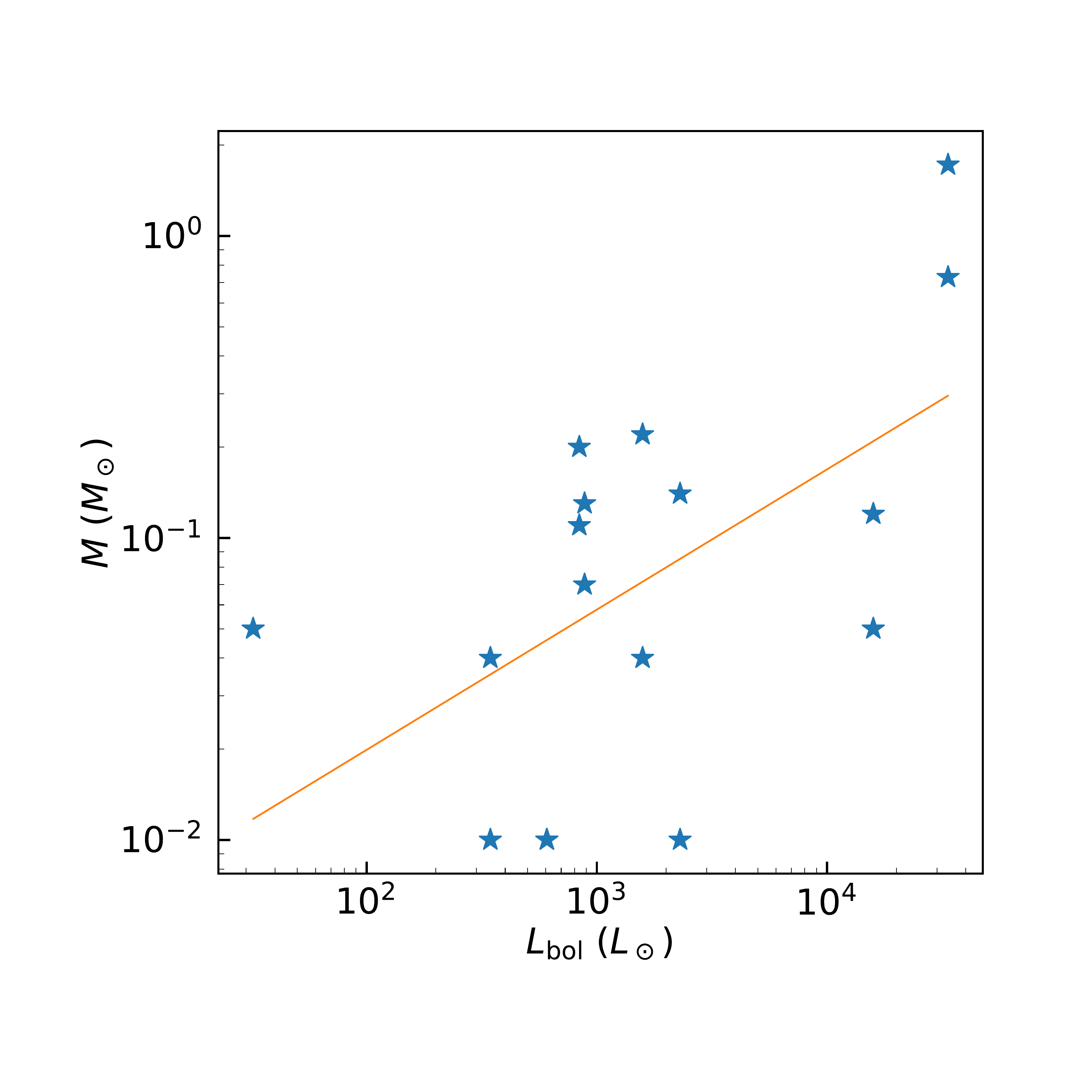}}
	\subfigure[\xco]{\includegraphics[width=6cm]{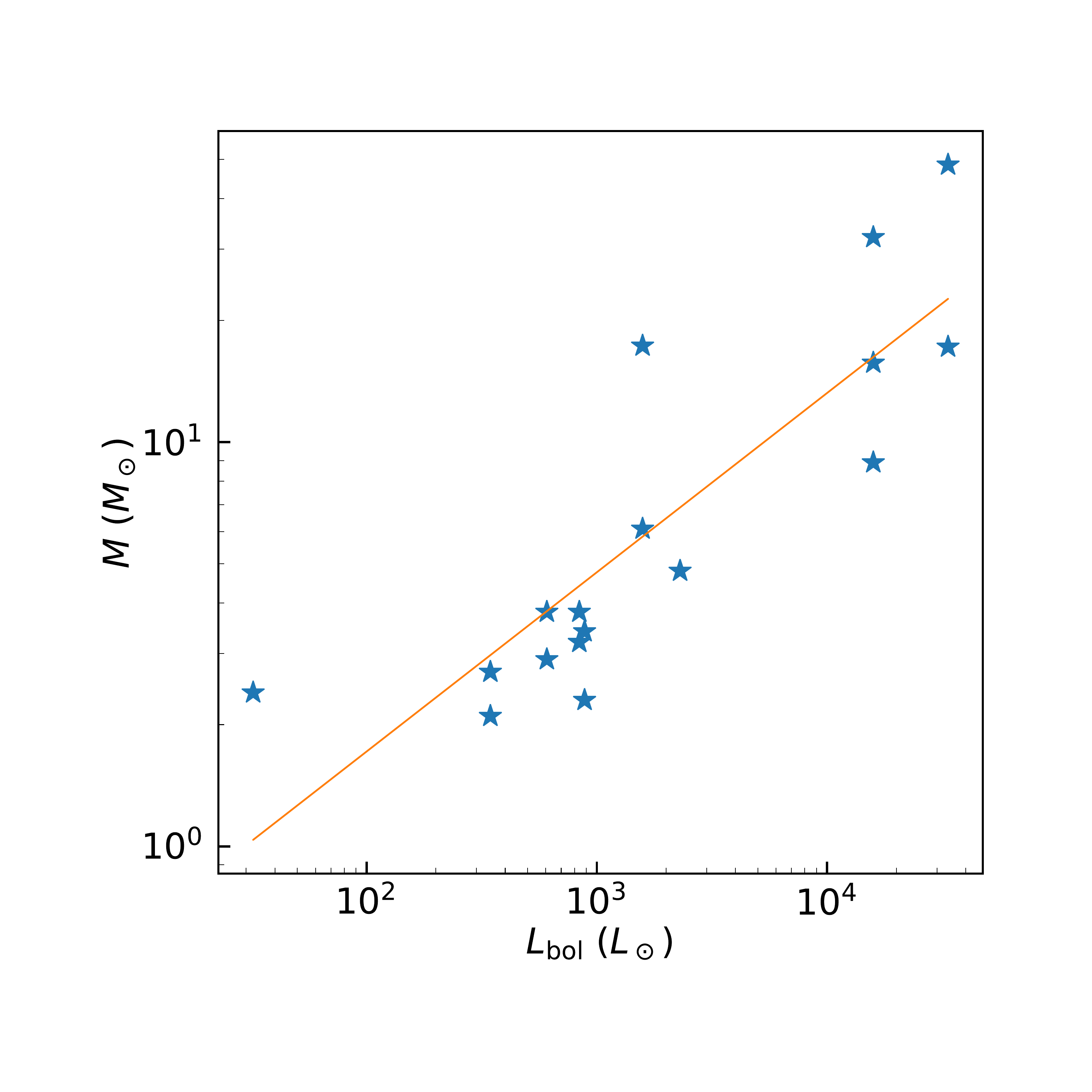}}
	\subfigure[\hco]{\includegraphics[width=6cm]{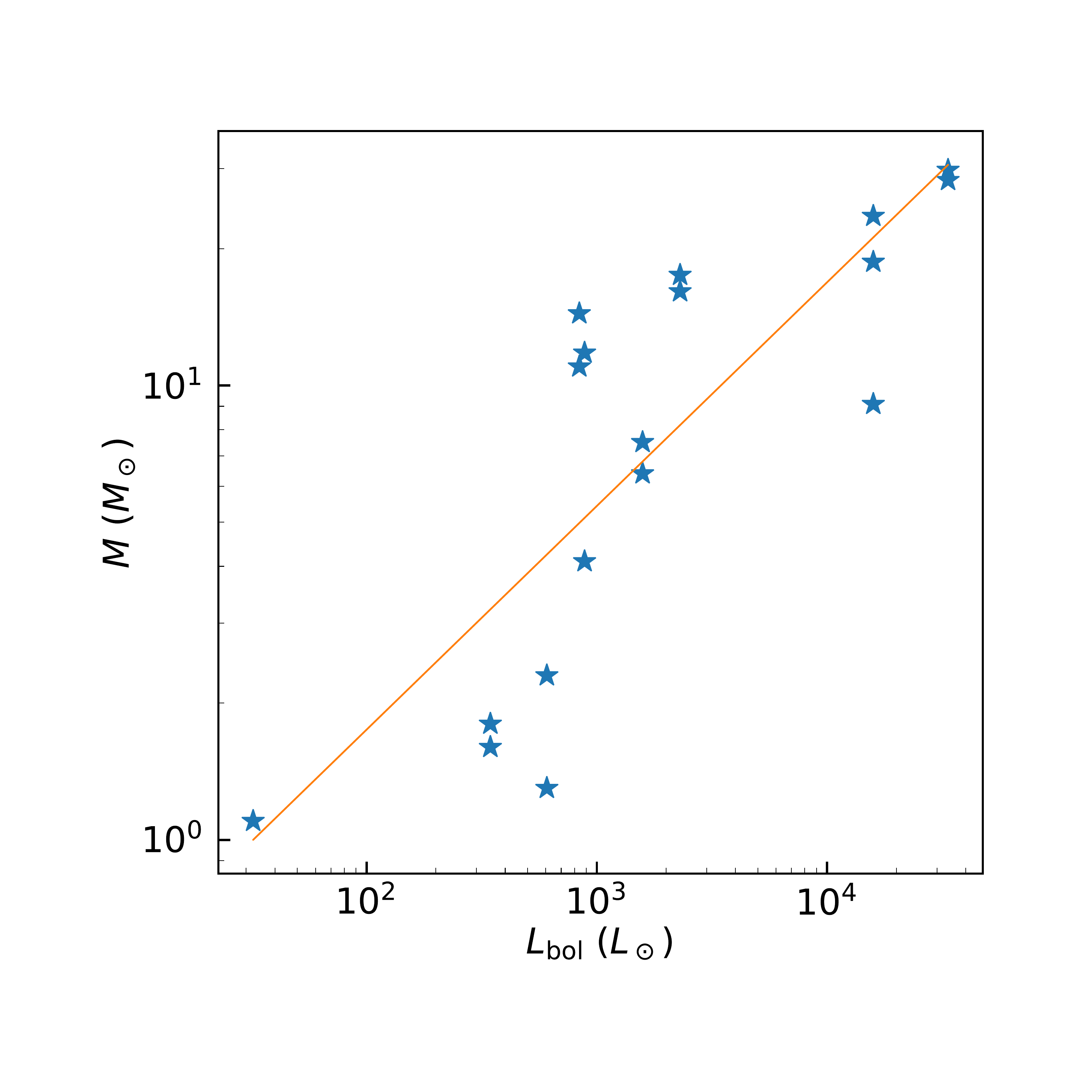}}
	\subfigure[CS]{\includegraphics[width=6cm]{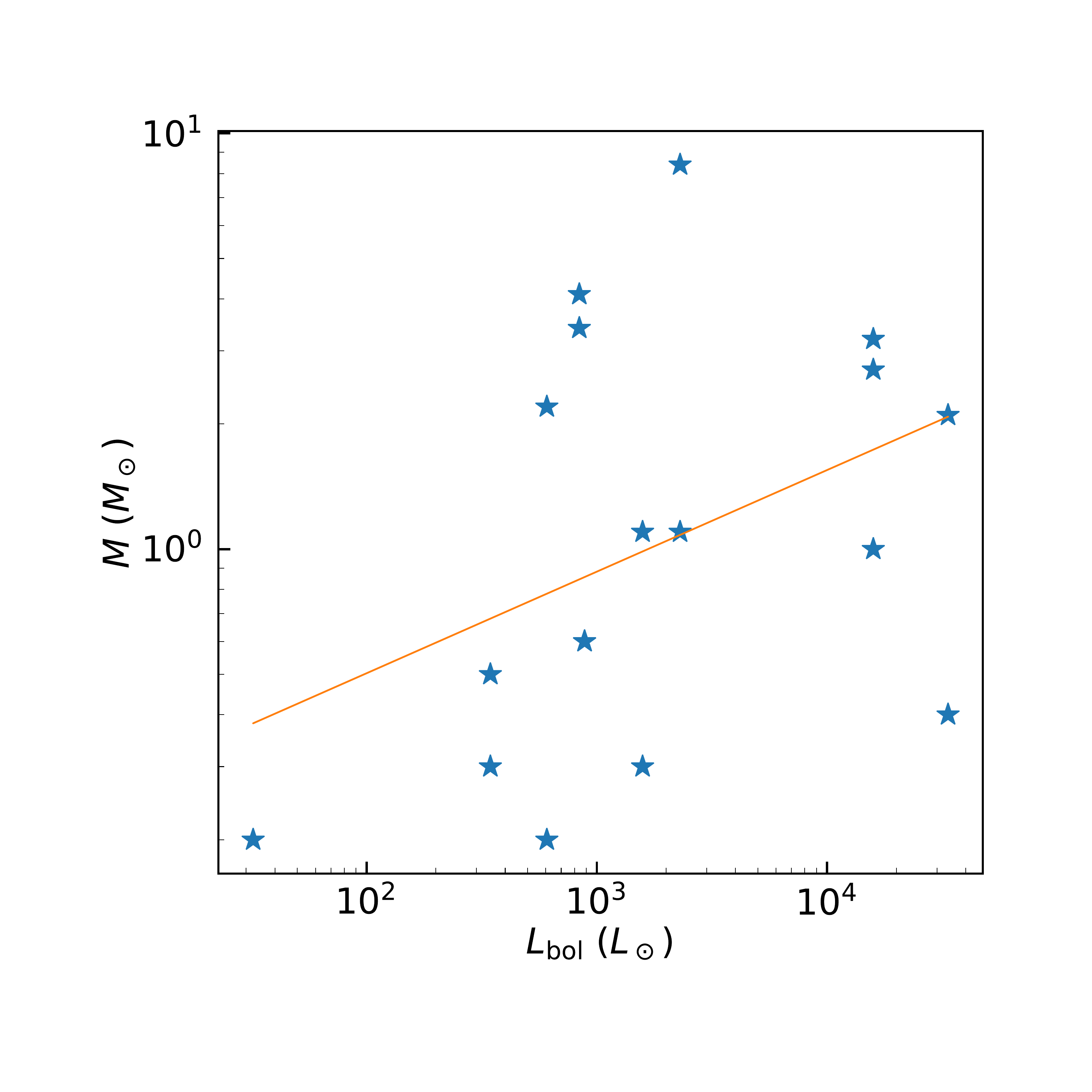}}
	\subfigure[Fitting lines]{\includegraphics[width=6cm]{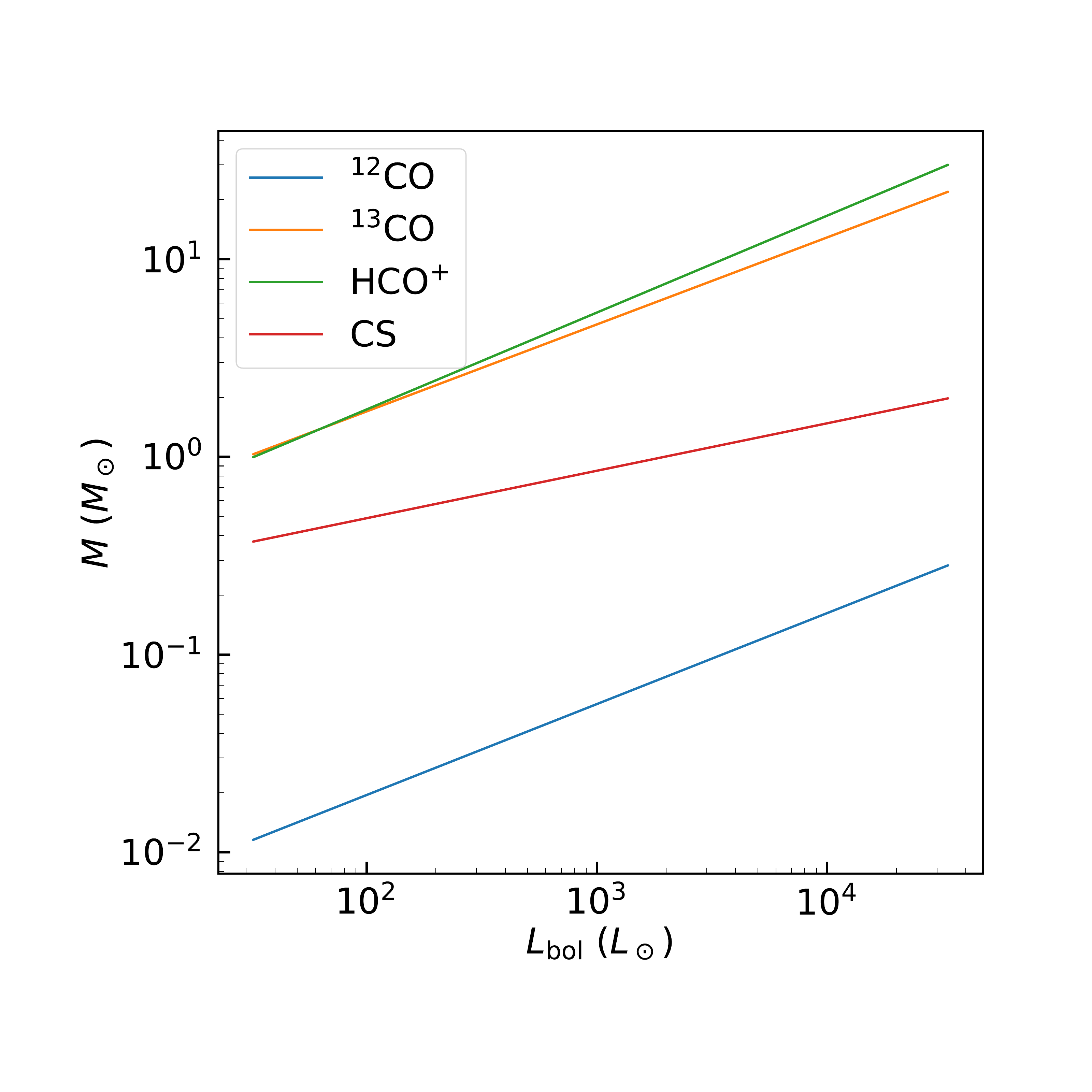}}
	\caption{The mass ($M$) of different molecular outflows as a function of the bolometric luminosity ($L_{\rm bol}$) of the central IRAS source. Each molecular line is given at the bottom of each panel. Panel (e) is the least-square fitting lines of the four molecular outflows.}
	\label{fig:LvsM}
\end{figure}

The mass, $M$, of the \coo, \xcoo, \hcoo, and CS outflows as a function of the bolometric luminosity, $L_{\rm bol}$, of the central IRAS source is as follows (see Figure~\ref{fig:LvsM}):

\begin{subequations}\label{equ:1}
	\begin{align}
	& \log M{\rm (^{12}CO)} = (0.46 \pm 0.17)\ \log L_{\rm bol} + (-2.63 \pm 0.57), r = 0.57; \label{equ:1a}\\
	& \log M{\rm (^{13}CO)} = (0.44 \pm 0.07)\ \log L_{\rm bol} + (-0.65 \pm 0.24), r = 0.85; \label{equ:1b}\\
	& \log M{\rm (HCO^{+})} = (0.49 \pm 0.09)\ \log L_{\rm bol} + (-0.74 \pm 0.29), r = 0.82; \label{equ:1c}\\
	& \log M{\rm (CS)} = (0.24 \pm 0.14)\ \log L_{\rm bol} + (-0.79 \pm 0.47), r = 0.40. \label{equ:1d}
	\end{align}
\end{subequations}

\begin{figure}
	\centering
	\subfigure[\co]{\includegraphics[width=6cm]{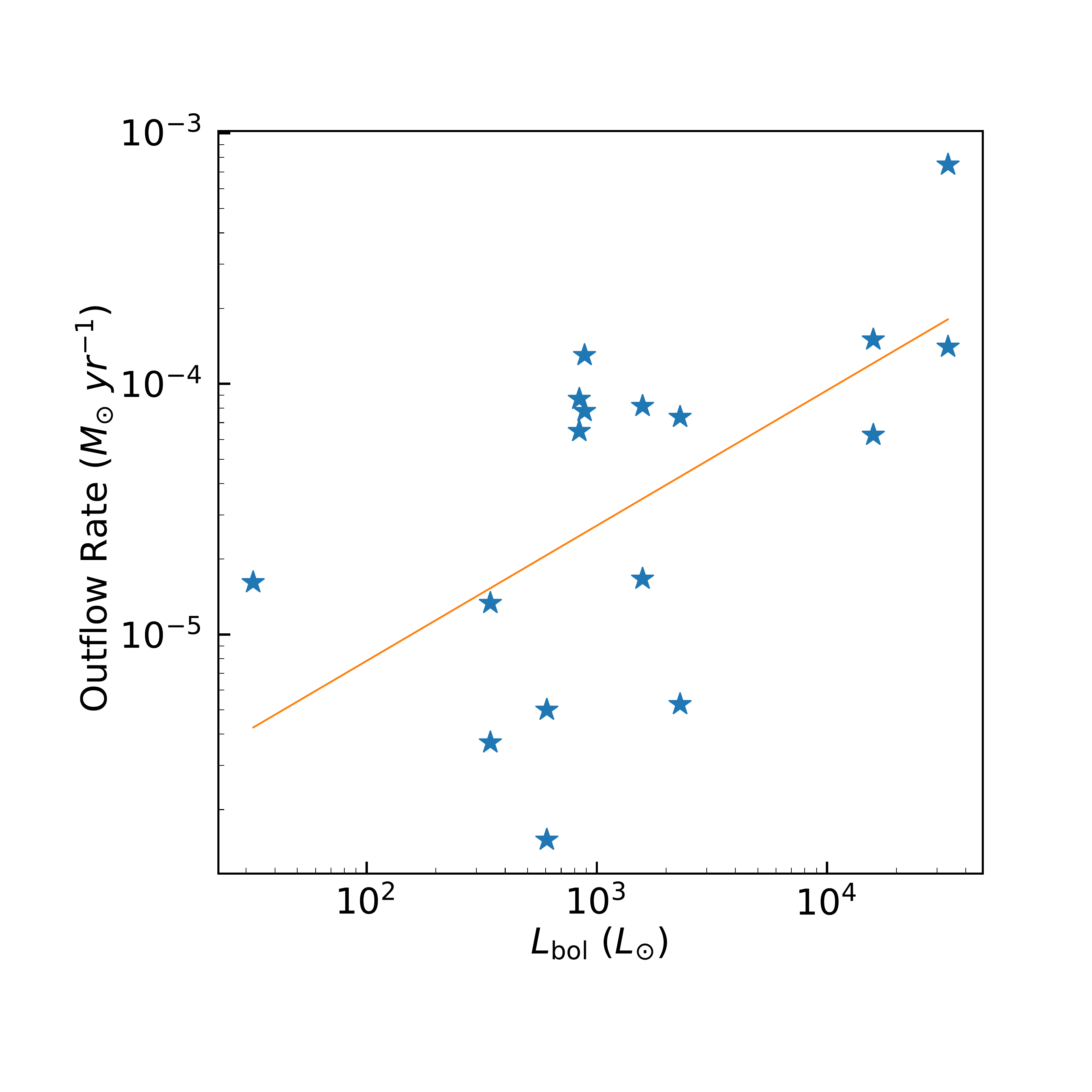}}
	\subfigure[\xco]{\includegraphics[width=6cm]{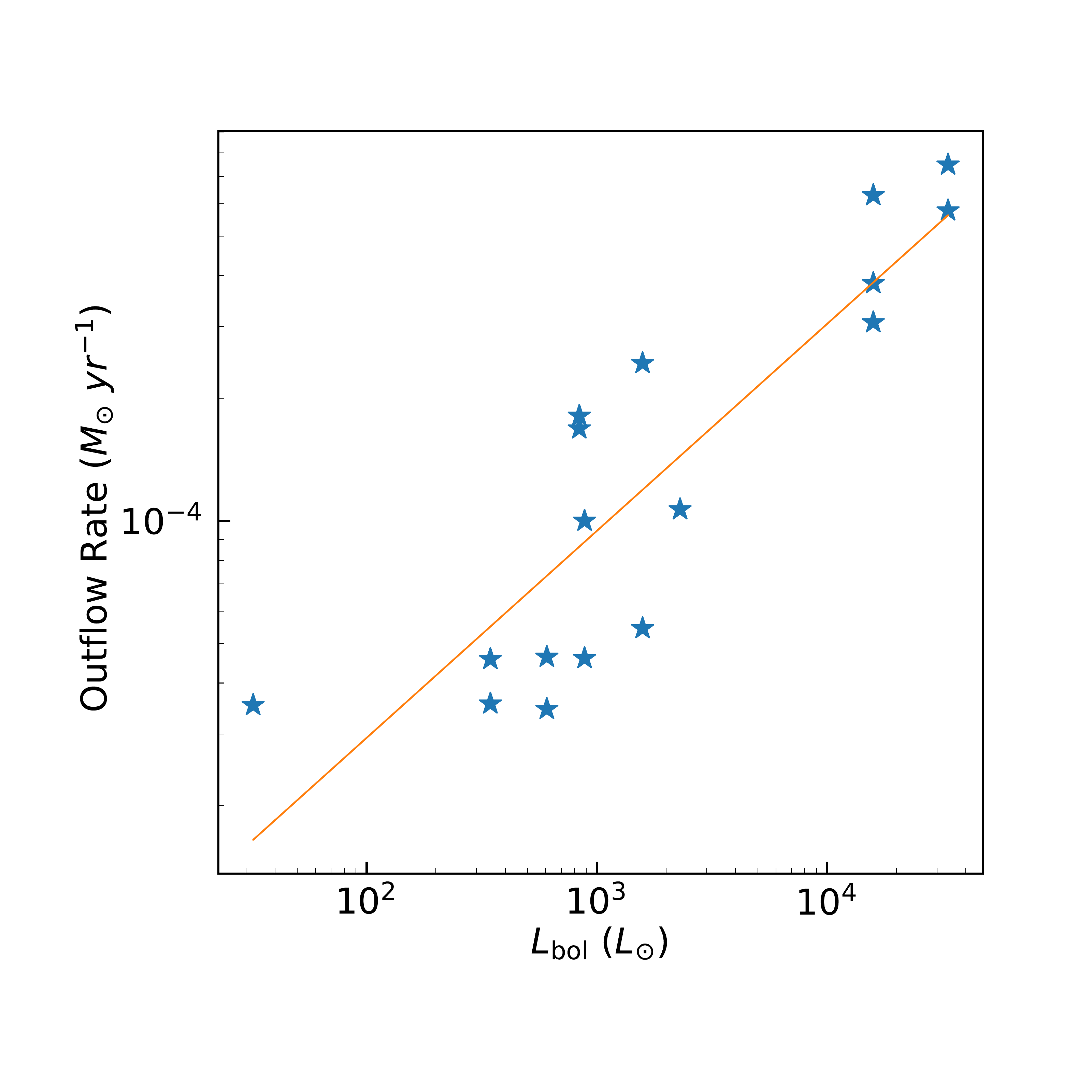}}
	\subfigure[\hco]{\includegraphics[width=6cm]{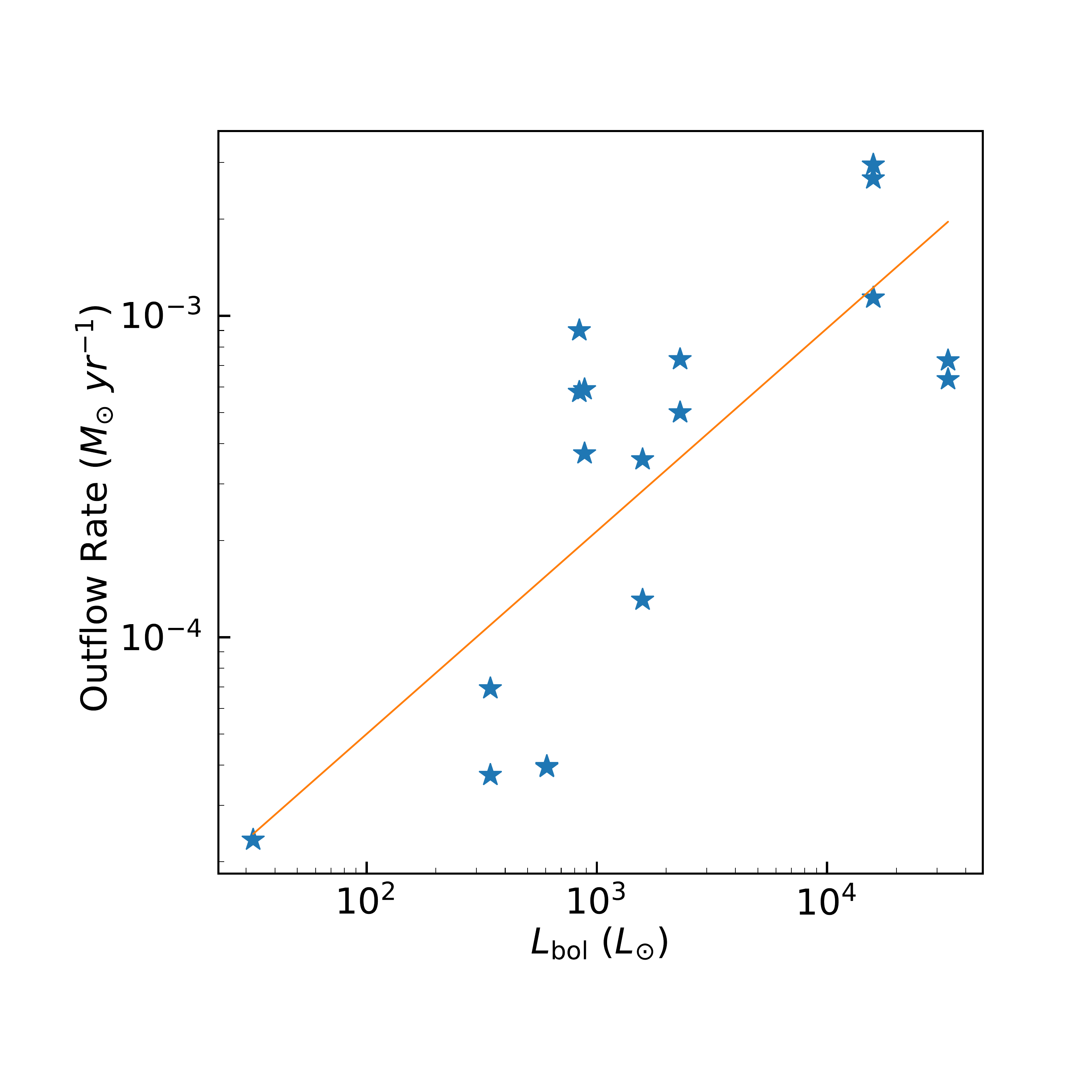}}
	\subfigure[CS]{\includegraphics[width=6cm]{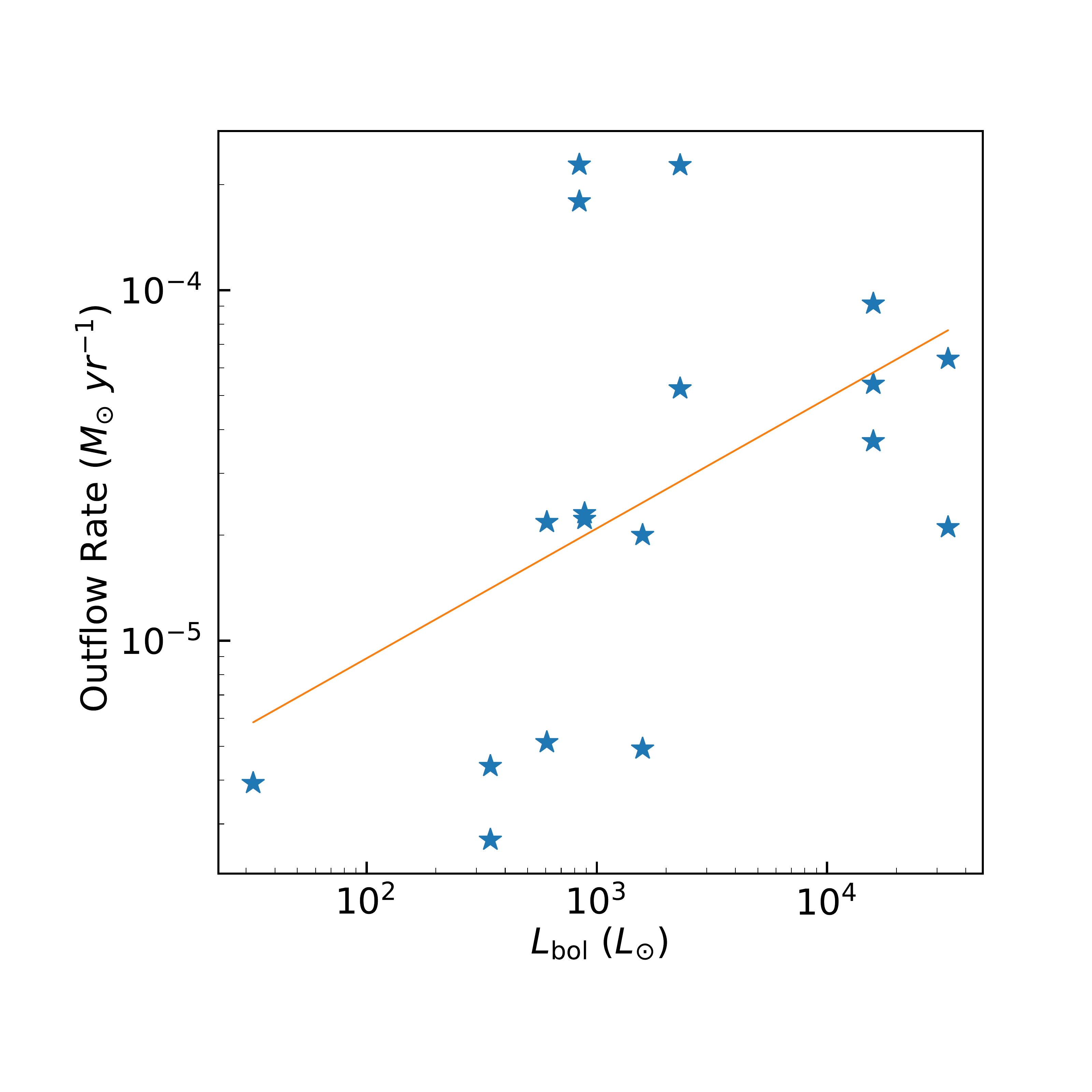}}
	\subfigure[Fitting lines]{\includegraphics[width=6cm]{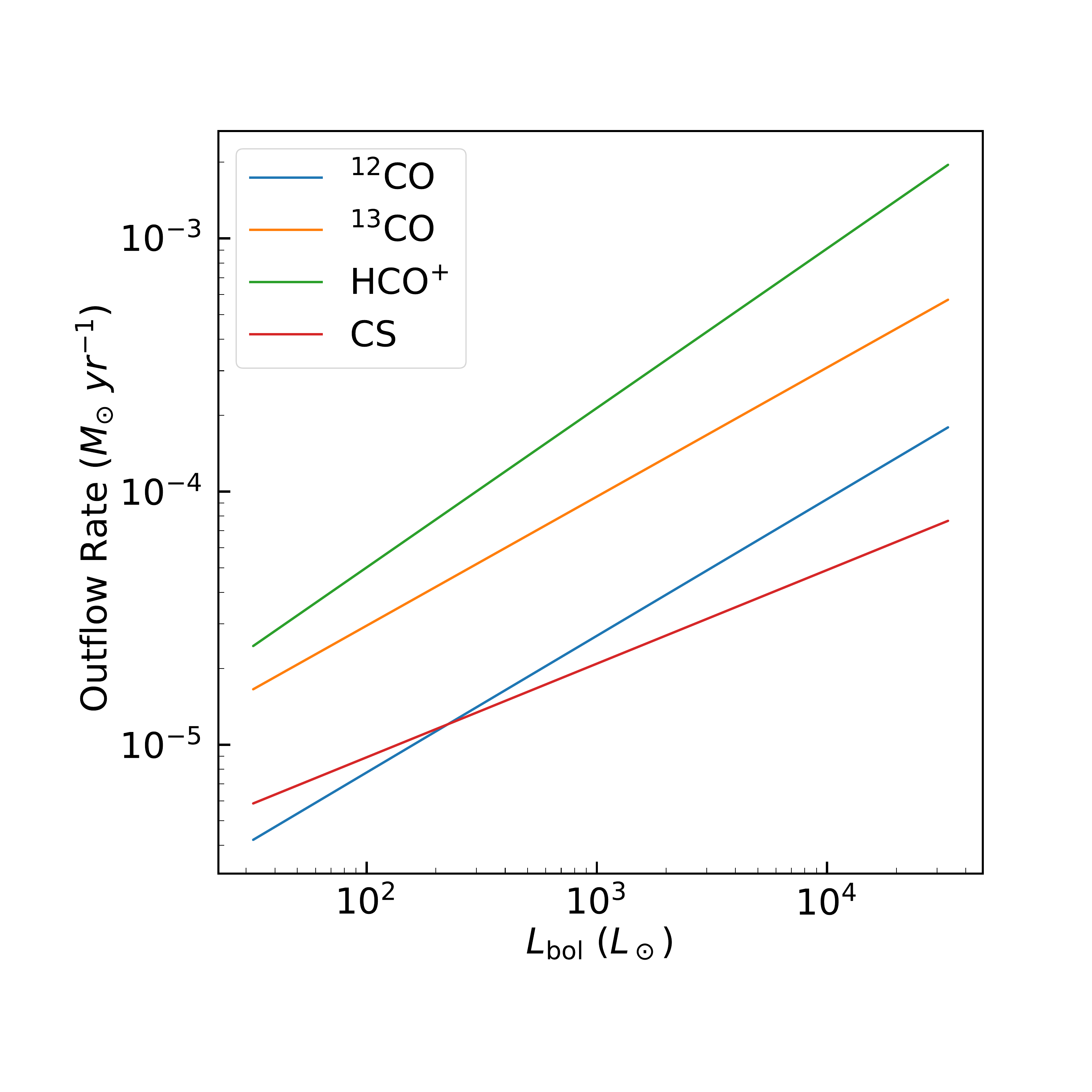}} 
	\caption{The mass outflow rate ($\dot{M}$) of different molecular outflows as a function of the bolometric luminosity ($L_{\rm bol}$) of the central IRAS source. Each molecular line is indicated at the bottom of each panel. Panel (e) is the least-square fitting lines of the four molecular outflows.}
	\label{fig:LvsR}
\end{figure}

The mass outflow rate, $\dot{M}$, of the \coo, \xcoo, \hcoo, and CS outflows as a function of the bolometric luminosity, $L_{\rm bol}$, of the central IRAS source is as follows (see Figure~\ref{fig:LvsR}):
\begin{subequations}\label{equ:2}
	\begin{align}
	& \log \dot{M}{\rm (^{12}CO)} = (0.54 \pm 0.19)\ \log L_{\rm bol} + (-6.19 \pm 0.62), r = 0.60; \label{equ:2a} \\
	& \log \dot{M}{\rm (^{13}CO)} = (0.51 \pm 0.07)\ \log L_{\rm bol} + (-5.55 \pm 0.24), r = 0.87; \label{equ:2b} \\
	& \log \dot{M}{\rm (HCO^{+})} = (0.63 \pm 0.13)\ \log L_{\rm bol} + (-5.56 \pm 0.43), r = 0.78; \label{equ:2c} \\
	& \log \dot{M}{\rm (CS)} = (0.37 \pm 0.17)\ \log L_{\rm bol} + (-5.79 \pm 0.56), r = 0.48. \label{equ:2d}
	\end{align}
\end{subequations}

\begin{figure}
	\centering
	\subfigure[\co]{\includegraphics[width=6cm]{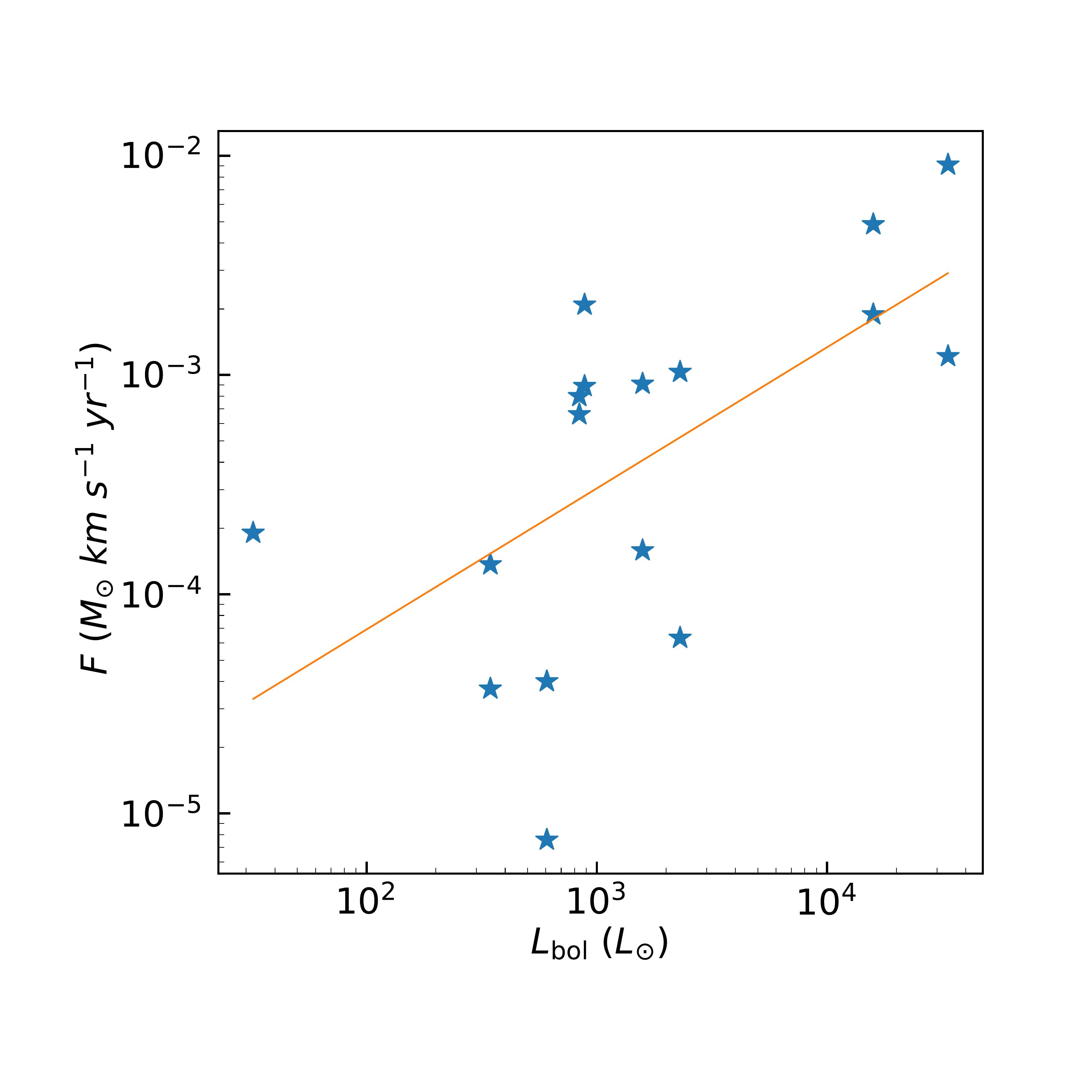}}
	\subfigure[\xco]{\includegraphics[width=6cm]{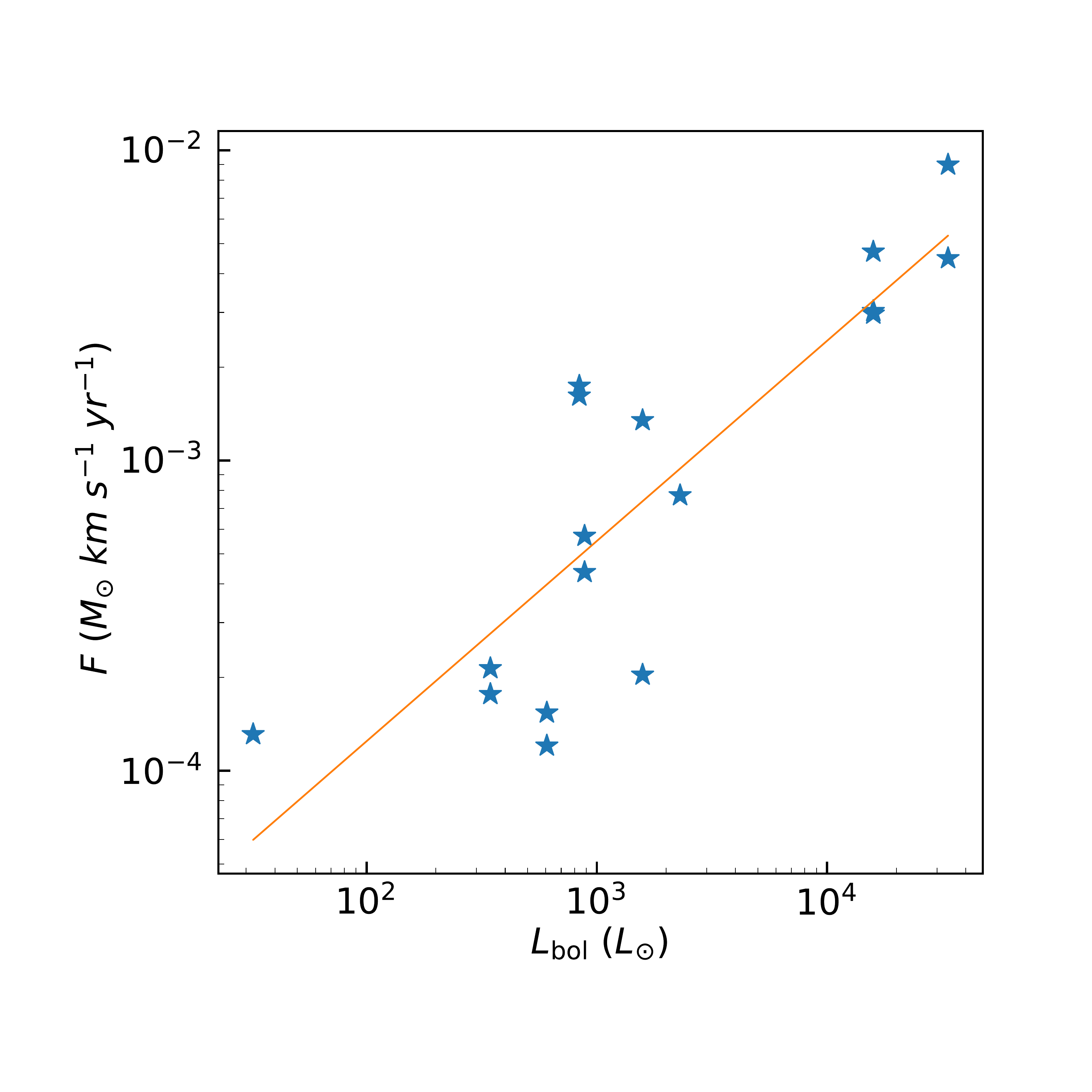}}
	\subfigure[\hco]{\includegraphics[width=6cm]{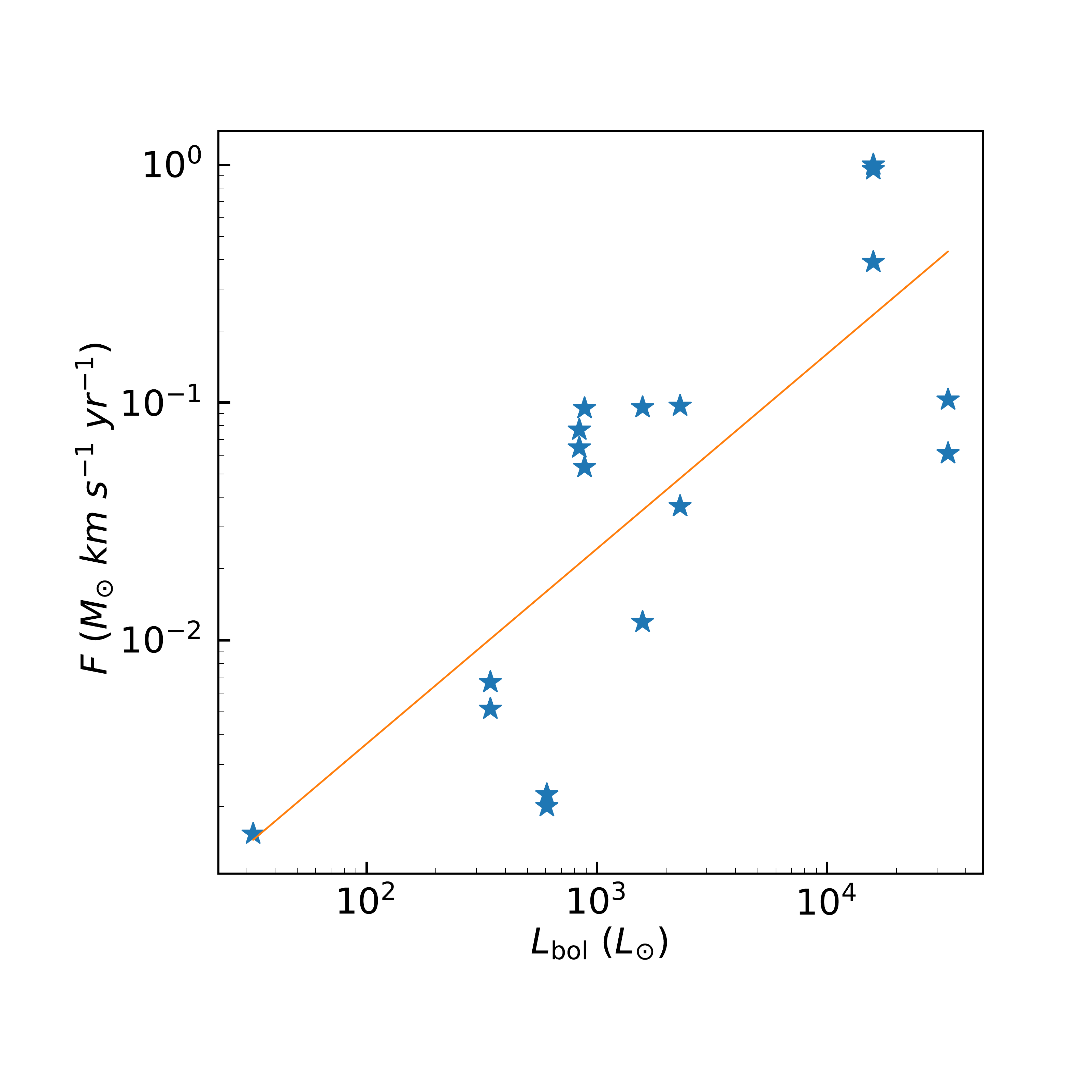}}
	\subfigure[CS]{\includegraphics[width=6cm]{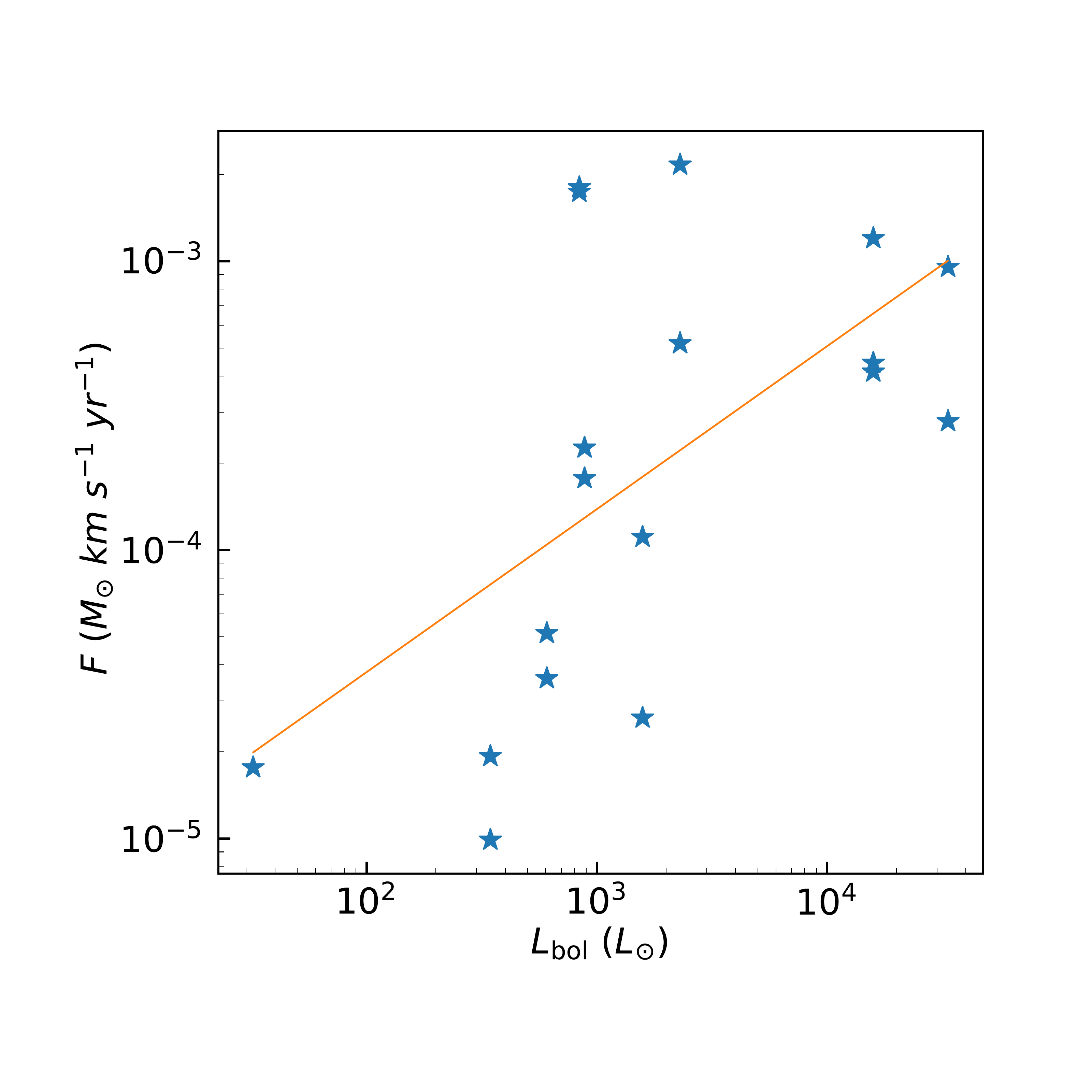}}
	\subfigure[Fitting lines]{\includegraphics[width=6cm]{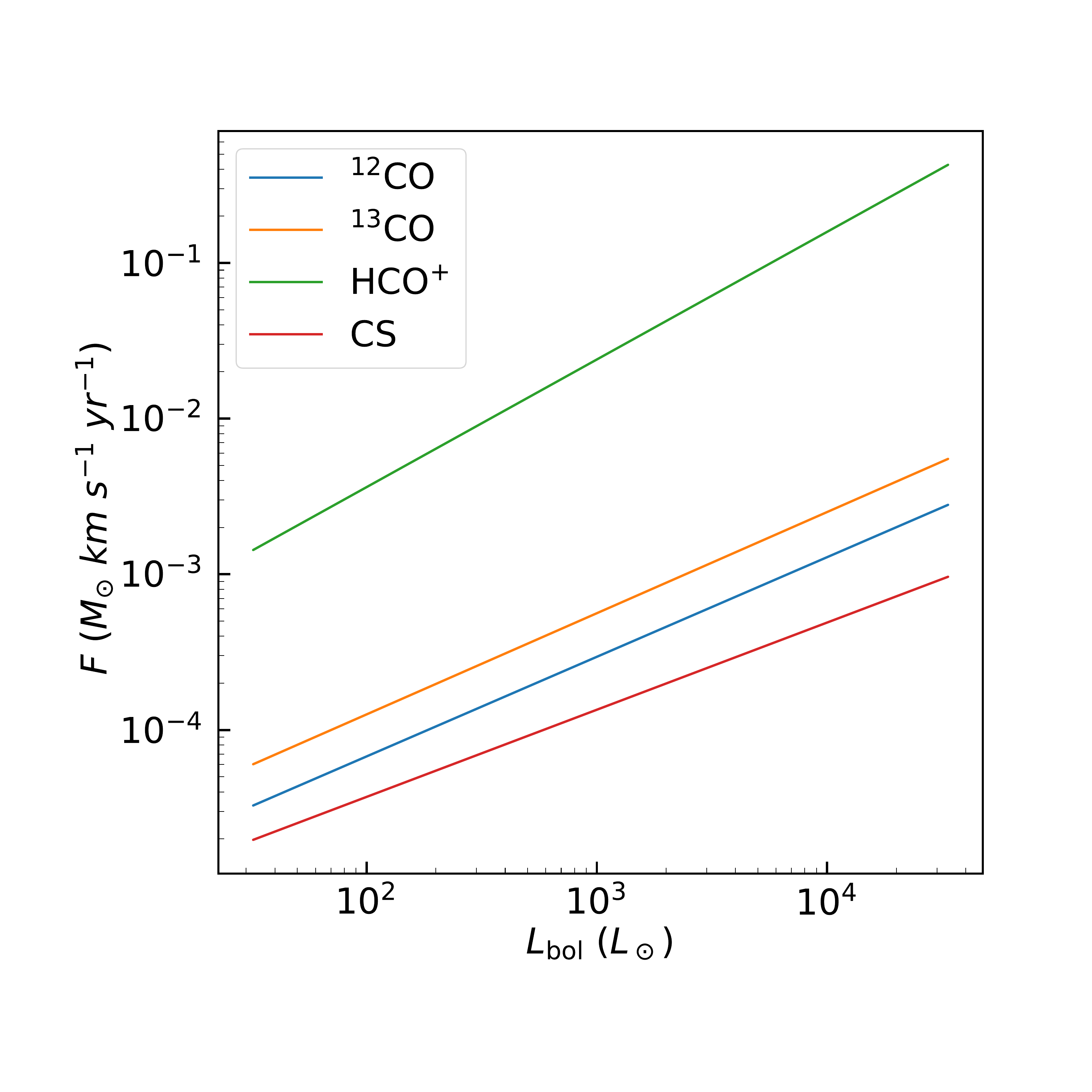}}
	\caption{The outflow force ($F$) of the different molecular outflows as a function of the bolometric luminosity ($L_{\rm bol}$) of each central IRAS source. The molecular line is given at the bottom of each panel. Panel (e) is the least-square fitting lines of the four molecular outflows.}
	\label{fig:LvsF}
\end{figure}

The outflow force, $F$, of the \coo, \xcoo, \hcoo, and CS outflows as a function of the bolometric luminosity, $L_{\rm bol}$, of the central IRAS source is as follows (see Figure~\ref{fig:LvsF}):

\begin{subequations}\label{equ:3}
	\begin{align}
	& \log F{\rm (^{12}CO)} = (0.64 \pm 0.21)\ \log L_{\rm bol} + (-5.45 \pm 0.70), r = 0.61; \label{equ:3a} \\
	& \log F{\rm (^{13}CO)} = (0.65 \pm 0.10)\ \log L_{\rm bol} + (-5.20 \pm 0.34), r = 0.86; \label{equ:3b} \\
	& \log F{\rm (HCO^{+})} = (0.82 \pm 0.17)\ \log L_{\rm bol} + (-4.08 \pm 0.58), r = 0.76; \label{equ:3c} \\
	& \log F{\rm (CS)} = (0.56 \pm 0.19)\ \log L_{\rm bol} + (-5.55 \pm 0.65), r = 0.59. \label{equ:3d}
	\end{align}
\end{subequations}

\begin{figure}
	\centering
	\subfigure[\co]{\includegraphics[width=6cm]{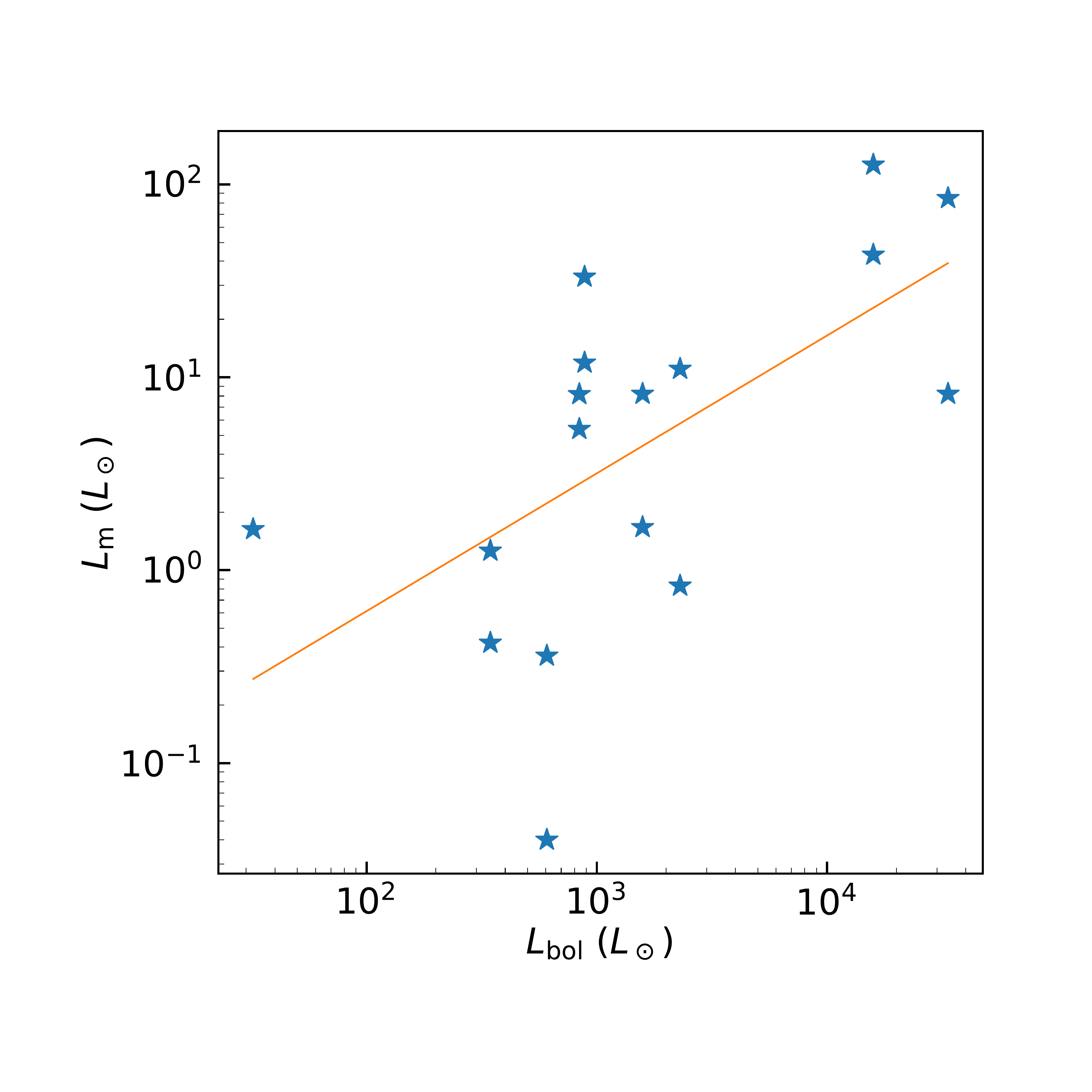}}
	\subfigure[\xco]{\includegraphics[width=6cm]{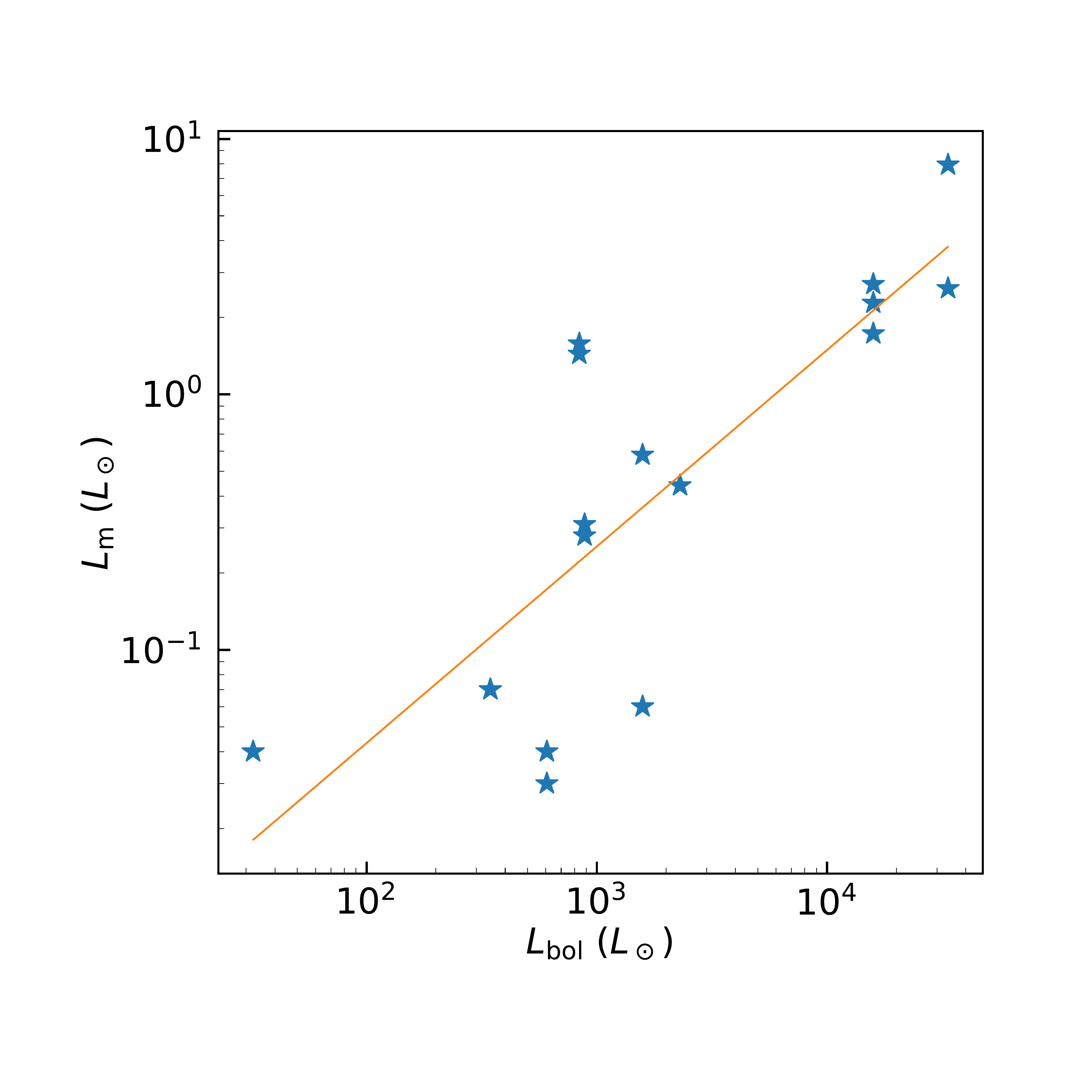}}
	\subfigure[\hco]{\includegraphics[width=6cm]{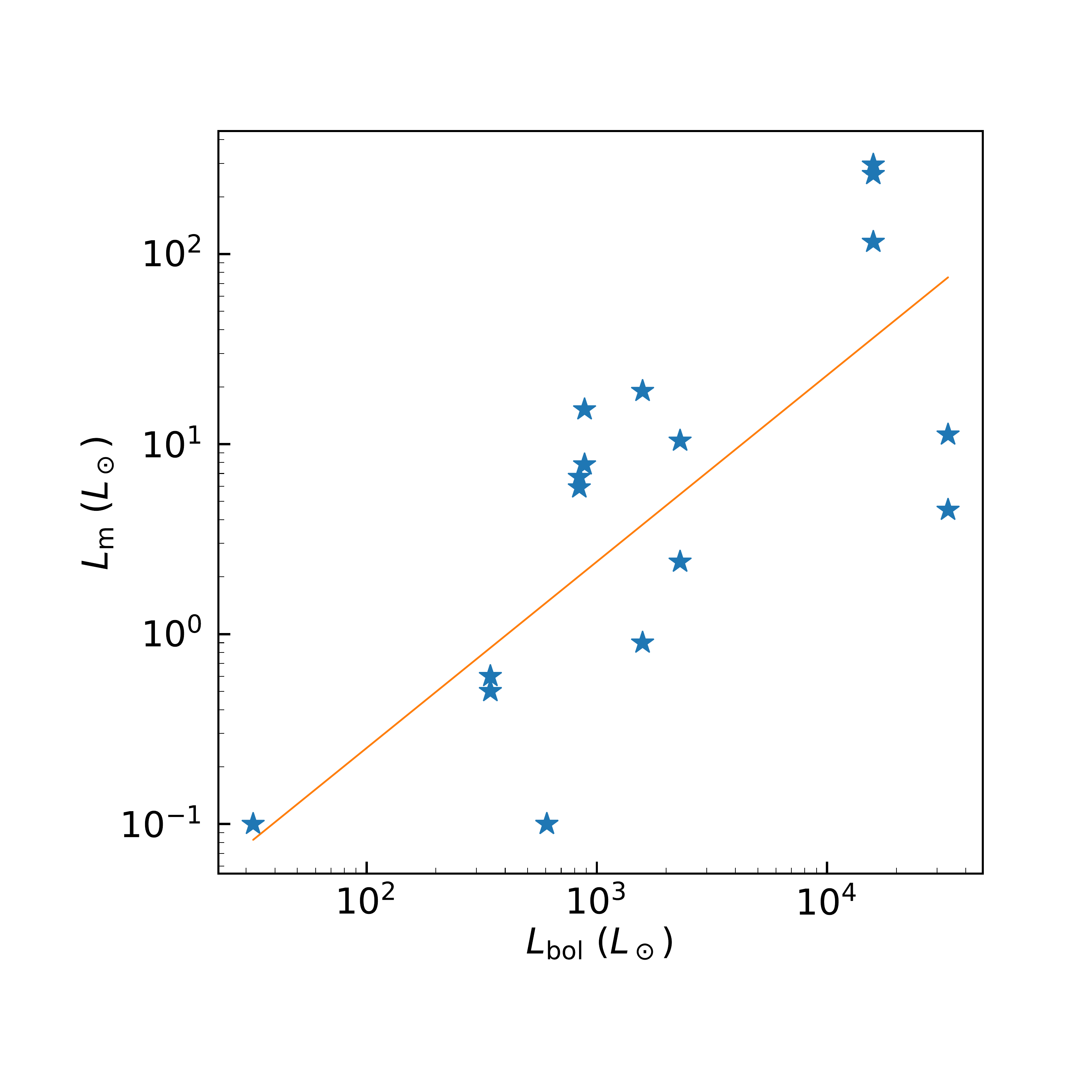}}
	\subfigure[CS]{\includegraphics[width=6cm]{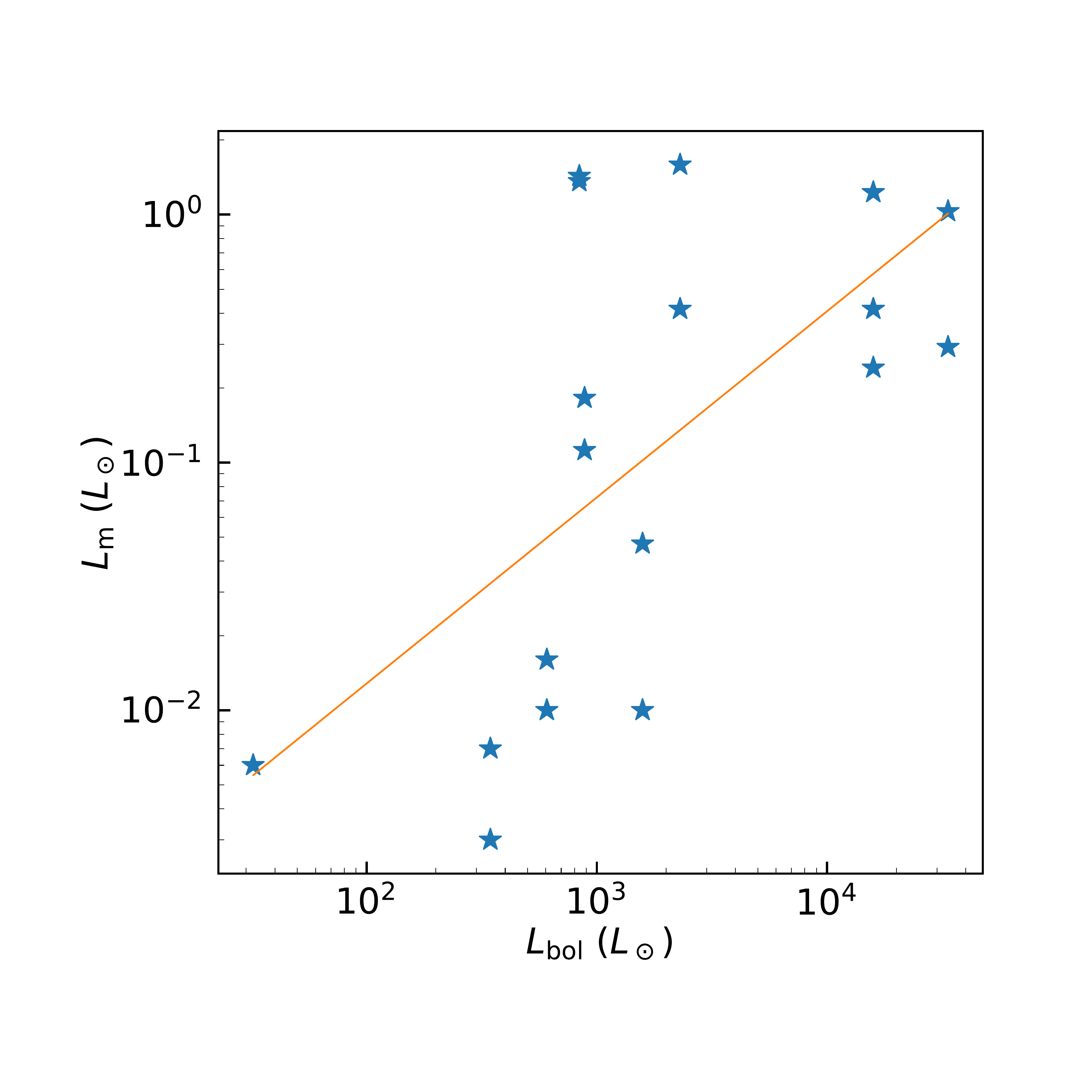}}
	\subfigure[Fitting lines]{\includegraphics[width=6cm]{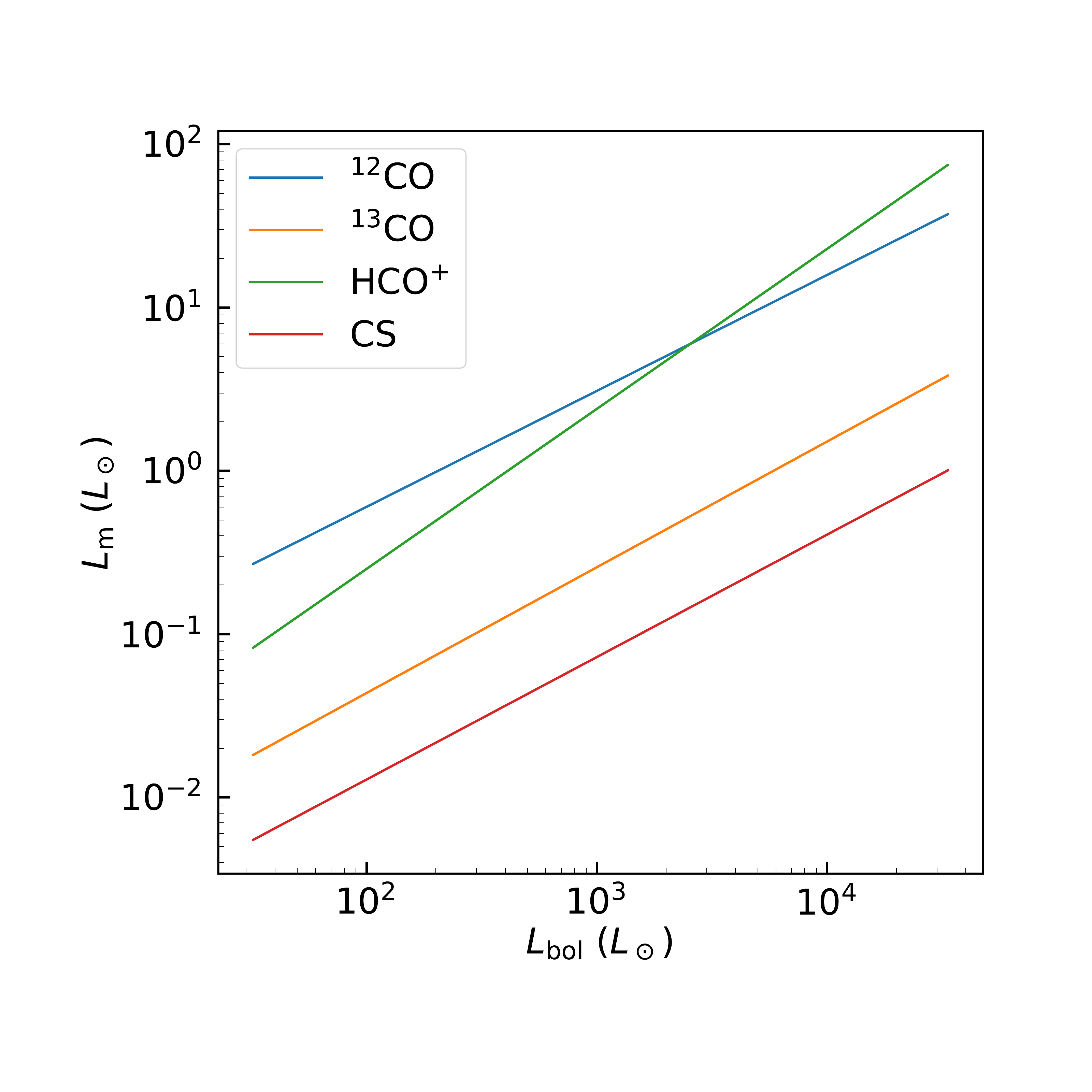}}
	\caption{The mechanical luminosity ($L_{\rm m}$) of the different molecular outflows as a function of the bolometric luminosity ($L_{\rm bol}$) of the central IRAS source. The molecular line is given at the bottom of each panel. Panel (e) is the least-square fitting lines of the four molecular outflows.}
	\label{fig:LvsL}
\end{figure}

The mechanical luminosity, $L_{\rm m}$, of the \coo, \xcoo, \hcoo, and CS outflows as a function of the bolometric luminosity, $L_{\rm bol}$, of the central IRAS source is as follows (see Figure~\ref{fig:LvsL}):

\begin{subequations}\label{equ:4}
	\begin{align}
	& \log L_{\rm m}{\rm (^{12}CO)} = (0.71 \pm 0.24)\ \log L_{\rm bol} + (-1.64 \pm 0.79), r = 0.61; \label{equ:4a} \\
	& \log L_{\rm m}{\rm (^{13}CO)} = (0.77 \pm 0.14)\ \log L_{\rm bol} + (-2.90 \pm 0.48), r = 0.81; \label{equ:4b} \\
	& \log L_{\rm m}{\rm (HCO^{+})} = (0.98 \pm 0.23)\ \log L_{\rm bol} + (-2.56 \pm 0.78), r = 0.73; \label{equ:4c} \\
	& \log L_{\rm m}{\rm (CS)} = (0.75 \pm 0.23)\ \log L_{\rm bol} + (-3.39 \pm 0.77), r = 0.63. \label{equ:4d}
	\end{align}
\end{subequations}

\begin{figure}
	\centering
	\subfigure[\co]{\includegraphics[width=6cm]{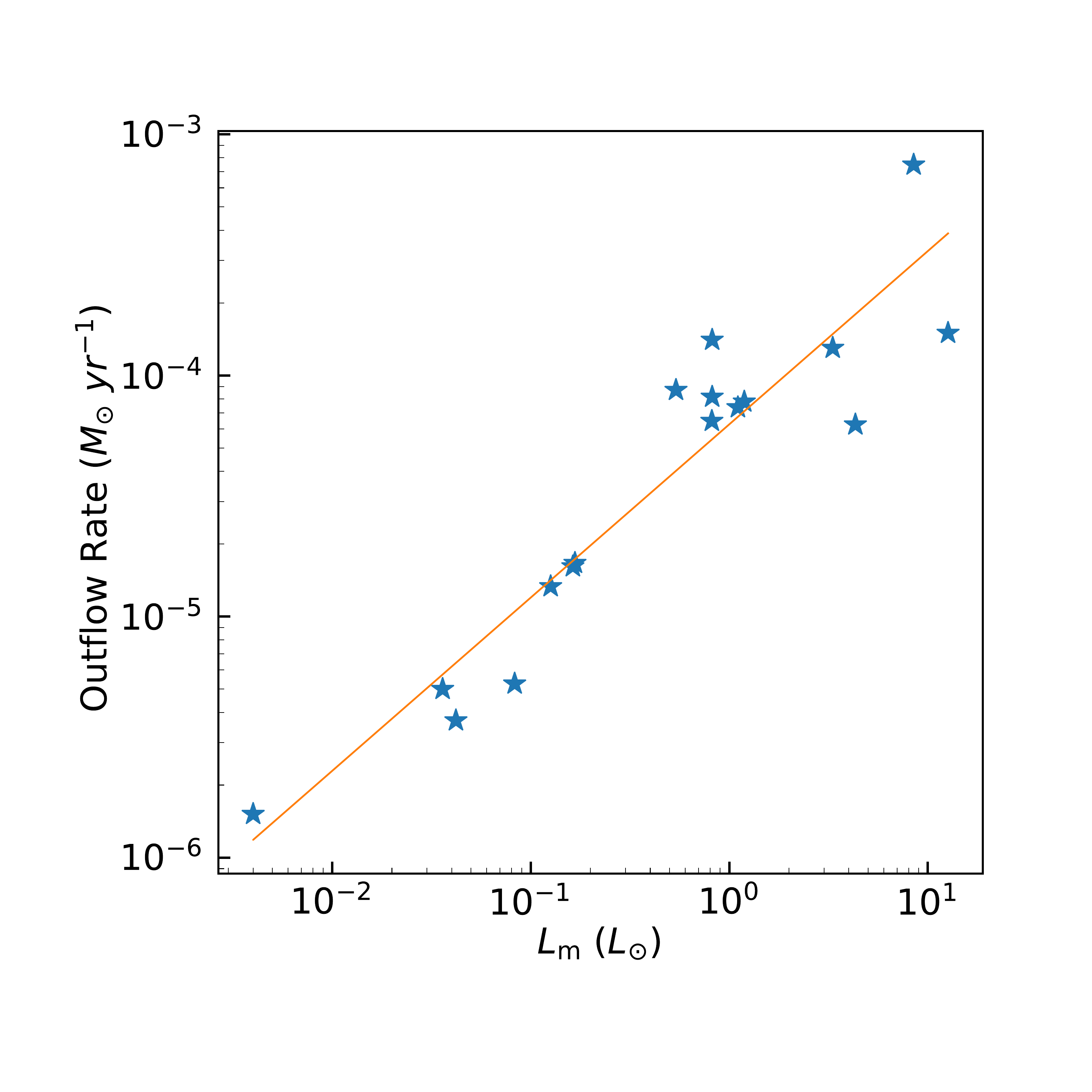}}
	\subfigure[\xco]{\includegraphics[width=6cm]{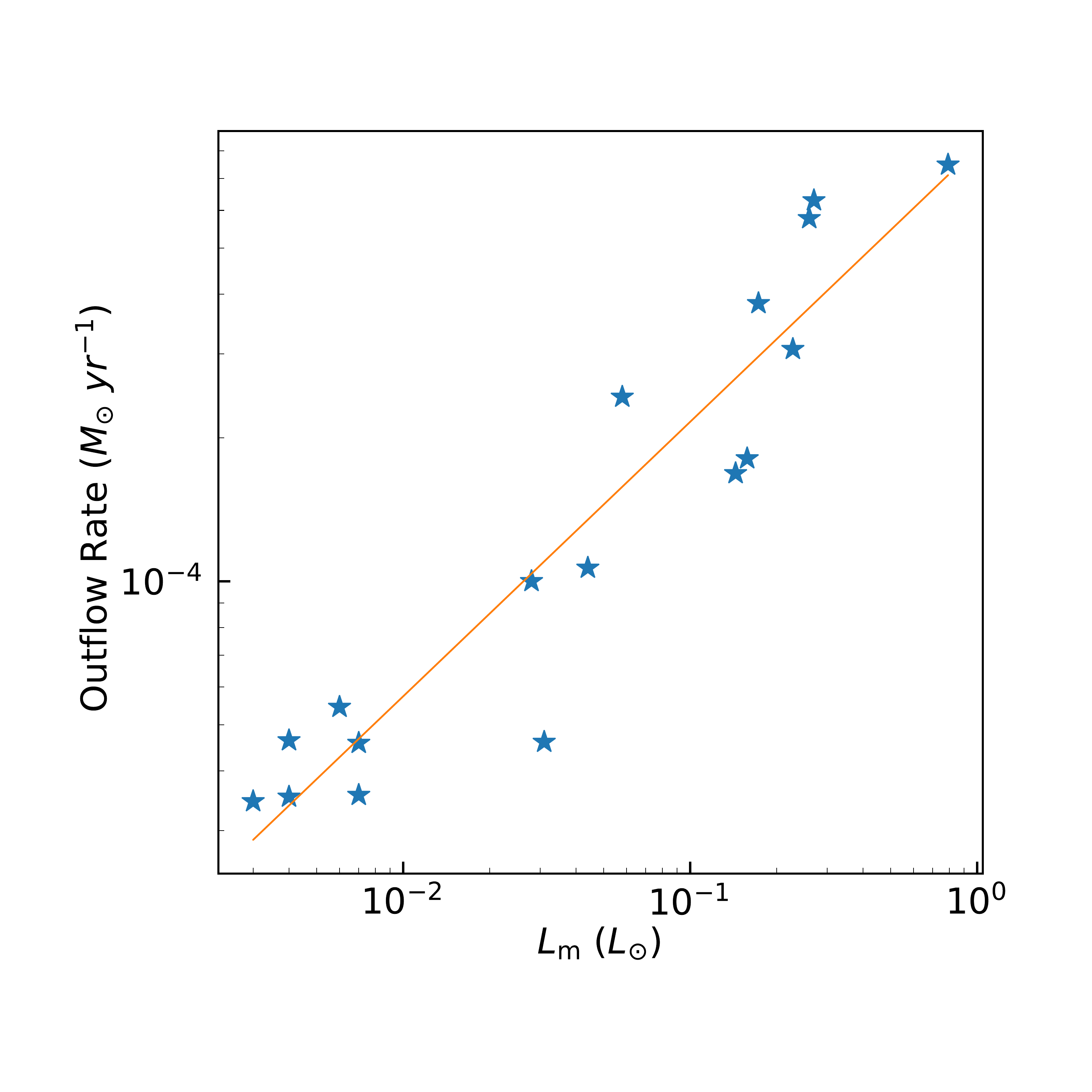}}
	\subfigure[\hco]{\includegraphics[width=6cm]{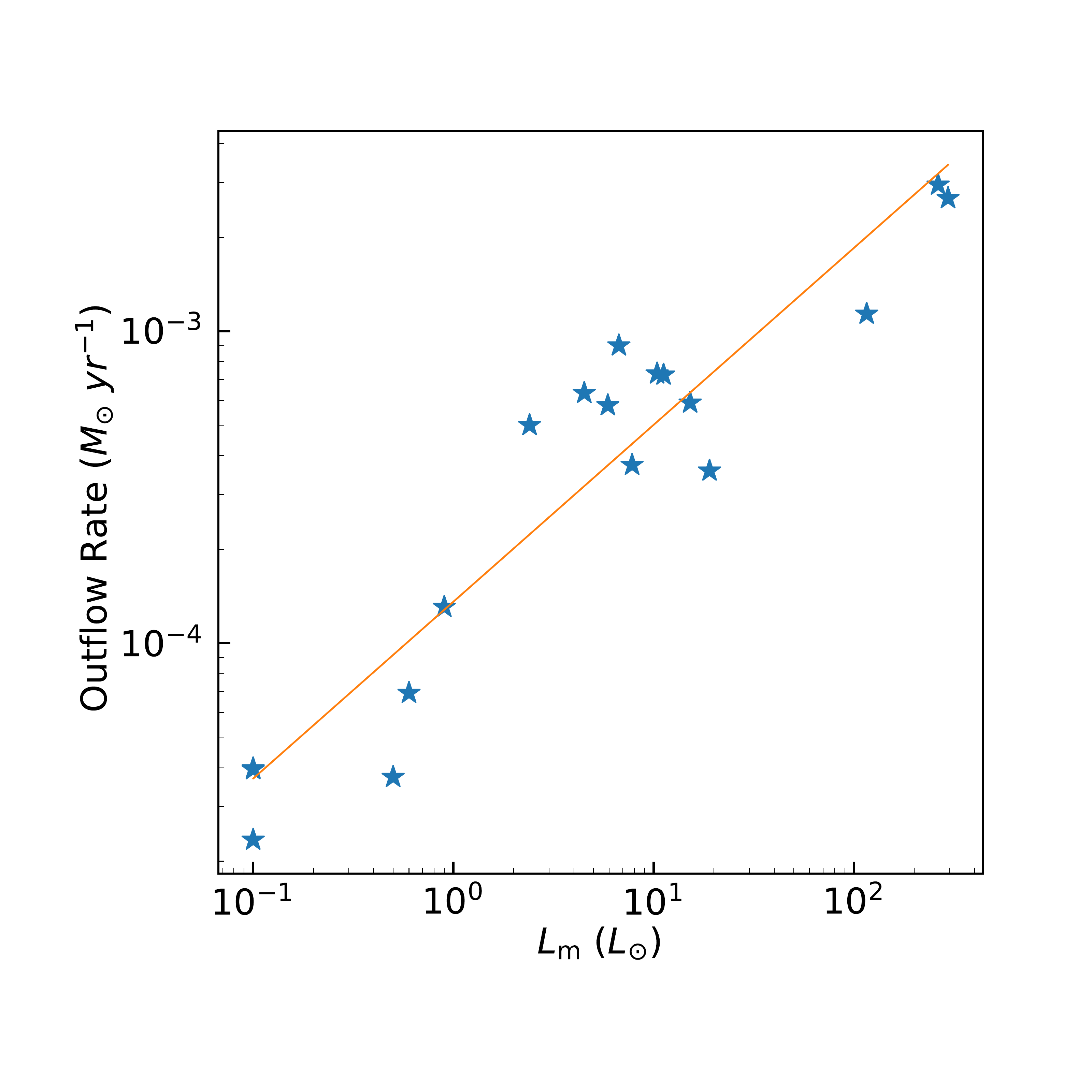}}
	\subfigure[CS]{\includegraphics[width=6cm]{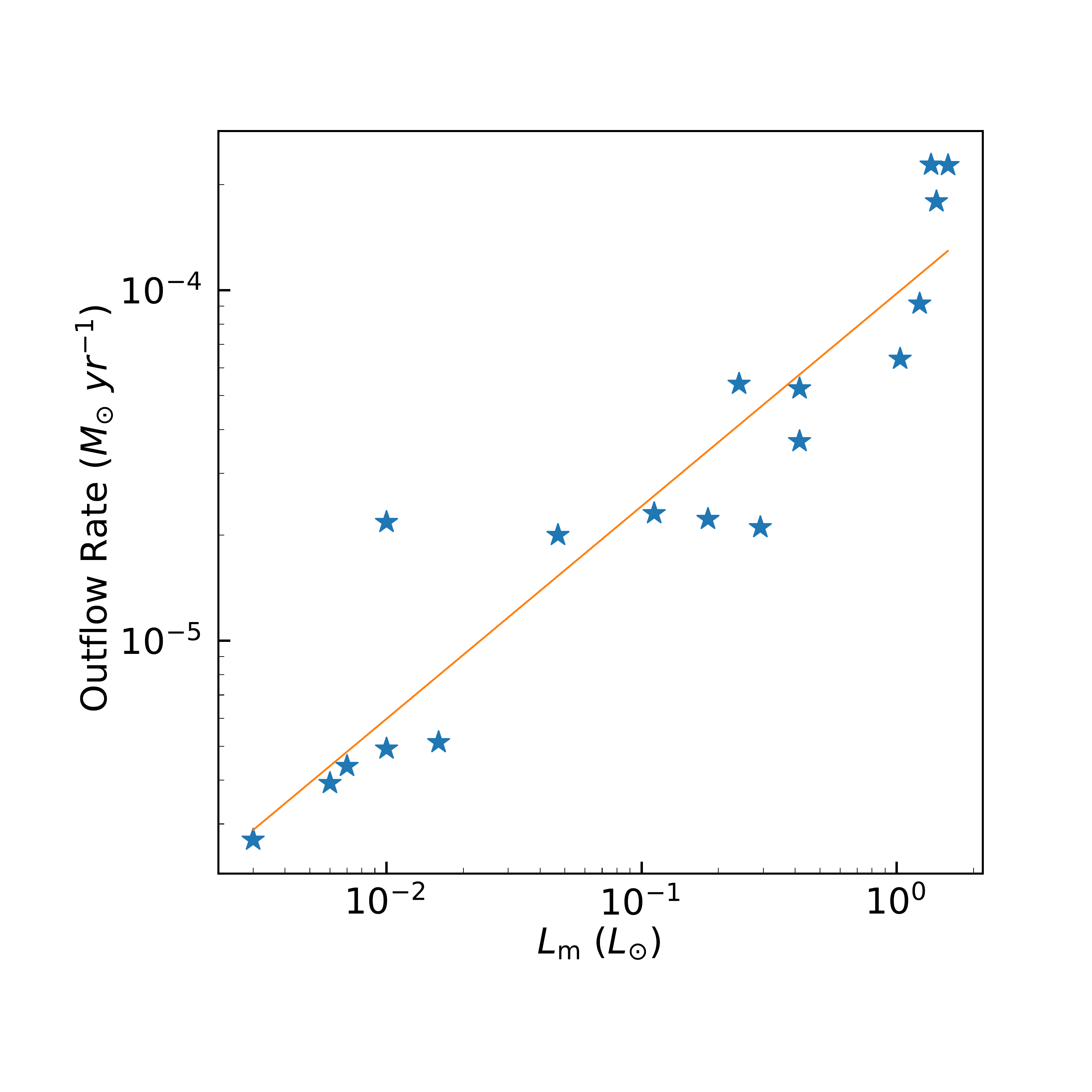}}	
	\subfigure[Fitting lines]{\includegraphics[width=6cm]{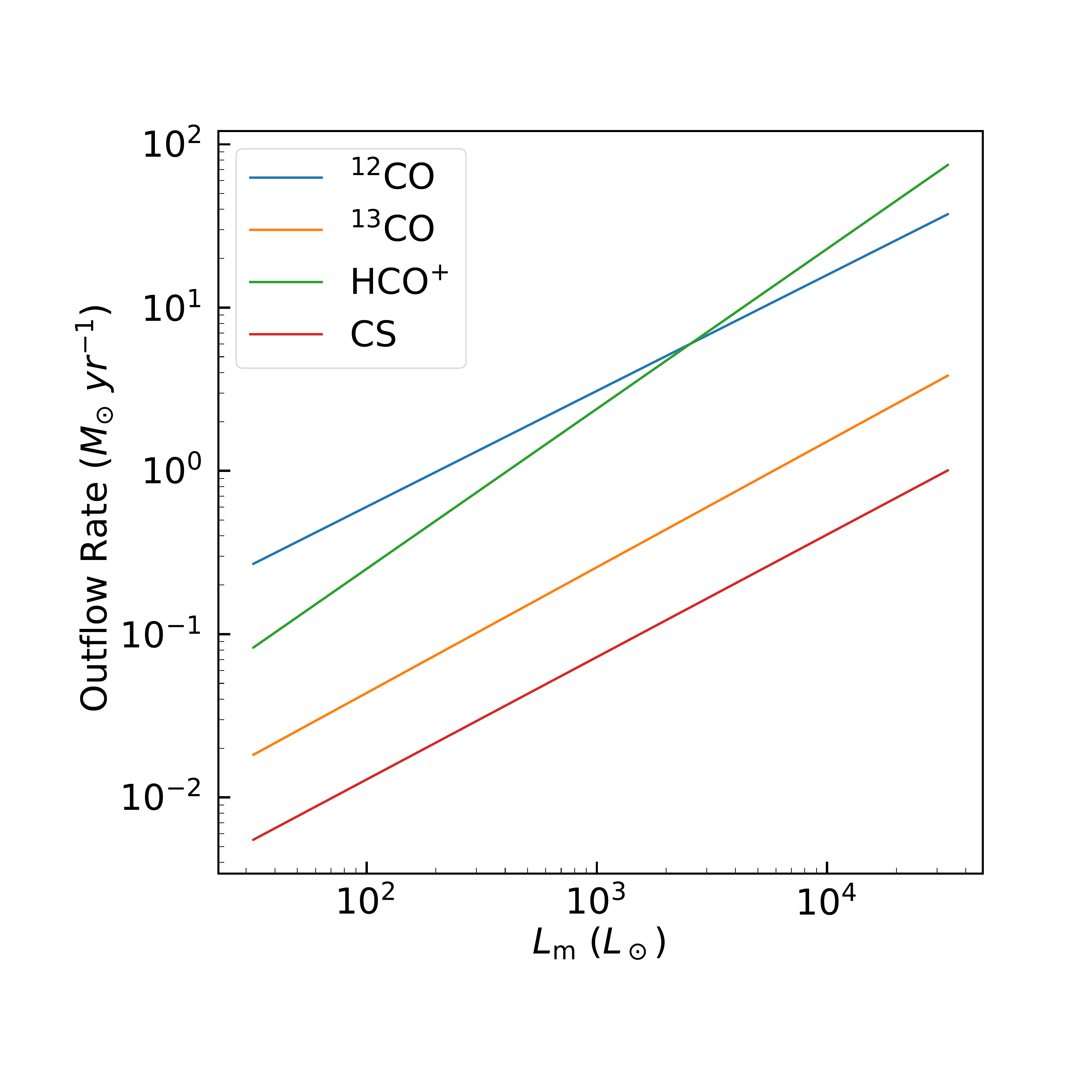}}
	\caption{The outflow rate ($\dot{M}$) of the different molecular outflows as a function of the mechanical luminosity ($L_{\rm m}$) of outflow. The molecular transition is given at the bottom of each panel. Panel (e) is the least-square fitting lines of the four molecular outflows.}
	\label{fig:LvsM_t}
\end{figure}

The mass outflow rate, $\dot{M}$, of the \coo, \xcoo, \hcoo, and CS outflows as a function of the mechanical luminosity, $L_{\rm m}$, of the outflows is as follows (see Figure~\ref{fig:LvsM_t}):

\begin{subequations}\label{equ:5}
	\begin{align}
	& \log \dot{M}{\rm (^{12}CO)} = (0.72 \pm 0.07)\ \log L_{\rm m} + (-4.20 \pm 0.07), r = 0.93; \label{equ:5a} \\
	& \log \dot{M}{\rm (^{13}CO)} = (0.58 \pm 0.05)\ \log L_{\rm m} + (-3.09 \pm 0.08), r = 0.94; \label{equ:5b} \\
	& \log \dot{M}{\rm (HCO^{+})} = (0.57 \pm 0.05)\ \log L_{\rm m} + (-3.87 \pm 0.06), r = 0.94; \label{equ:5c} \\
	& \log \dot{M}{\rm (CS)} = (0.61 \pm 0.06)\ \log L_{\rm m} + (-4.01\pm 0.08), r = 0.94. \label{equ:5d}
	\end{align}
\end{subequations}

\clearpage

\bibliography{sample63}{}
\bibliographystyle{aasjournal}


\end{CJK*}
\end{document}